\documentclass{article}
\usepackage{graphicx}
\usepackage{amsmath}
\usepackage{amssymb}
\usepackage{amsfonts}
\usepackage{amsthm}
\usepackage{color}
\usepackage{cases}
\usepackage{setspace}
\usepackage{fullpage}

\def\theequation{{\arabic{section}}.{\arabic{equation}}}

\DeclareMathAlphabet{\itbf}{OML}{cmm}{b}{it}

\newcommand{\mb}[1]{\mathbf{#1}}
\usepackage{hyperref}

\usepackage{subcaption}

\usepackage{algorithm}
\usepackage{algorithmic}

\newcommand{\nc}{\newcommand}
\nc{\dsp}{\displaystyle}
\nc{\R}{\Bbb{R}}
\nc{\Z}{\Bbb{Z}}
\nc{\Pp}{\Bbb{P}}
\nc{\Ap}{\Bbb{A}}
\nc{\Wp}{\Bbb{W}}
 \nc{\brho}{\boldsymbol \rho}
\nc{\va}{\vec{\boldsymbol \mu}}
\nc{\ve}{\vec{\boldsymbol \epsilon}}
\nc{\bk}{{\bf k}}
\nc{\vrho}{\vec{\brho}} 
\nc{\vr}{\vec{\bf r}}
\nc{\bx}{{\bf x}}
\nc{\vx}{\vec{\bf x}}
\nc{\om}{\omega}
\nc{\brhoi}{\brho^{\cal I}}
\nc{\bzetai}{\bzeta^{\cal I}}
\nc{\vzeta}{\vec{\bzeta}}
\nc{\vzetai}{\vec{\bzeta}^{\cal I}}
\nc{\cI}{{\cal I}}
\nc{\cJ}{{\cal J}}
\nc{\cF}{{\cal F}}
\nc{\cW}{{\cal W}}
\nc{\cA}{{\cal A}}
\nc{\cL}{{\cal L}}
\nc{\cS}{{\cal S}}
\nc{\cC}{{\cal C}}
\nc{\cN}{{\cal N}}
\nc{\vrhoi}{\vec{\brho}^{\,\cal I}}
\nc{\xii}{\xi^{\cI}}
\nc{\etai}{\eta^{\cI}}
\nc{\la}{\lambda}
\nc{\de}{\delta}
\nc{\ep}{\varepsilon}
\nc{\vu}{\vec{\bf u}}
\nc{\bu}{{\bf u}}
\nc{\vui}{\vec{\bf u}^{\cI}}
\nc{\bui}{{\bf u}^{\cI}}
\nc{\bt}{{\bf t}}
\nc{\vt}{\vec{\bt}}
\nc{\bn}{{\bf n}}
\nc{\vn}{\vec{\bn}}
\nc{\bm}{{\bf m}}
\nc{\vm}{\vec{\bm}}
\nc{\vrp}{\vec{{\bf r}'}}
\nc{\vrc}{\vec{{\bf r}^c_p}}
\nc{\ts}{\tilde s}
\nc{\os}{\overline s}
\nc{\tom}{\tilde \om}
\nc{\tO}{\tilde \Omega}
\nc{\tS}{\tilde S}
\nc{\oS}{\overline S}
\nc{\vrhos}{\vrho_{\star}}
\nc{\vrhosi}{\vrho_{\star}^{\cI}}
\nc{\brhos}{\brho_{\star}}
\nc{\brhosi}{\brhos^{\cI}}
\nc{\vms}{\vm_{\star}}
\nc{\vmi}{\vm_{_{\cI}}}
\nc{\vM}{\vec{\bf M}}
\nc{\vMi}{\vM_{_{\cI}}}
\nc{\Ppi}{\Pp_{\cI}}
\nc{\vxi}{\vec{\bxi}}
\nc{\bK}{{\bf K}}
\nc{\bmi}{\bm_{_{\cI}}}
\nc{\Pppi}{\Ppi^p}
\nc{\be}{{\bf e}}
\nc{\bep}{{\bf e}^p}
\renewcommand{\hat}{\widehat}

\begin{document}
	\title{Correlation based Imaging for rotating satellites}\author{Matan Leibovich\footnotemark[1], George Papanicolaou\footnotemark[2], and Chrysoula Tsogka\footnotemark[3]
	}\renewcommand{\thefootnote}{\fnsymbol{footnote}}
	\footnotetext[2]{Institute for Computational and Mathematical Engineering, Stanford University, Stanford, CA 94305. \\\   (matanle@stanford.edu)} 
	\footnotetext[2]{Department of
		Mathematics, Stanford University, Stanford, CA 94305.
		(papanicolalou@stanford.edu)}\footnotetext[3]{Department of Applied Mathematics,
		University of California, Merced, 5200 North Lake Road, Merced, CA
		95343 (ctsogka@ucmerced.edu)}\date{}
	\maketitle  \date{}
	
	% ------------------------------
	\begin{abstract}
		We consider imaging of fast moving small objects in space, such as low earth orbit satellites, which are also rotating around a fixed axis. The imaging system consists of ground based, asynchronous sources of radiation and several passive receivers above the dense atmosphere. We use the cross-correlation of the received signals to reduce distortions from ambient medium fluctuations. Imaging with correlations also has the advantage of not requiring any knowledge about the probing pulse and depends weakly on the emitter positions. We account for the target's orbital velocity by introducing the necessary Doppler compensation. To image a fast rotating object we also need to compensate for the rotation. We show that the rotation parameters can be extracted directly from the auto-correlation of the data before the formation of the image. We then investigate and analyze an imaging method that relies on backpropagating the cross-correlation data structure to two points rather than one, thus forming an interference matrix. The proposed imaging method consists of estimating the reflectivity as the top eigenvector of the migrated cross-correlation data interference matrix. We call this the rank-1 image and show that it provides superior image resolution compared to the usual single-point migration scheme for fast moving and rotating objects. Moreover, we observe a significant improvement in resolution due to the rotation leading to a diffraction limited resolution. We carry out a theoretical analysis that illustrates the role of the two point migration method as well as that of the inverse aperture and rotation in improving resolution. Extensive numerical simulations support the theoretical results.
	\end{abstract}

\section{Introduction}
\label{sec:intro}
There is a growing need for reliably imaging and tracking objects in low earth orbit (LEO). This is the result of the proliferation of satellites operating in those altitudes (200$km$-2000$km$) \cite{sandia}. Following satellite collisions, and satellites impacted by other space debris, the number of debris has dramatically increased in recent years\cite{amos15}, raising concerns that without efficient tools to track debris and mitigate their effect, certain parts of LEO would cease to be operational \cite{wormnes, kessler1978collision}.

\begin{figure}[htbp]
	\centering
	\begin{subfigure}[t]{0.25\textwidth}
		\includegraphics[width=\textwidth]{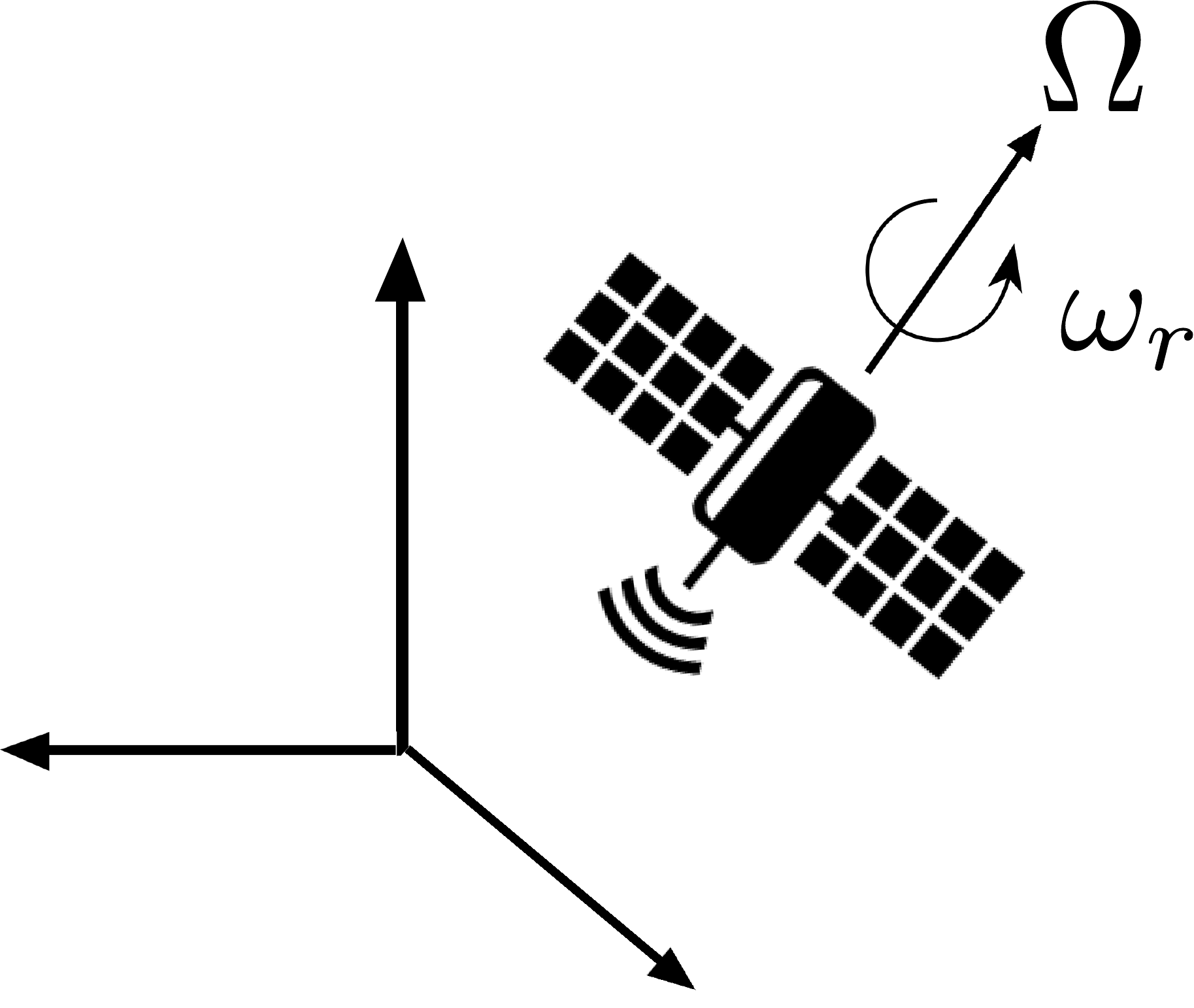}
		\caption{ }
		\label{fig:layout_sat}
	\end{subfigure}
	\begin{subfigure}[t]{0.24\textwidth}
		\includegraphics[width=1\textwidth]{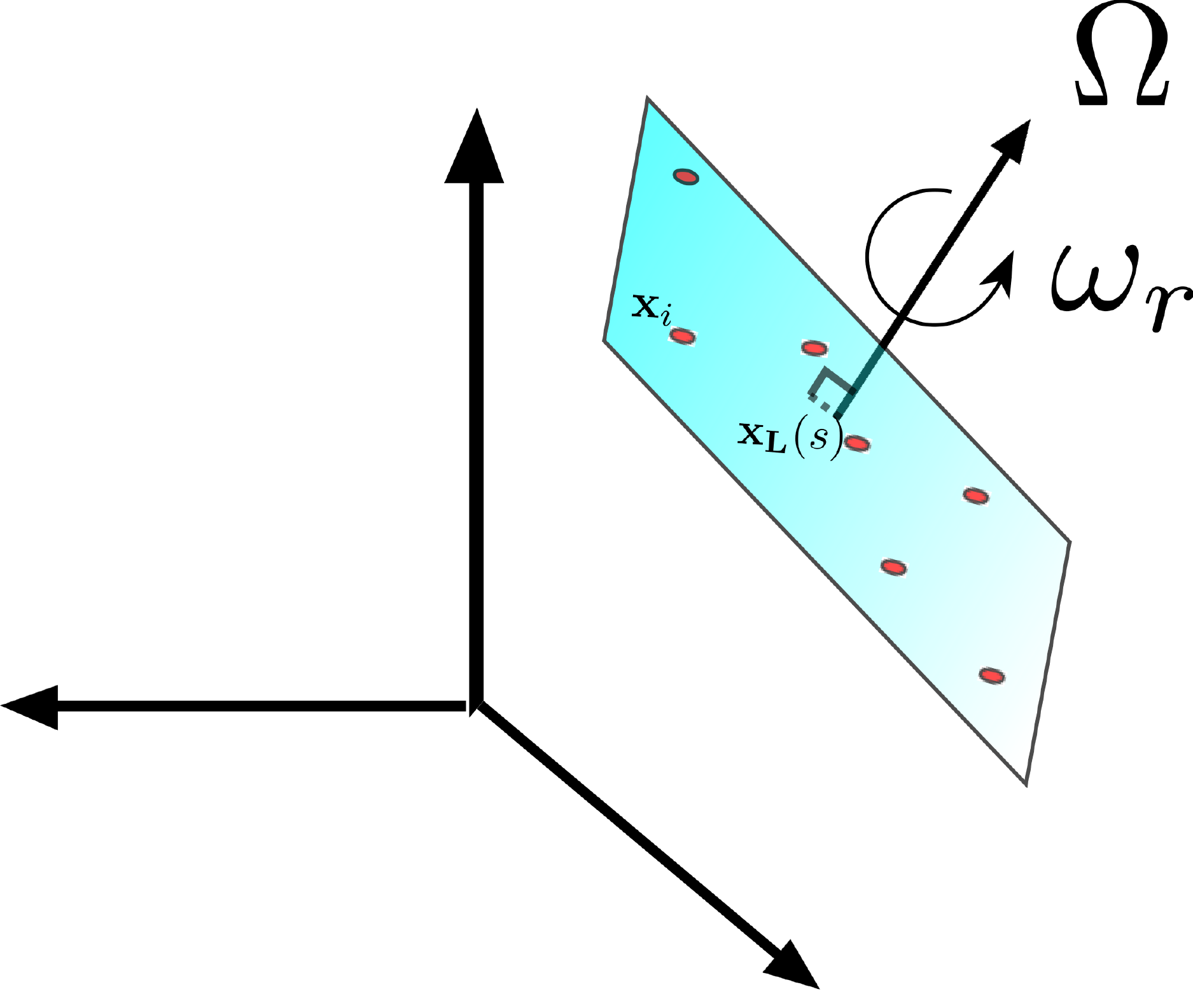}
		\caption{ }
		\label{fig:layout_sat_2}
	\end{subfigure}
	\caption{$(a)$ The object to be imaged is rotating with an angular velocity $\omega_r$ with respect to a direction $\Omega=(\theta_{\text{rot}},\phi_{\text{rot}})$. $(b)$ The object is modeled as a collection of closely spaced point scatterers rotating in a plane \textcolor{black}{ perpendicular to the axis of rotation.}}
	\label{fig:layout_rot}
\end{figure}

In \cite{leibovich2020generalized}, we proposed a method for imaging small fast moving objects in orbit. In this paper,  we consider a more complex scenario, where the objects are bigger and they can also rotate. We model the fast moving debris as a cluster of point-like reflectors moving with constant velocity, $\mb v_\mb T$, as well as having a rotational velocity $\omega_r$, with respect to a fixed axis with a solid angle $\Omega$ with respect to the $\hat{z}$ axis as illustrated in Figure~\ref{fig:layout_rot}. \textcolor{black}{To simplify the analysis, we assume that the scatterers are within a single plane, perpendicular to the axis of rotation. This can be easily generalized to a three dimensional rigid body.}

	The data used for imaging are the signals scattered from the targets, induced by a ground based transmitter. The receivers are assumed to be at a height of 15$km$ above the ground, with positions uniformly distributed on an area with diameter $a$, which defines the physical aperture of the imaging system. We take $a$ to be approximately 200$km$, which provides high cross range resolution for LEO objects. Taking advantage of the targets' motion, the different scattered signals are synthesized to improve resolution, generating an inverse synthetic aperture radar (iSAR) configuration. The transmitted signals are assumed to have carrier frequency $f_o$ and bandwidth $B$, as well as a known pulse repetition frequency. We focus in our analysis on sources operating at X-band (8-12 GHz), with a relatively high bandwidth (600 MHz). \textcolor{black}{For simplicity, we assume that the receiver array is planar and that the target has a linear speed, rather than angular. This is a reasonable assumption since the size of the array is much smaller than the Earth's radius.}%, and this factors would only affect small changes in the travel time model}

	Rather than using the data directly for imaging we first cross-correlate the echoes received at different pairs of receivers and use these cross-correlations as the input of the imaging function. Imaging using cross-correlations has several advantages. First, while the pairs of receivers need to be synchronized with respect to each other, they do not need to be synchronized with the emitter. Conditioned on a high enough sampling rate \cite{garnier17}, this alleviates most of the dependence on the emitter location in the imaging function, and allows the use of opportunistic sources whose direct profile is unknown, such as global navigation satellite systems (GNSS) \cite{amos16_leo}. Second, correlation data is more robust to medium fluctuations \cite{lawrence,mcmillan}, which distort the travel time. 
	For receivers located above the turbulent atmosphere, correlation mitigates the effect of the medium fluctuations. 
	This allows to get a stable image, considering an effective unperturbed medium \cite{garnier12a,garnier16,virtual2}. 

Correlation based imaging for satellites was considered in \cite{fournier2017matched}. The imaging function that was proposed and analyzed is a generalization of Kirchhoff migration (KM), which forms an image by superposing coherently the recorded signals after translating them by the travel time to a single search point in the image window. In \cite{leibovich2020generalized}, we presented a generalization of this migration method for correlations, which is motivated by the structure of the correlation data. The result of the generalized migration is not an image but rather a matrix \textbf{two-point interference pattern}, attained by migrating the data to two separate search points. We showed that there are several possible ways to derive an image from the matrix two-point interference pattern, with one in particular that we call the \textbf{rank-1 image}. This rank-1 image is the result of taking the first eigenvector of the matrix interference pattern. It provides superior resolution compared to the single point migration function, which was demonstrated both analytically and with numerical simulations. 
This generalization was also considered in interferometric synthetic aperture imaging of random media \cite{borcea2019twopoint}.

In this paper we consider the more complex scenario in which, in addition to linear motion, the object also rotates with respect to a fixed axis. 
The rotational periods of objects in LEO can vary greatly in duration from a few seconds to hundreds of seconds \cite{yanagisawa2012shape},\cite{choi2016determining},\cite{vsilha2018apparent}. Targets in orbit can have a rotation period of 5 to 10 seconds. Debris tend to decelerate and they do not rotate as fast.  When the target is rotating slowly, its aspect does not change greatly during a single pass through the area spanned by the receiving apertures, and we can ignore that rotation. Rapid rotation of the object during the acquisition time requires the appropriate registration of the data collected. An important step is, therefore, the estimation of the rotational parameters as a prerequisite for imaging  \cite{solodyna_MIT}. 
In this paper we introduce an imaging method using the correlation data of ground receivers for rapidly rotating objects. We assume the object rotates around a fixed axis, without precession. This is a reasonable assumption for inverse synthetic apertures on the order of a few periods of the object's rotation. For $a\approx 200$ km, the effective synthetic aperture where objects in LEO are visible is $\approx$400$km$, or $\approx$ 60 seconds. We show that the rotational degrees of freedom can be extracted from the data, and introduce a method to estimate the rotational degrees of freedom of the object from the autocorrelation of the receivers. We then show, by simulations and by analysis of the structure of the rank-1 image, that when introducing rotation the achieved resolution can be dramatically improved when using the rank-1 image. Thus rotation, while generating more complex data,  provides diversity which can improve the resolution when properly considered.

The rest of the paper is as follows. In Section~\ref{sec:corr_imaging} we briefly review the model for our data, and introduce the cross-correlation data structure. We also present the algorithm for rotation parameter estimation, expanded in more detail in Appendix~\ref{app:rot_est}. In Section~\ref{sec:generalized_migration} we review the results of \cite{leibovich2020generalized}, and introduce the rank-1 image for rotating objects. In Section~\ref{sec:numerical_simulations} we present numerical simulations which confirm the superior performance of the rank-1 image. We motivate the results by further simulations, as well as analysis in Appendix~\ref{app:stat_phase}. We conclude with a summary and conclusions in Section~\ref{sec:summary}.
\section{Migration imaging and rotation}
\label{sec:corr_imaging}
In this section we present the general form of the recorded signals and then review briefly the basic algorithms of correlation based imaging accounting for Doppler compensation. We present the cross-correlation data structure introduced in \cite{leibovich2020generalized},  that uses a constant Doppler factor over limited imaging regions, and is more efficient for imaging. We then discuss the estimation of rotation parameters, detailed in Appendix~\ref{app:rot_est}.
\subsection{The forward problem: scattering from fast moving objects}
In this problem we are looking to image an object rotating as a rigid body. We model the object as a cluster of point scatterers; $\bx_L$, is located on the axis of rotation and represents the scattering from the bulk of the object that is insensitive to rotation, and $N_{T}$ rotating point scatterers $\bx_{\mathcal R}^i, i=1,...,N_T$, representing glints, or sharp reflectors at the object's extremities (e.g. satellite solar panels). 

\begin{figure}[htbp]
	\centering
	\begin{subfigure}[t]{0.4\textwidth}
		\includegraphics[width=\textwidth]{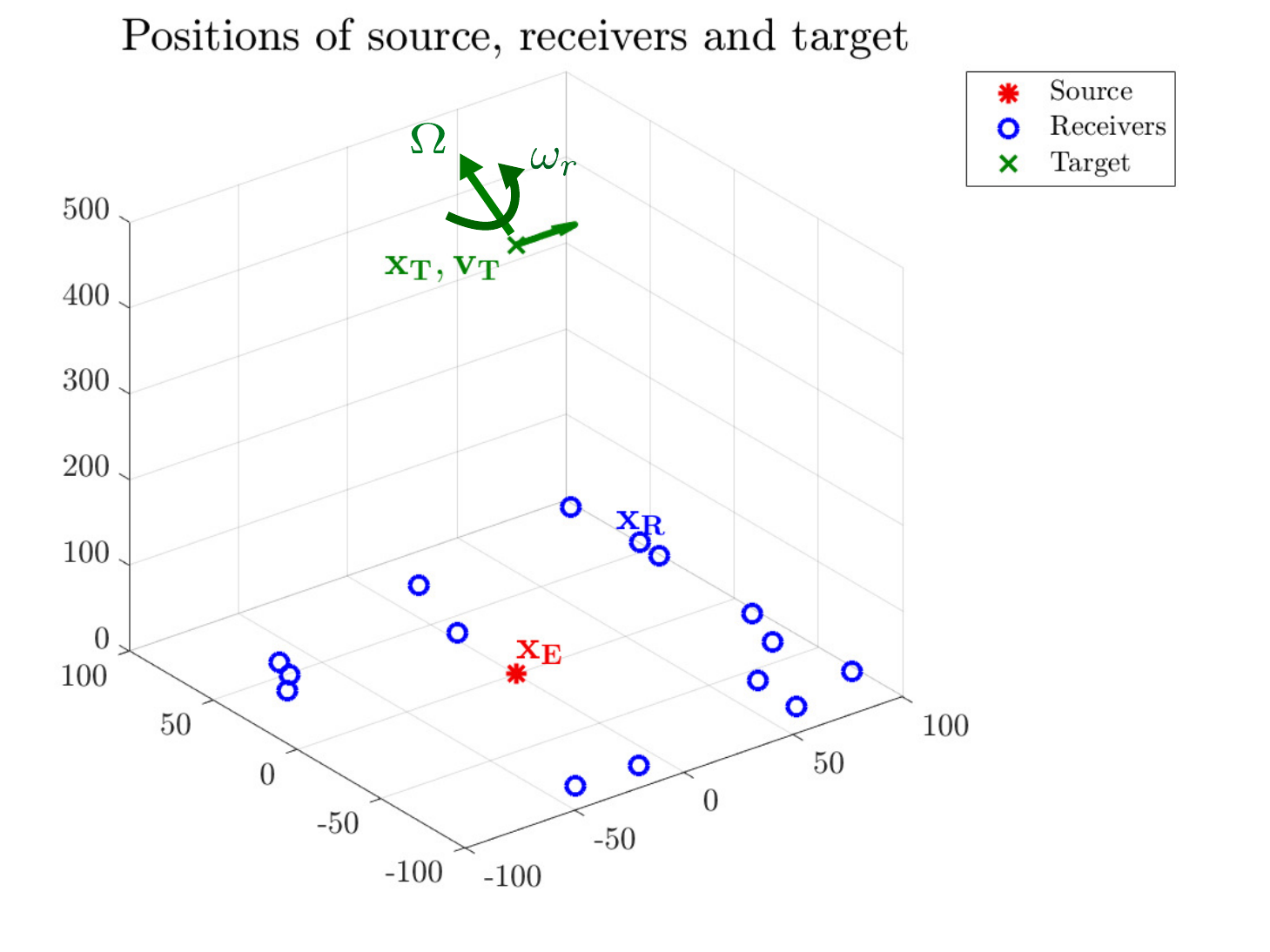}
		\caption{}
			\end{subfigure}
			\begin{subfigure}[t]{0.25\textwidth}
			\includegraphics[width=\textwidth]{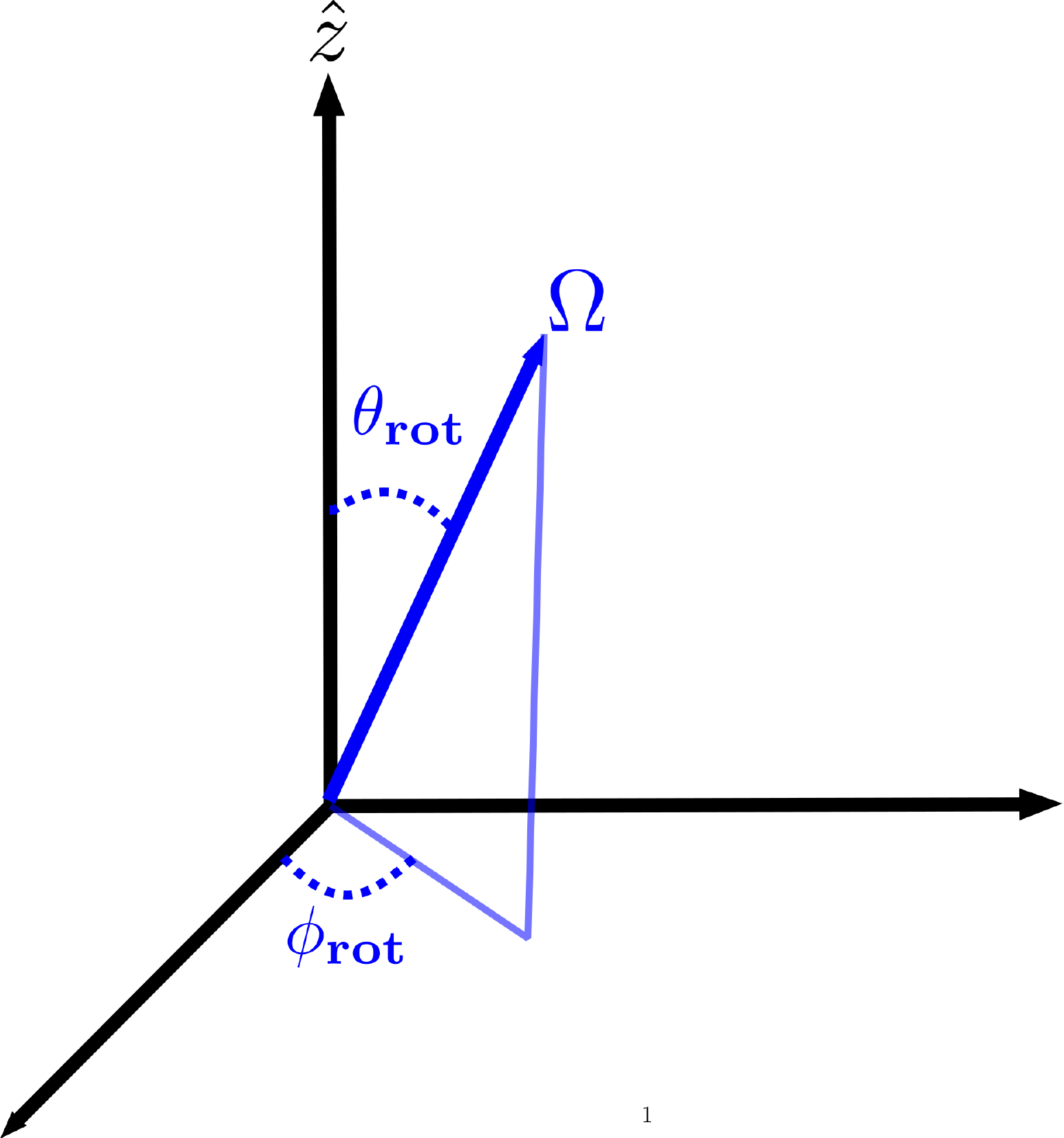}
\caption{ }
	\end{subfigure}
\caption{$(a)$ iSAR imaging schematic. We assume a network of 15 receivers, randomly distributed over an area of \textcolor{black}{200$\times$ 200 km}.  The receivers are located at 15 km height, and the target is at 500 km, moving at $7$ km/s. $(b)$ The solid angles $\theta_{\text{rot}},\phi_{\text{rot}}$ define the direction of the axis of rotation with respect to the $z$ axis in the frame of reference.}
\label{fig:layout}
\end{figure}

We assume an object whose motion is comprised of both a linear velocity and rotation, such that
\begin{equation}
\bx^i_\mb T(s)=\mb x_\mb L(s)+\mb x_i^{\mathcal{R}}(s),\quad\mb x_\mb L(s)=\bx_\mb T+s\mb v_\mb T,\quad  \bx^{\mathcal{R}} _i(s)=\mathcal{R}(s)\mb x_i. 
\label{eq:point_scatter}
\end{equation}
\textcolor{black}{Here $\mb x_\mb L(s)$ is moving on a linear trajectory with velocity $\mb v_\mb T$, its position at slow time $s=0$ is $\bx_\mb T$. 
The rotation of the point scatterers, $\bx^{\mathcal{R}} _i(s)$, is modeled with the rotation matrix $\mathcal{R}(s)$ 
\begin{equation}
\label{eq:rot_op}
\mathcal{R}(s)=\mathcal{R}_{\Omega}(\phi_{\text{rot}},\theta_{\text{rot}})  \mathcal{R}_{z}(\omega_r s),
\end{equation}
with $\theta_{\text{rot}}\in[0,\pi],\phi_{\text{rot}}\in[0,2\pi]$ the solid angles that define the direction of the axis of rotation with respect to the $z$ axis and  
$\omega_r \ge0$ the counter-clockwise angular velocity (see Figure \ref {fig:layout}). The matrix $\mathcal{R}_{z}(\omega_r s)$, describes the rotation 
about the z-axis through an angle determined by the angular velocity and the elapsed time 
\begin{equation}
\mathcal{R}_{z}(\varphi)=\begin{pmatrix} \cos\varphi&-\sin\varphi&0 \\\sin\varphi&\cos\varphi&0\\0&0&1\end{pmatrix}. 
\end{equation}
and $\mathcal R_{\Omega}(\phi_{\text{rot}},\theta_{\text{rot}})$, takes the frame of reference back to the $xyz$ coordinates
\begin{equation}
\mathcal{R}_{\Omega}(\phi,\theta)=\begin{pmatrix} \cos\phi&-\sin\phi&0 \\\sin\phi&\cos\phi&0\\0&0&1\end{pmatrix} 
\begin{pmatrix}\cos\theta&0&-\sin\theta\\ 0&1&0\\ \sin\theta&0&\cos\theta \end{pmatrix}
\label{eq:rot_mat}
\end{equation}
}

The data is the collection of signals recorded at ground based or at low elevation receivers, with positions $\mb x_\mb R$. Successive pulses are emitted at a slow time $s$ by a source located at  $\mb x_\mathbf{E}$ on the ground,  as illustrated in Figure~\ref{fig:layout}. The receiver emits a series of pulses $f(t)=\cos(\omega_o t)e^{-B^2t^2/2}$, at slow time intervals of $\Delta s$, with a total aperture size $S$, such that the recorded signal at the receiver location $\mb x_\mb R$ due to a pulse, $f(s+t)$, emitted at slow time $s\in[-S/2,S/2]$, is  
\begin{equation}
u_{\mathbf{R}}(s,t;\mathcal{R})=-\rho \frac{f''(s+\gamma_{\mathbf{R}}(\mb x_\mathbf{T}(s),\mb x_\mathbf{E},\mathbf{v}_\mb T) t-t_{\mathbf{R}}(\mb x_\mathbf{T}(s),\mb x_\mathbf{E},\mathbf{v}_\mb T))}{(4\pi|\mb x_{\mathbf{T}}(s)-\mb x_\mathbf{R}|)^2}.
\label{eq:scattered_field}
\end{equation}
The derivation of  \eqref{eq:scattered_field} \textcolor{black}{can be found in Appendix~A in \cite{leibovich2020generalized}}.
Here $c_0$ denotes the speed of light,  \textcolor{black}{$\rho$ is the unknown reflectivity we seek to determine},  $\gamma_{\mathbf{R}}$ is the Doppler scale correction factor and $t_\mathbf{R}$ is the signal travel time, which to first order in $|\mb v_{\mb T}|/c_0$ are given by
\begin{equation}
\begin{split}
&\gamma_{\mathbf{R}}(\mb x_\mathbf{T},\mb x_\mathbf{E},\mathbf{v}_\mb T)=1-\frac{\mathbf{v}_\mb T}{c_0}\cdot\left(\frac{\mb x_\mathbf{T}-\mb x_\mathbf{E}}{|\mb x_\mathbf{T}-\mb x_\mathbf{E}|}+\frac{\mb x_\mathbf{T}-\mb x_\mathbf{R}}{|\mb x_\mathbf{T}-\mb x_\mathbf{R}|}\right),\\
&t_{\mathbf{R}}(\mb x_\mathbf{T},\mb x_\mathbf{E},\mathbf{v}_\mb T)=\frac{|\mb x_\mathbf{T}-\mb x_\mathbf{E}|}{c_0}+\frac{|\mb x_\mathbf{T}-\mb x_\mathbf{R}|}{c_0}\gamma_{\mathbf{R}}(\mb x_\mathbf{T},\mb x_\mathbf{E},\mathbf{v}_\mb T).
\end{split}
\label{eq:gamma_t}
\end{equation}
The formula (\ref{eq:scattered_field}) for the recorded data will be used in the theoretical analysis.
\subsection{The cross-correlation data structure in iSAR}
Expression \eqref{eq:gamma_t} shows that $\gamma_{\mb R}$ depends on the target's position and velocity. Having an accurate value for $\gamma_{\mb R}$ is extremely important in calculating cross-correlations, as small changes in the frequency support would greatly affect the outcome of the correlation integral and hence the resolution of the image. In \cite{leibovich2020generalized}, it was shown that, for a limited image search region centered  around $\mb x_0,\mb v_0$, we have 
\begin{equation}
\label{eq:gamma_var}
\frac{\gamma_{\mathbf{R}}(\mb x_\mathbf{T},\mb x_\mathbf{E},\mathbf{v_T})}{\gamma_{\mathbf{R}}(\mb x_\mathbf{0},\mb x_\mathbf{E},\mathbf{v}_0)}\approx 1, 
\end{equation}
so that the constant $\gamma_{\mathbf{R}}(\mb x_\mathbf{0},\mb x_\mathbf{E},\mathbf{v}_0)$ can be used to account for the Doppler effect. 
Note that the rotational velocities, which are on the order of $1m/s$ are negligible with respect to any Doppler effects. 
Relying upon the weak variation of $\gamma_\mb R$,  we use a rescaled signal to image over the image window, rather than rescaling the signal for every point in the image window as was done in \cite{fournier2017matched}(eq. 3.6). 
Specifically, we construct $C_{\mathbf{RR}'}(s,\tau)$ by scaling the signal received by a reference Doppler factor,

\begin{equation}
\tilde{u}_{\mb R,\mathbf{x}_0,\mb v_0}(s,t;\mathcal{R})=u_\mathbf{R}\left(s,\frac{t}{\gamma_\mb R(\mb x_0,\mb x_\mb E,\mathbf{v_0})};\mathcal{R}\right).
\end{equation}
Note that for a reflector with a trajectory $\mb x_{\mb T}, \mb v_\mb T$, the received signal has the form
\begin{equation}
\tilde{u}_{\mb R,\mathbf{x}_0,\mb v_0}(s,t;\mathcal{R})=-\rho \frac{f''(s+\frac{\gamma_{\mathbf{R}}(\mb x_\mathbf{T},\mb x_\mathbf{E},\mathbf{v_T})}{\gamma_{\mathbf{R}}(\mb x_0,\mb x_\mathbf{E},\mathbf{v_0})} t-t_{\mathbf{R}})}{(4\pi|\mb x_{\mathbf{T}}-\mb x_\mathbf{R}|)^2}\approx -\rho \frac{f''(s+t-t_{\mathbf{R}})}{(4\pi|\mb x_{\mathbf{T}}-\mb x_\mathbf{R}|)^2}.
\label{eq:field_scaled}
\end{equation}

We define the cross-correlation function as,
\begin{equation}
\begin{split}
C_{\mathbf{RR}'}(s,\tau;\mathcal{R})=\int dt \tilde u_{\mb R,\mb x_0,\mb v_0}(s,t+t_\mb R(\mb x_0+s \mb v_0, \mb x_ \mb E, \mb v)) \tilde u_{\mb R',\mb x_0,\mb v_0}(s,t+t_{\mb R'}(\mb x_0+s \mb v_0, \mb x_ \mb E, \mb v)+\tau).
\end{split}
\label{eq:scaled_CC}
\end{equation}
\textcolor{black}{The continuous in time cross-correlation function $C_{\mathbf{RR}'}(s,\tau;\mathcal{R})$ is stored as a data structure, by sampling at discrete times $s_i,\tau_j$}.
As we shall see in the following sections, this data structure can be used in migration schemes to image.
Migration imaging is based on a travel time estimate from the receivers to a search point in the image domain, which in general includes both the targets' positions and velocities. We can use prior information at our disposal to limit the image domain. For example, $\mb v_\mb T$  can be estimated from the fact that the objects are assumed to be in a Keplerian orbit, i.e., $|\mb v_\mb T|\approx \textcolor{black}{\sqrt{\frac{G M_{\text{Earth}}}{\mb R_\mb T}}}$, \textcolor{black}{with $G$ denoting the gravitational constant, $M_{\text{Earth}}$ the mass of Earth and $\mb R_\mb T$ the target's orbital radius}. Accurate range measurements based on the signal's bandwidth can be used to estimate $\mb R_\mb T$, and subsequently $\mb v_\mb T$. 
\subsection{Rotation parameter estimation}
\label{sec:rot_estimate}
\begin{figure}[htbp]
	\centering
	\begin{subfigure}[t]{0.32\textwidth}
		\includegraphics[width=\textwidth]{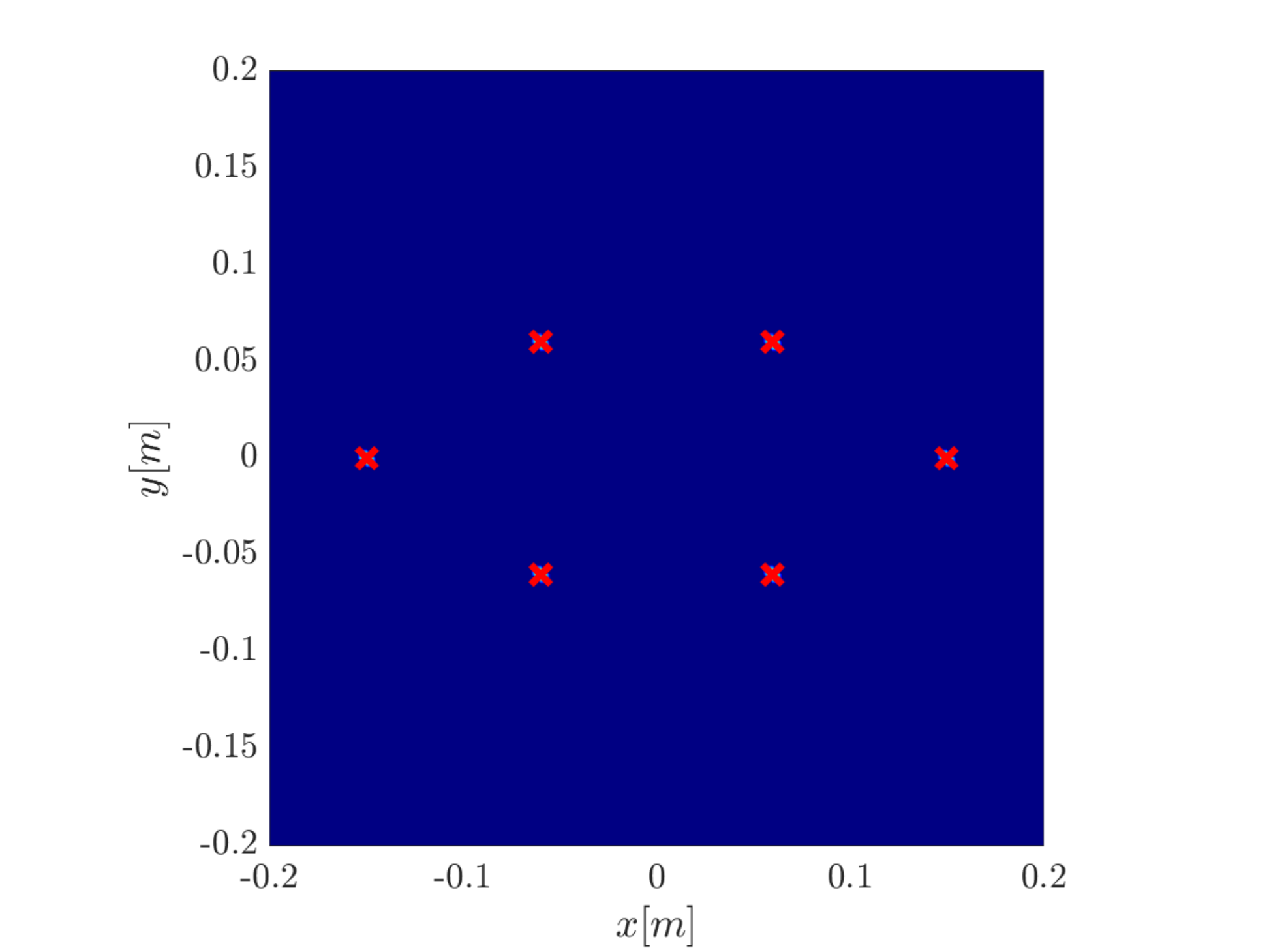}
	\caption{}
		\label{fig:sat_demo_layout}
	\end{subfigure}
	\begin{subfigure}[t]{0.3\textwidth}
		\includegraphics[width=\textwidth]{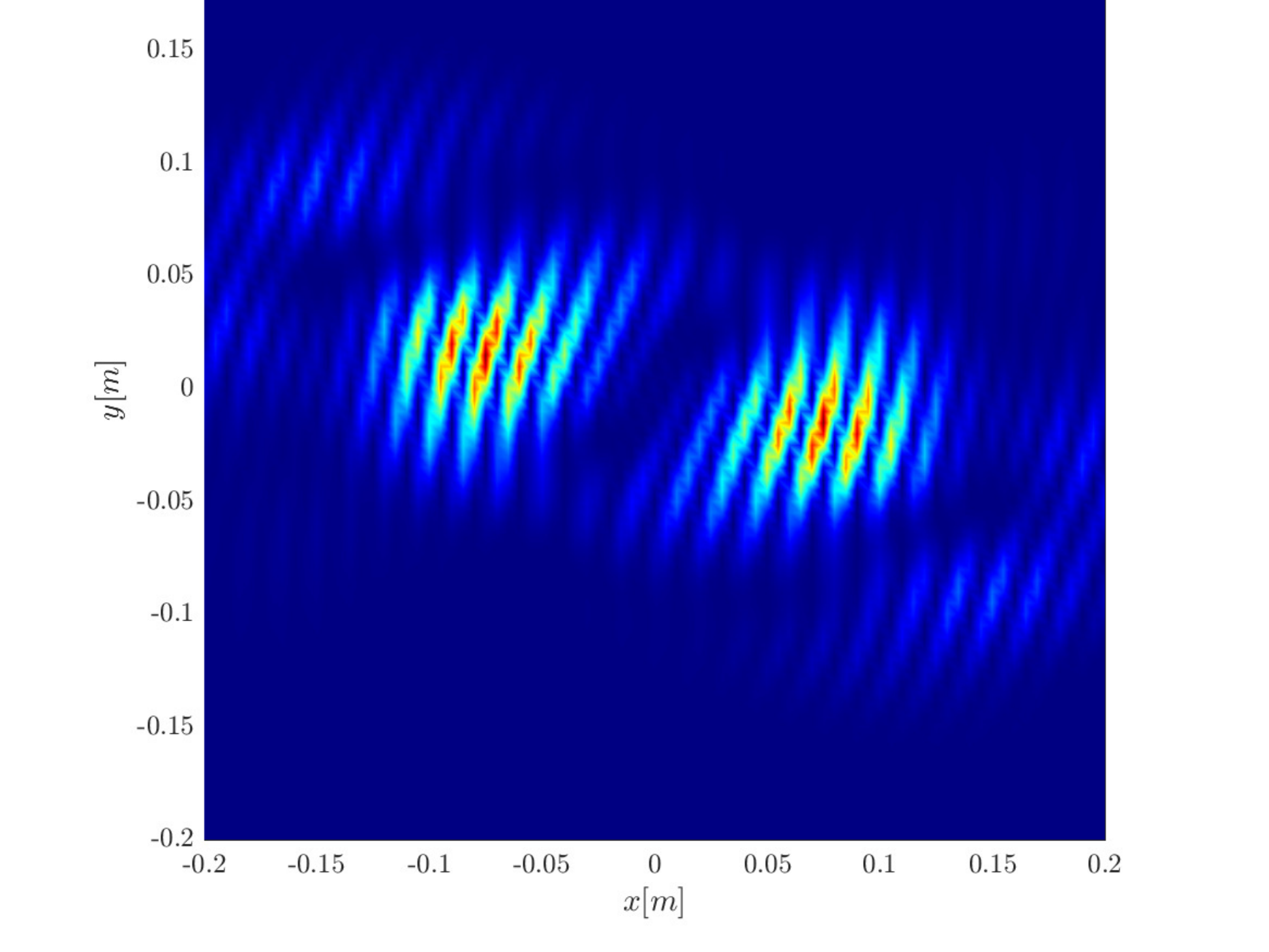}
		\caption{}
		\label{fig:I_rot_example_norot}
	\end{subfigure}
	\begin{subfigure}[t]{0.3\textwidth}
		\includegraphics[width=\textwidth]{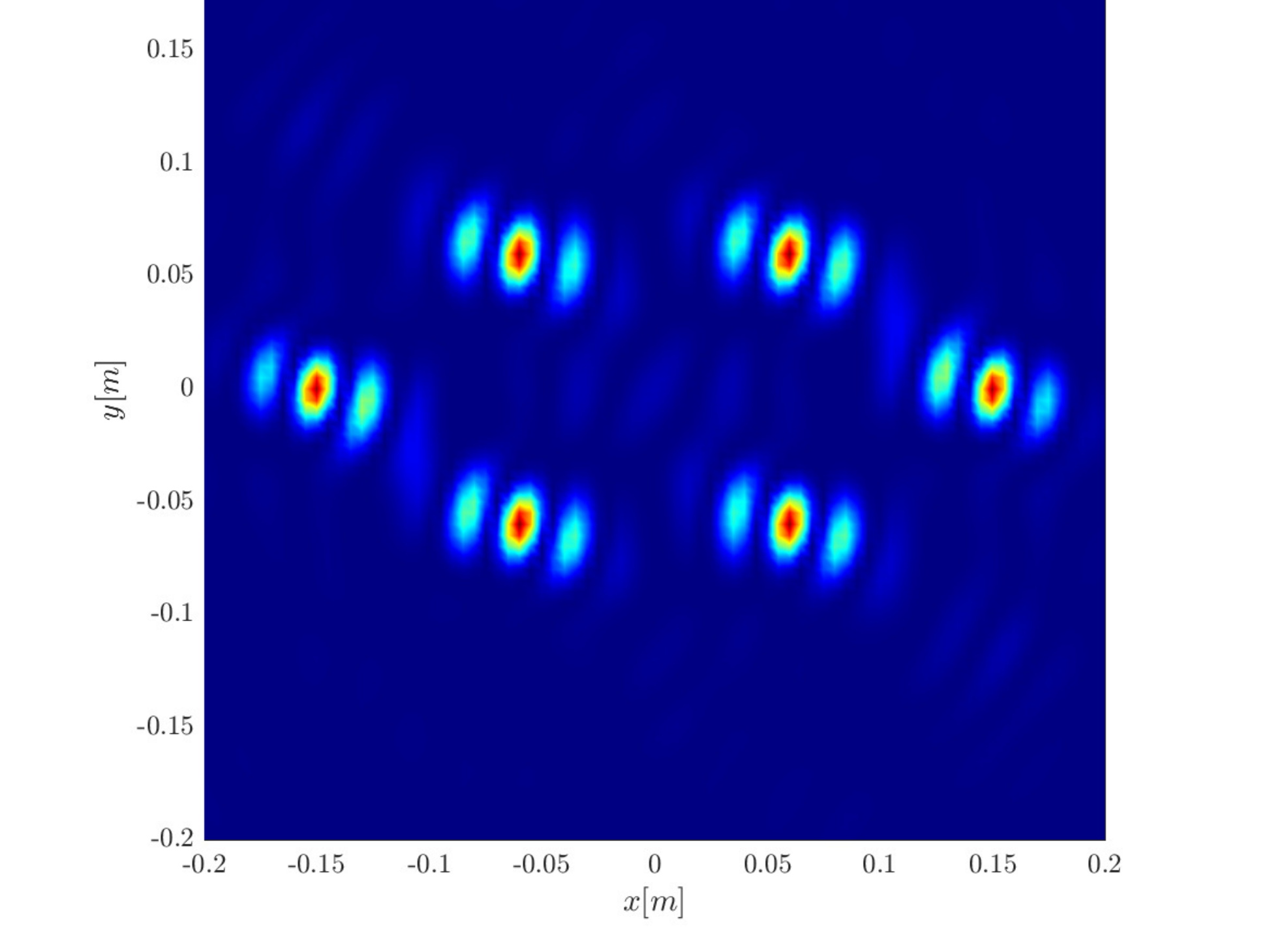}
		\caption{}
		\label{fig:I_rot_example}
	\end{subfigure}	
	\caption{\textcolor{black}{(a) The object to be imaged. It is modeled as a cluster of 6 point scatterers rotating as a rigid body. The two extremal points are used in estimating the target's rotation parameters.}
	(b) Example of image obtained without correctly accounting for the rotational velocity; the features are blurred.  $(b)$ Example of image obtained when accounting for the rotational velocity parameters. The  \textcolor{black}{point scatterers are well resolved in this case}. }
	\label{fig:rot_estimate}
\end{figure}

A prerequisite for any migration scheme is that if an object is rotating, as well as moving linearly, one has to compensate for the rotation as well. Without the correct rotational parameters, it would be impossible to synthesize the migrated data coherently from different measurements. This is illustrated in Figure~\ref{fig:rot_estimate} where we see that the image, \textcolor{black}{i.e., the reconstruction of the reflectivity of $\rho$ in the image window,} is blurry when it is constructed without correctly accounting for the rotational velocity (see Figure~\ref{fig:rot_estimate}-(b)) while the targets are well resolved when accounting for the rotational velocity parameters (see Figure~\ref{fig:rot_estimate}-(c)). 
Under the model assumptions of \eqref{eq:rot_op}, we need to determine four parameters, the two angles $\theta_{\text{rot}},\phi_{\text{rot}}$ and the rotational velocity $\omega_r$, as well as the center of the axis of rotation. 

Assuming the object is indeed rigid, the center of rotation can be evaluated by only compensating for the linear motion, as in \cite{leibovich2020generalized}. For the other three parameters, one could apply a brute force approach,  looking for the parameters which yield the sharpest image (as in auto-focus algorithms). This has the advantage of not making any prior assumptions. However, the computational cost of resolving the three additional parameters can be prohibitive. 

We present here a solution which resolves the rotational parameters directly from the correlation measurements. This algorithm applies to eccentric objects by which we mean that there is a specific direction in which the object has the largest diameter. In this case, one can use the autocorrelation data to determine the rotational parameters as we explain next.

\begin{figure}[htbp]
	\centering
	\begin{subfigure}[t]{0.3\textwidth}
		\includegraphics[width=\textwidth]{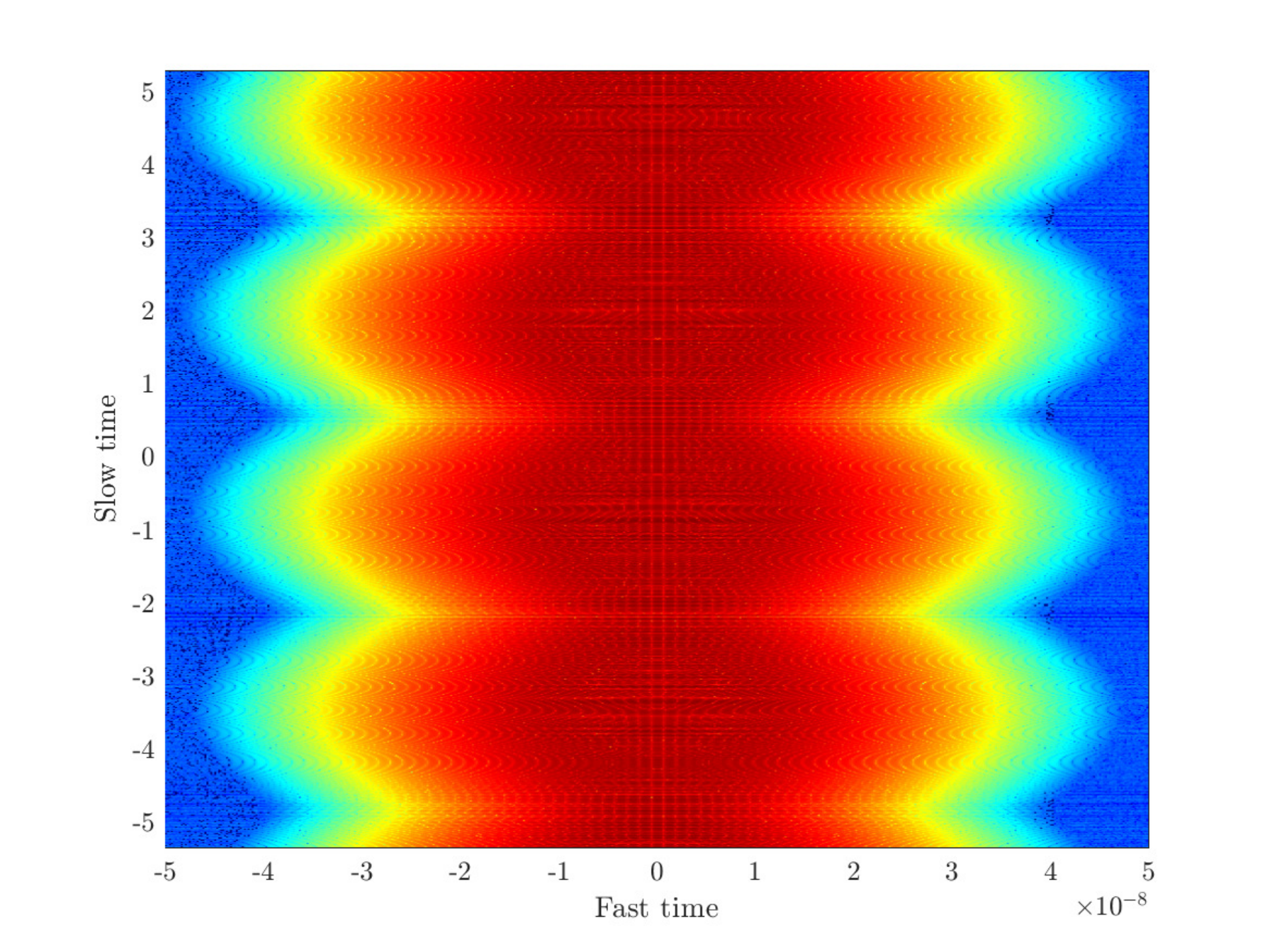}
		\caption{}
		\label{fig:AC_example}
	\end{subfigure}
\hspace{4em}
	\begin{subfigure}[t]{0.3\textwidth}
		\includegraphics[width=\textwidth]{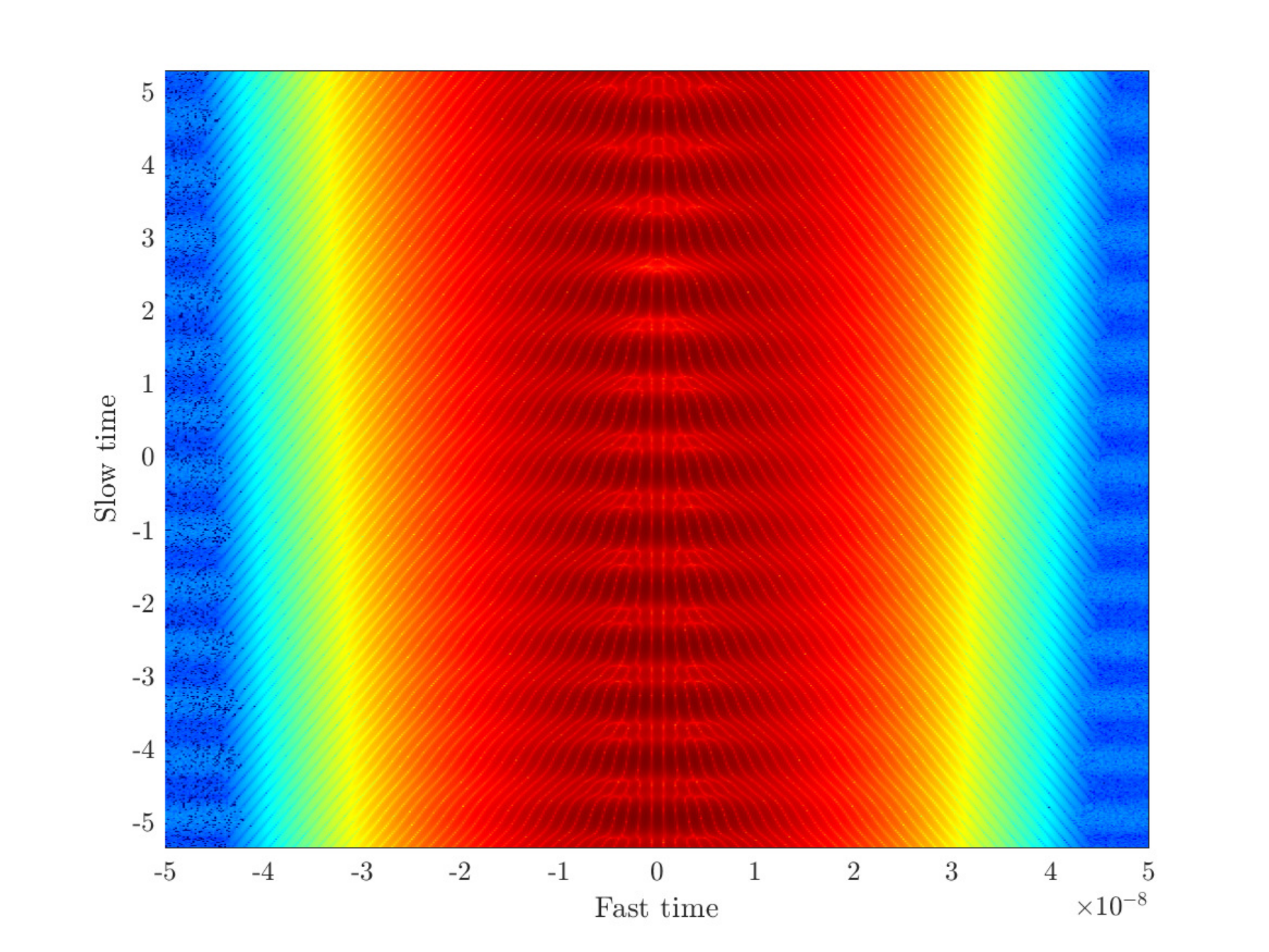}
		\caption{}
		\label{fig:AC_example_no_rot}
	\end{subfigure}
	\caption{ $(a)$ $C_R(\tau,s)$ for receiver $R=1$ and for reflectors positioned as in Fig.~\ref{fig:sat_demo_layout}. We can see the periodic expansion and contraction of the signal's support in $\tau$;  $(b)$ $C_R(\tau,s)$ for receiver $R=1$ when $\omega_r$ is set to 0. We can see that the oscillatory behavior is gone when there is no rotation.}
		\label{fig:rot_corr_ex}
\end{figure}

\textcolor{black}{
We are looking at the autocorrelation data defined as
\begin{equation}
C_{\mathbf{R}}(\tau,s)=\int dt \tilde u_{\mathbf{R},\mb x _{\mb L}(s)}(t) \tilde u_{\mathbf{R},\mb x _{\mb L}(s)}(t+\tau).
\end{equation}}
As illustrated in Figure~\ref{fig:rot_corr_ex}, rotation causes the support of the autocorrelation data, $C_{\mathbf{R}}(\tau,s)$, to oscillate with the slow time $s$. We look for the peaks in the spread of the autocorrelation data, as a function of the slow time $s$, and use them as data points. We then solve a regression problem, where we match these data points to a model that depends on the rotation parameters. 

\begin{figure}[htbp]
	\centering
	\begin{subfigure}[t]{0.4\textwidth}
		\includegraphics[width=\textwidth]{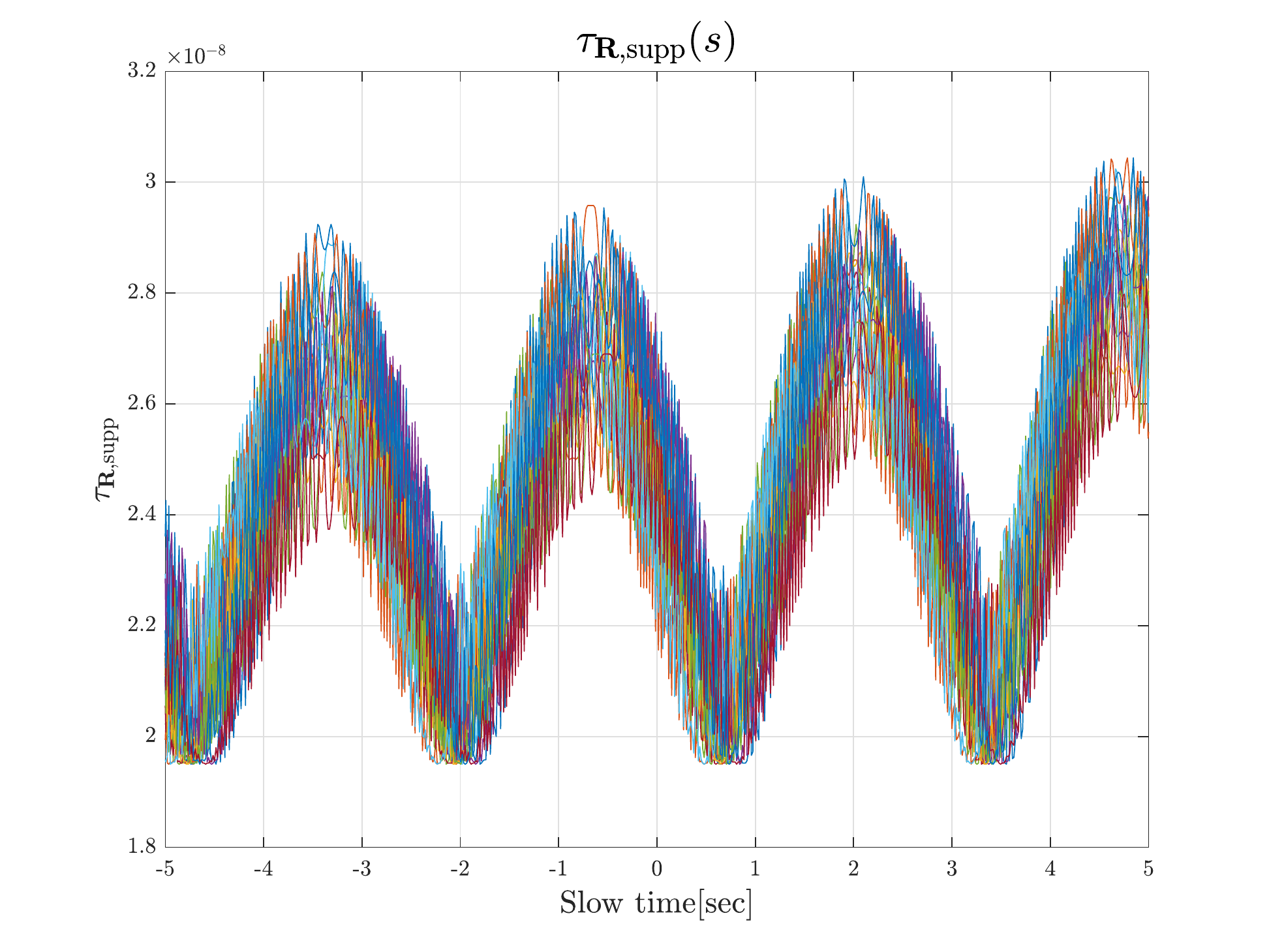}
		\label{fig:t_max_ex}
		\caption{}
	\end{subfigure}
	\begin{subfigure}[t]{0.4\textwidth}
		\includegraphics[width=\textwidth]{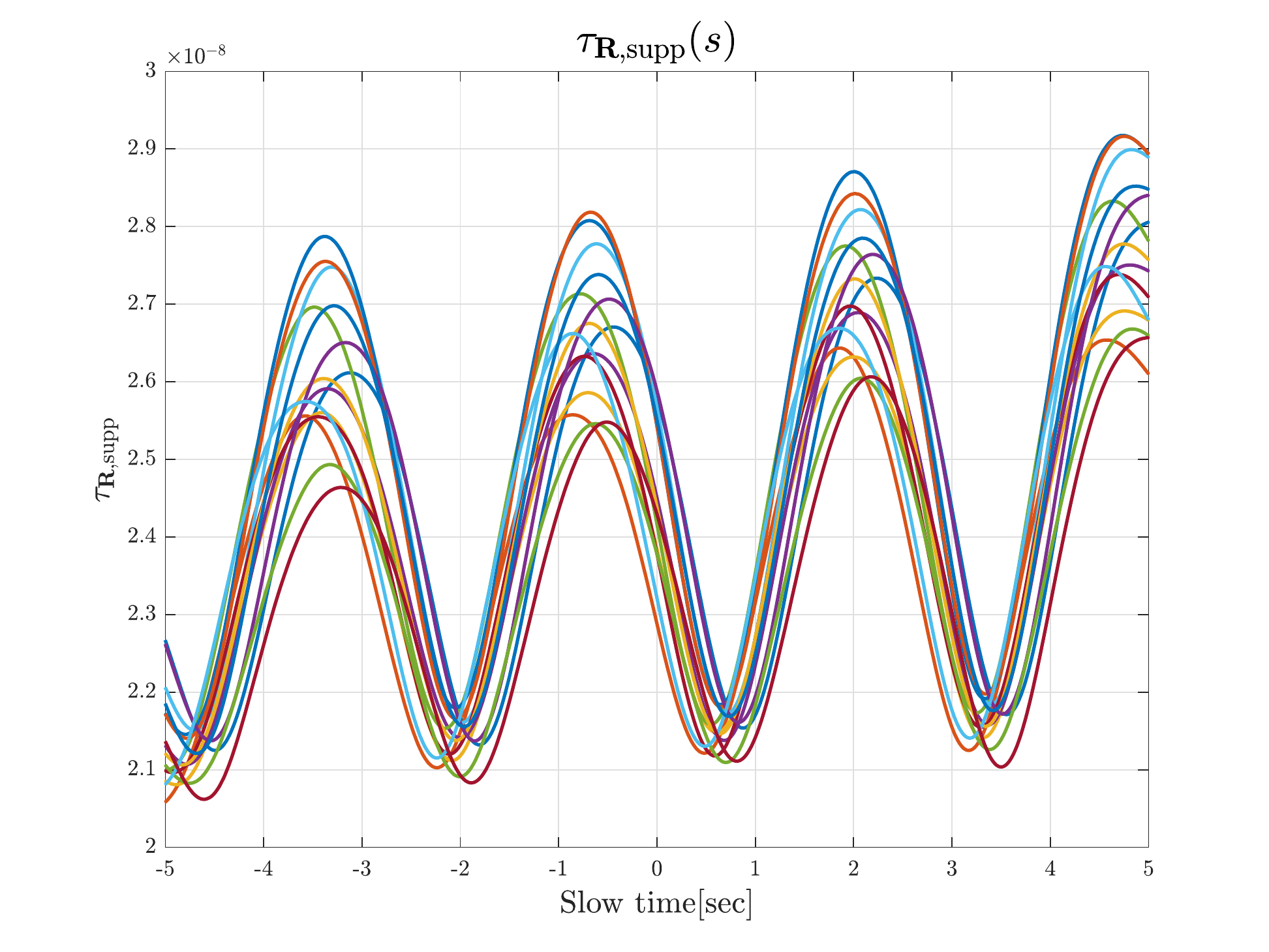}
		\label{fig:t_max_ex_smooth}
		\caption{}
	\end{subfigure}
	\caption{$(a)$ $\tau_{\mb R,\text{supp} }(s)$ $(b)$ $\tau_{\mb R,\text{supp} }(s)$ after Gaussian smoothing with a 100 time steps window size. \textcolor{black}{There are 15 curves in the plot, one for each receiver.}}
	\label{fig:time_smoot}
\end{figure}

\paragraph{Algorithm for estimating the rotational parameters}
The algorithm for estimating the rotational parameters is comprised of the following steps~
\begin{enumerate}
	\item Measure the total support of the autocorrelation at each receiver, as a function of time $s$, denoted as $\tau_{\mb R,\text{supp} (s)}$,
	\begin{equation}
	\tau_{\mb R,\text{supp} }(s)=2 \max\left\{\tau \hspace{0.2em} \Big|\hspace{0.2em} |C_{\mb R }(s,\tau)|\ge \alpha \max \limits_{\tau}|C_{\mb R }(s,\tau)|\right\}.
	\label{eq:tau_R}
	\end{equation}
	\textcolor{black}{$\alpha$ is adapted to the ambient noise level of the received signal. For a signal to noise ration of $0$ dB we found $\alpha=0.075$ to be adequate. For noiseless measurements $\alpha$ can be much smaller. We used $\alpha=0.001$ in the noiseless case.}
	\item Retrieve the times $s^*_{\mb R}$ for which the support of the autocorrelation at receiver $\mb R$ has a local maximum. The signal is smoothed with a Gaussian filter to extract the peaks automatically, as illustrated in Figure~\ref{fig:time_smoot}.
	
	\item Use the rigid body rotation model to define an implicit objective function, relating $s^*_{\mb R}$ to the rotation parameters $\theta_{\text{rot}},\phi_{\text{rot}}$ and $\omega_r$ 
	\begin{equation}
	\label{defg}
	g(\theta_{\text{rot}},\phi_{\text{rot}},\mb d_\mb R(s^*_\mb R))=\tan(\omega_r s^*_\mb R+\varphi).
	\end{equation}
	\textcolor{black}{
	Here $\mb d_\mb R(s)$ is defined using the unit vectors pointing from the center of rotation to the source and receiver, 
	$$
	\mb d_\mb R(s)=\frac{\mb x_{\mb L}(s)-\mb x_{\mb E}}{|\mb x_{\mb L}(s)-\mb x_{\mb E}|}+\frac{\mb x_{\mb L}(s)-\mb x_{\mb R}}{|\mb x_{\mb L}(s)-\mb x_{\mb R}|}\gamma_{\mathbf{R}}(\mb x_\mb L(s),\mb x_{\mb E},\mb v_\mb L),$$ 
and $g(\theta,\phi,\mathbf{d}_\mb R)$ is the function
\begin{equation}
g(\theta,\phi,\mathbf{d}_\mb R)=\frac{-(d_{\mb R})_1 \sin\phi + (d_{\mb R})_2 \cos \phi }{(d_{\mb R})_1 \cos\theta\cos\phi +(d_{\mb R})_2 \cos\theta\sin\phi + (d_{\mb R})_3 \sin\theta  }
\end{equation}
with $(d_{\mb R})_i$, $i=1,2,3$ denoting the $i$th component of the vector $d_{\mb R}$. 
The angle $\varphi$ in \eqref{defg} denotes the orientation in which the object has nontrivial eccentricity, i.e, we assume that there is a certain direction in which the object is further spread out from the center of rotation. This angle is assumed to exist but we do not need to know it so as to determine the rotational parameters because the loss function \eqref{eq:rot_loss_obj} defined below is independent of $\varphi$.}

	\item Collect all the data points from the different receivers $(s^*_{\mb R_i},\mb d_{\mb R_i})$. Define the loss function $L(\theta,\phi,\omega_r)$ as the mean square error of the model using all data points $s^*_{\mb R}$, 
	\begin{equation}
L(\theta,\phi,\omega_r)=\sum\limits_i \left(\tan^{-1} g(\theta,\phi,\mathbf{d}_{\mb R_i}(s_{\mb R_i}^*))-\tan^{-1} g(\theta,\phi,\mathbf{d}_{\mb R_{i-1}}(s_{\mb R_{i-1}}^*))-(\omega_r (s^*_{\mb R_i}-s^*_{\mb R_{i-1}}))\right)^2,
\label{eq:rot_loss_obj}
\end{equation}
and find its argmin. 
\end{enumerate}

This problem can be solved inexpensively by any number of readily available solvers, without the need to form the preliminary image. \textcolor{black}{This is because the angular parameter space is limited, so we can find the global minimizer of a smooth non convex loss function by function evaluation. For the angular velocity, we also have a good prior from the periodicity of $\tau_{\mb R,\text{supp} }$. In practice, we used MATLAB's fminsearch function for minimizing the loss function, near an initial guess. }An illustration of the estimation results is shown in Figure~\ref{fig:L_ex_sec}. \textcolor{black}{The derivation of the loss function \eqref{eq:rot_loss_obj} is detailed in Appendix~\ref{app:rot_est}.}

\begin{figure}[htbp]
	\centering
	\begin{subfigure}[t]{0.32\textwidth}
		\includegraphics[width=\textwidth]{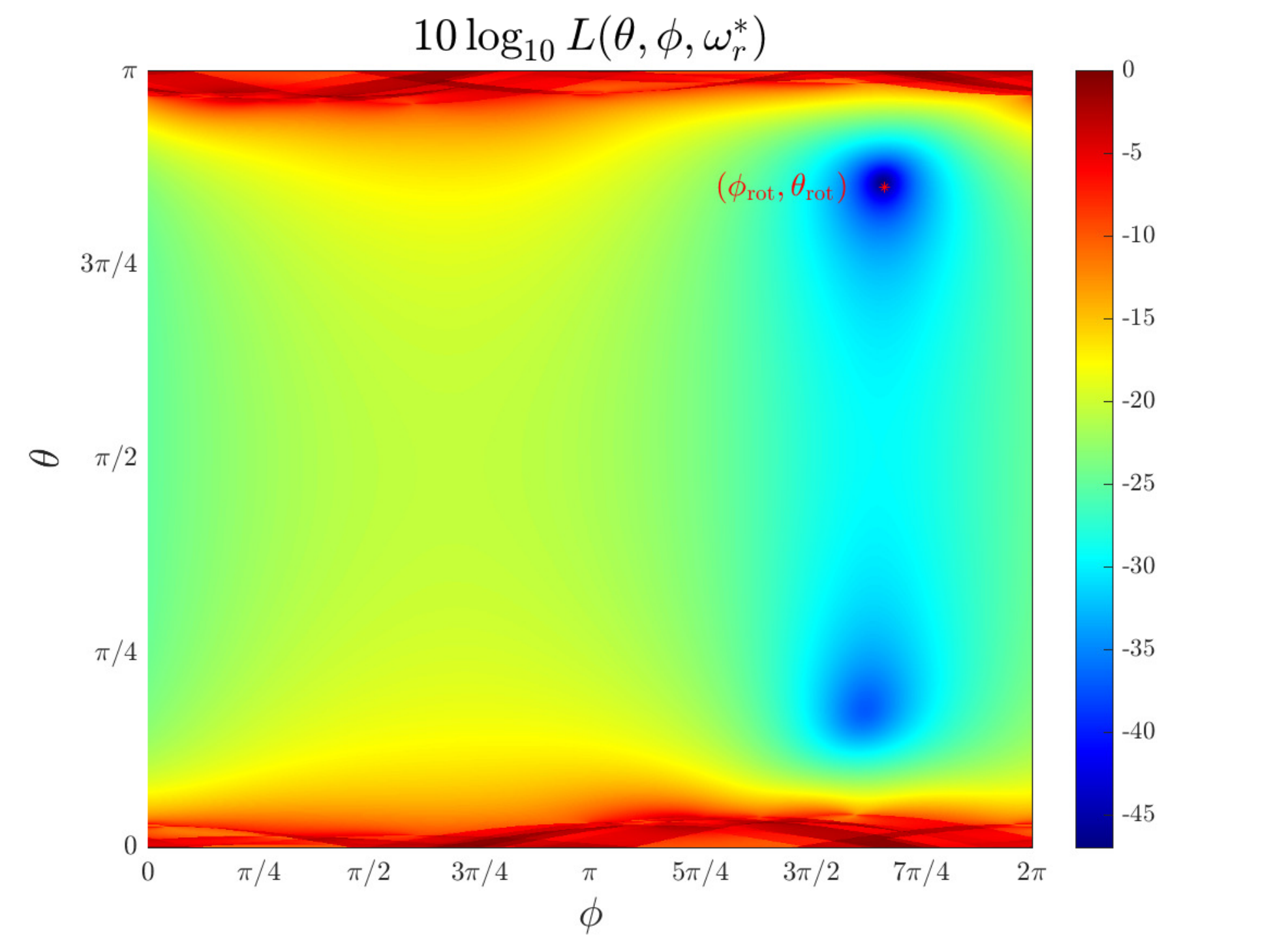}
		\caption{}
		\label{fig:L_ex}
	\end{subfigure}
	\begin{subfigure}[t]{0.32\textwidth}
			\includegraphics[width=\textwidth]{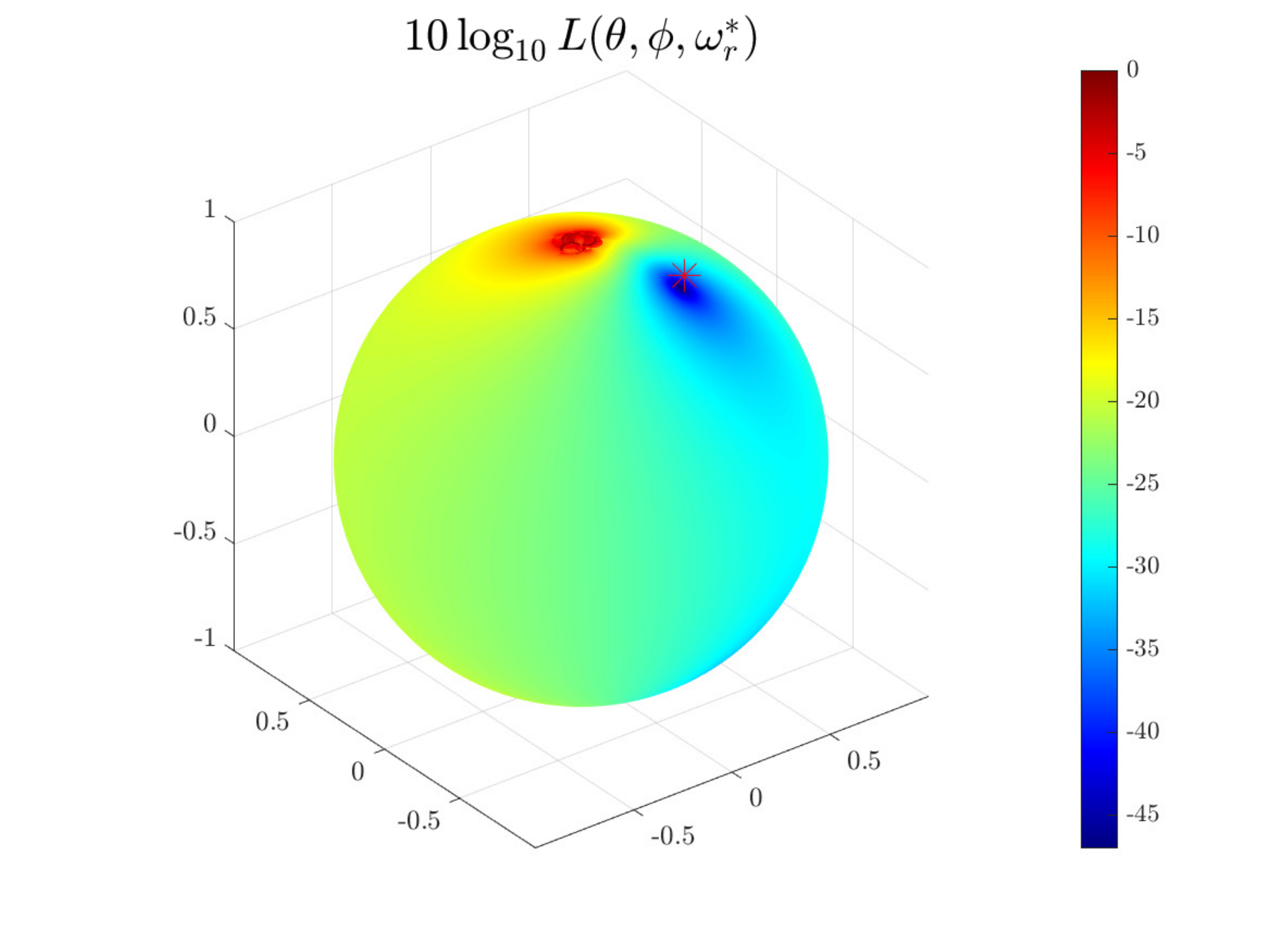}
		\label{fig:L_ex_sphere}
		\caption{}
	\end{subfigure}
\begin{subfigure}[t]{0.32\textwidth}
	\includegraphics[width=\textwidth]{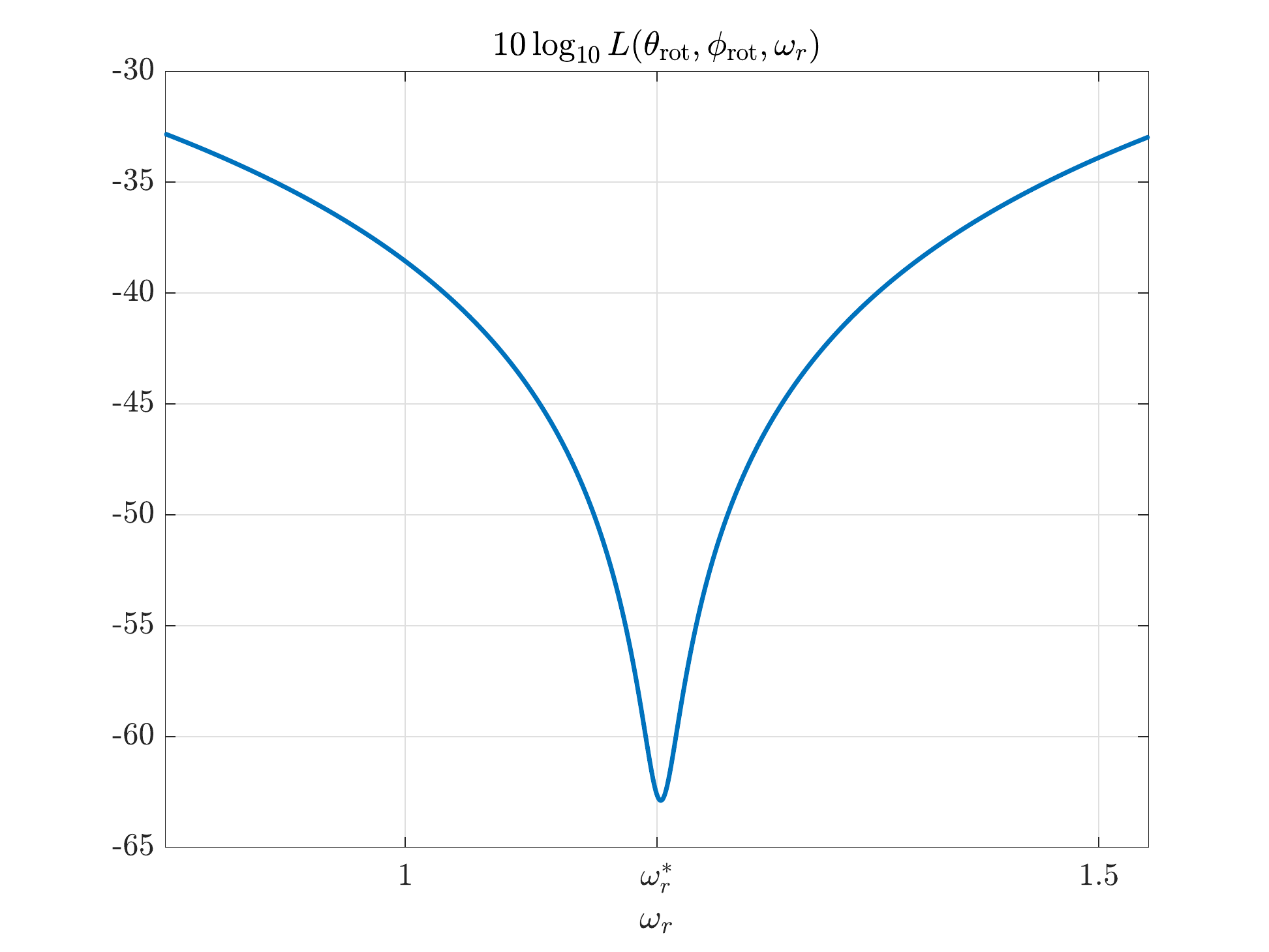}
	\caption{}
	\label{fig:L_ex_w}
\end{subfigure}
	\caption{Example of $L(\theta,\phi,\omega_r)$ (a) $L(\theta,\phi,\omega_r^*)$ the loss function for estimating the rotation parameters. The minimum is very close to the actual value $(\phi_{\text{rot}}, \theta_{\text{rot}})$ $(b)$ Projection of $L(\theta,\phi,\omega_r^*)$ on the sphere. $(c)$ Plot of $L(\theta^*,\phi^*,\omega_r)$, as we can see the angular velocity can also be inferred from the measurements.}
	\label{fig:L_ex_sec}
\end{figure}

\textcolor{black}{
\paragraph{Noise robustness of the estimation algorithm}
We consider here that the measurements are not exact and we study the effect of the noise on the estimation of the rotational parameters. We observe that by relying on the autocorrelation rather than on direct measurements, the proposed algorithm is robust to measurement noise. This is illustrated in Figure~\ref{fig:CT_noise}, where we see that the noise doesn't affect the main lobe of the autocorrelation signal. When using the noisy data to infer the rotational degrees of freedom, the angles are resolved with a similar degree of accuracy as illustrated in Figure~\ref{fig:L2d_noise} (to be compared with Figure~\ref{fig:L_ex}).
\begin{figure}[htbp]
	\centering
	\begin{subfigure}[t]{0.3\textwidth}
		\includegraphics[width=\textwidth]{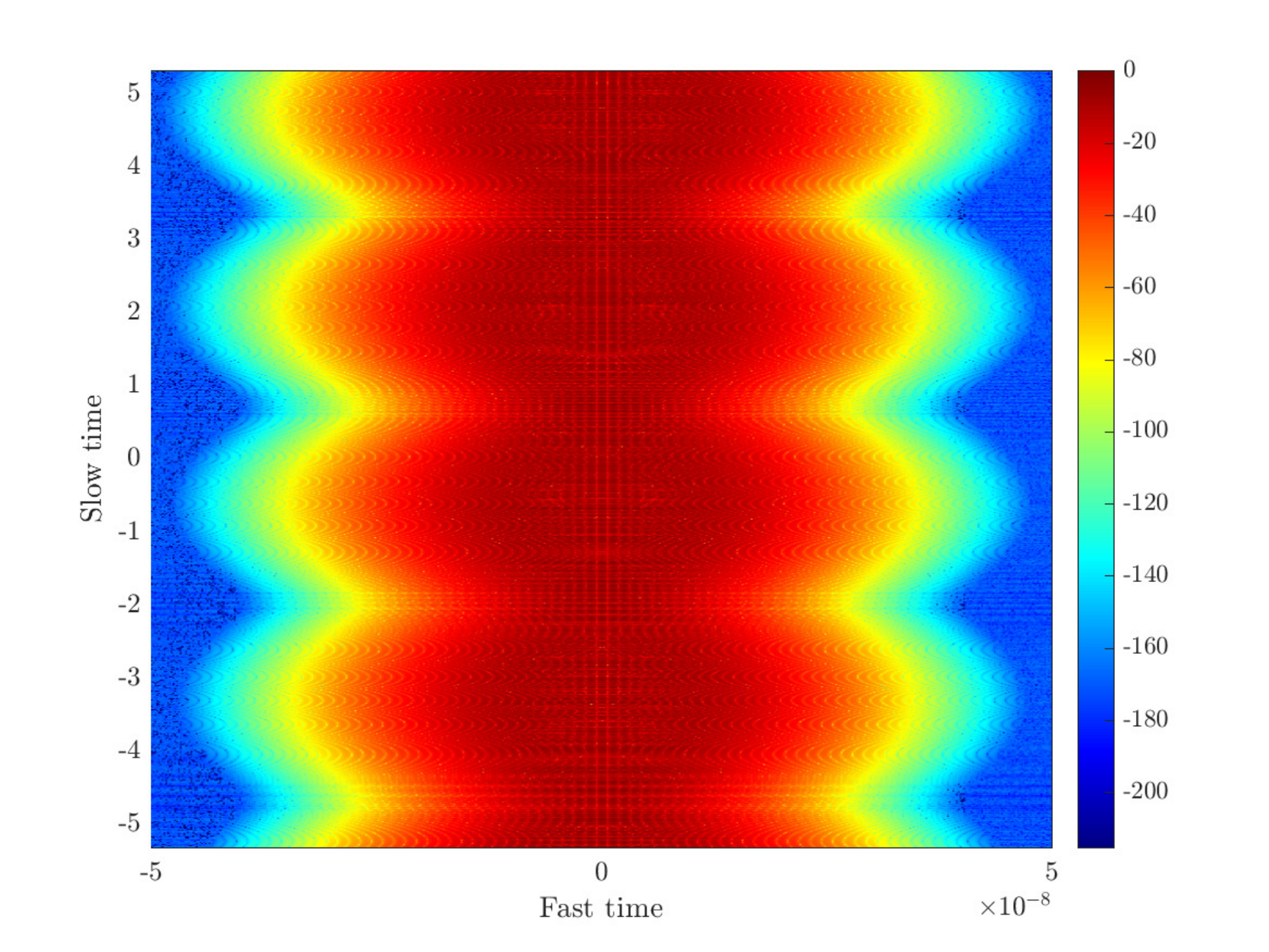}
		\caption{}
		\label{fig:CT_nonoise}
	\end{subfigure}
	\begin{subfigure}[t]{0.3\textwidth}
		\includegraphics[width=\textwidth]{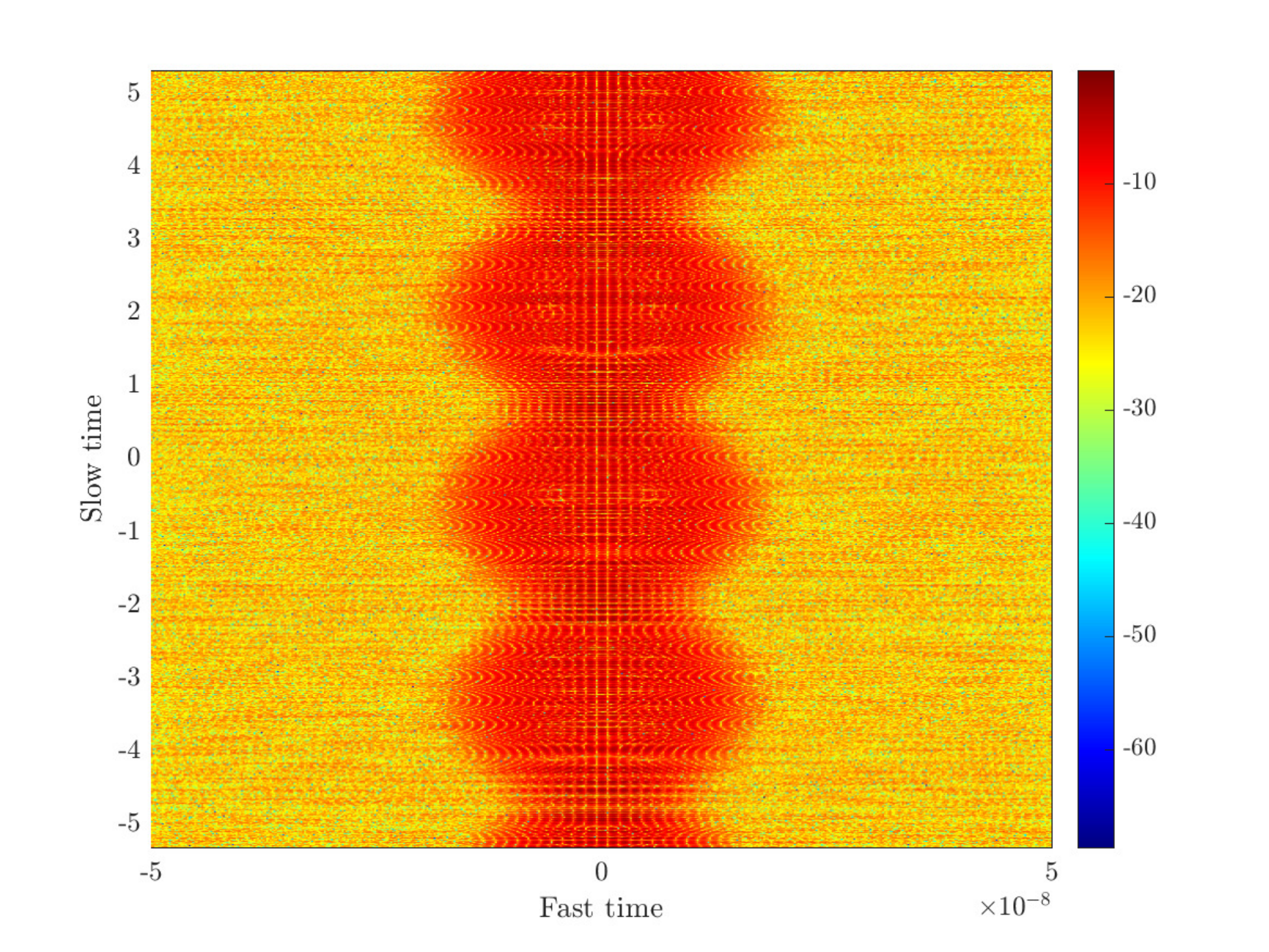}
		\caption{}
		\label{fig:CT_noise}
	\end{subfigure}
	\begin{subfigure}[t]{0.3\textwidth}
		\includegraphics[width=\textwidth]{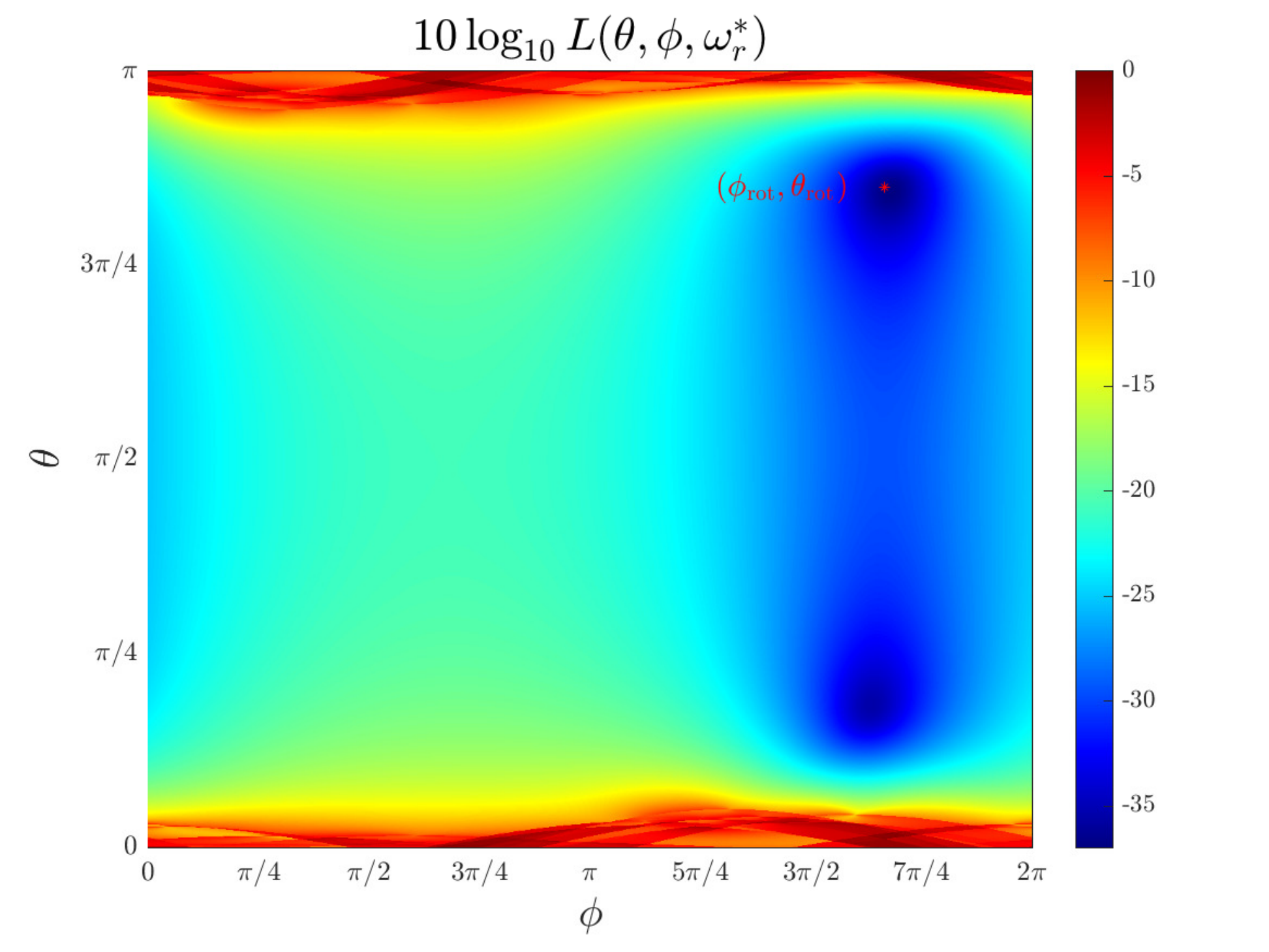}		
		\caption{}
		\label{fig:L2d_noise}
	\end{subfigure}		
	\caption{Effect of additive noise on the autocorrelation. $(a)$ Time domain autocorrelation data without noise. $(b)$ Time domain autocorrelation when additive white Gaussian noise is added to the data. The signal to noise ratio is $0$dB. We can see that the effect of rotation on the signal is still well observed, with the difference being that now the signal decays to an ambient noise level rather than to $0$. (c) We plot $L(\theta,\phi,\omega_r^*)$, the loss function for estimating the rotation parameters, as a function of $\theta$ and $\phi$ for the noisy data with $0$dB SNR.  We observe that while the noise can alter the shape of the loss function (compare with Figure~\ref{fig:L_ex}), the angular parameters can still be extracted with the same degree of accuracy.}
	\label{fig:C_noise}
\end{figure}
}

In the next section we explain how to build the imaging functional for the rotating object from the cross-correlation measurements assuming the rotation parameters have been estimated.

%\begin{figure}[htbp]
%	\centering
%	\begin{subfigure}[t]{0.3\textwidth}
%		\includegraphics[width=\textwidth]{LPT_Figures_accepted/L_2d_nonoise_2-eps-converted-to.pdf}
%		\caption{}
%		\label{fig:L2d_nonoise}
%	\end{subfigure}
%	\begin{subfigure}[t]{0.3\textwidth}
%		\includegraphics[width=\textwidth]{LPT_Figures_accepted/L_2d_noise_2-eps-converted-to.pdf}
%		\label{fig:L2d_noise}
%		\caption{}
%	\end{subfigure}
%	\caption{Effect of noise on rotational parameters' estimation. We plot $L(\theta,\phi,\omega_r^*)$, the loss function for estimating the rotation parameters, as a function of $\theta$ and $\phi$. (a) for the noiseless data. (b) for the noisy data with $0$dB SNR.
%	 We observe that while the noise can alter the shape of the loss function, the angular parameters can still be extracted with the same degree of accuracy.}
%	\label{fig:L_noise}
%\end{figure}

\section{Migration imaging for cross-correlation data}
\label{sec:generalized_migration}
In \cite{leibovich2020generalized}, we presented a generalization of the linear Kirchhoff migration for cross-correlation data. For completeness, we summarize here the main results. We introduce a forward model for the received data; We then recast the model in matrix form, introduce the migration of the cross-correlation data, and, the imaging functions derived from it. The reader is referred to \cite{leibovich2020generalized} for a detailed presentation. In addition, to keep the notation succinct, we drop the dependence on $\mathcal{R}$. Note that in general, as in \eqref{eq:u_freq}, all quantities are implicitly dependent on rotation.

Assume we have discretized the medium in a small image window relative to its moving center $\mb x _\mb L(s)$, as in (\ref{eq:point_scatter}), with grid points $\mb y_k$ and $\mb y_k= \mb0$ corresponds to the center of the window. The unknown reflectivity is discretized by its values on this grid $$\rho_k=\rho(\mb y_k),\hspace{.1em}k=1,\dots,K.$$
The unknown reflectivity vector has dimension $K$, which is the number of pixels in the image window. Most of these reflectivities are zero because there are usually few relatively strong reflectors that can be imaged. Imaging refers to estimating the location of the reflectors and their strength. As explained in Section~\ref{sec:corr_imaging}, we assume the image window is small enough so that the Doppler term $\gamma_{\mathbf{R}}$ is constant over it. 

We saw in \eqref{eq:field_scaled} that the signal recorded at receiver location $\mb x_\mb R$ can be written, after an appropriate scaling, as
\begin{equation}
\tilde{u}_{\mb R,\mathbf{x}_\mb L(s),\mb v_\mb L}(s,t)\approx-\sum\limits_{k=1}^K\rho_k \frac{f''(s+ t-(t^k_{\mathbf{R}}(s)-t_\mb R(s)))}{(4\pi|\mb x_{\mathbf{L}(s)}-\mb x_\mathbf{R}|)^2},
\end{equation}
where 
\begin{equation}
\begin{split}
t_\mb R^k(s) &=t_{\mathbf{R}}(\mb x_\mb L(s)+\mb y_k,\mb x_\mathbf{E},\mathbf{v})=\frac{|\mb x_\mathbf{L}(s)+\mathcal{R}(s)\mb y_k-\mb x_\mathbf{E}|}{c_0}+\frac{|\mb x_\mathbf{L}(s)+\mathcal{R}(s)\mb y_k-\mb x_\mathbf{R}|}{c_0}\gamma_{\mathbf{R}}(\mb x_\mathbf{L}(s)+\mathcal{R}(s)\mb y_k,\mb x_\mathbf{E},\mathbf{v}),\\
t_\mb R(s) &=t_{\mathbf{R}}(\mb x_\mb L(s),\mb x_\mathbf{E},\mathbf{v}).
\end{split}
\label{eq:t_rot}
\end{equation}
In the frequency domain, \textcolor{black}{with respect to the fast time $t$}, the recorded signal is
\begin{equation}
\label{eq:u_freq}
\hat{u}_{\mb R,\mathbf{x}_L(s),\mb v_L}(s,\omega)\approx\sum\limits_{k=1}^K\rho_k \omega^2 \frac{\hat f( \omega)}{(4\pi|\mb x_{\mathbf{L}}(s)-\mb x_\mathbf{R}|)^2}e^{i\omega (t_\mb R^k(s)-t_\mb R(s))}\equiv \sum\limits_{k=1}^K \frac{\omega^2 \hat{f} (\omega)}{(4\pi|\mb x_{\mathbf{L}}(s)-\mb x_\mathbf{R}|)^2}A_{\mb R,k}(s,\omega)\rho _k,
\end{equation}
with 
\begin{equation}
\label{def:Ark}
A_{\mb R,k}(s,\omega)=e^{i\omega (t_\mb R^k(s)-t_\mb R(s))}.
%\approx \omega^2 A_{\hat f (\omega),\mb R}e^{i\omega (t_\mb R^k-t_\mb R)}
\end{equation}
The phase $A_{\mb R,k}(s,\omega)$ comes from the reduced travel time from the target to the receiver, relative to that of the image window center. 

We assume the distance from the reflectors to the different receivers doesn't vary greatly, hence we can approximate
\begin{equation}
\frac{\omega^2 \hat{f} (\omega)}{(4\pi|\mb x_{\mathbf{L}}(s)-\mb x_\mathbf{R}|)^2}\approx \textcolor{black}{\xi(s,\omega)}.
\end{equation}
By neglecting the dependence of the amplitude factor on a specific receiver,  the accuracy with which the amplitude of the reflector can be retrieved is compromised but not its support. Using this notation, and the fact that correlation in time is equivalent to multiplication in frequency, we get that
\begin{equation}
\begin{split}
\hat C_{\mb R\mb R'}(s,\omega)=&\hat u_{\mb R,\mb x_L(s),\mb v_L}(s,\omega)\overline{\hat u_{\mb R',\mb x_L(s),\mb v_L}(s,\omega)}=|\textcolor{black}{\xi(s,\omega)}|^2\sum\limits_{k,k'=1}^K A_{\mb R,k}(s,\omega)\overline{A_{\mb R',k'}(s,\omega)}\rho_k \rho_{k'}\\
&=|\textcolor{black}{\xi(s,\omega)}|^2\sum\limits_{k,k'=1}^K A_{\mb R,k}(s,\omega)\overline{A_{\mb R',k'}(s,\omega) }\rho_k \rho_{k'}.
\end{split}
\label{eq:C_form_element}
\end{equation}
This is our model for the cross-correlation data in the frequency domain. 

\subsection{Matrix formulation of the forward model}

We introduce here matrix notation that relates our model for reflectivities to the cross-correlation data. 
Denote by $\pmb \rho$ the unknown reflectivities in vector form $$\pmb \rho=[\rho_1,\dots,\rho _K]^T\in\mathbb{R}^K.$$
Denote $\mb A (s,\omega)$, our model for the sensing matrix. It has dimensions $N_R \times K$ and entries $A_{\mb R,k}(s,\omega)$ defined in \eqref{def:Ark}. This matrix $\mb A(s,\omega)$ acts on the reflectivities and returns data.  
Denote the recorded signal data as an $N_R$ vector vector $\hat{\mb u}_R(s,\omega)$ whose entries are given by (\ref{eq:u_freq}).
The cross-correlation data is also a matrix, of dimension $N_R \times N_R$, $\hat {\mb C}(s,\omega)$ with entries 
$\hat C_{\mb R\mb R'}(s,\omega)$ as in \eqref{eq:C_form_element}.

Combining these,  we have in matrix form the following model for the recorded signal data vector $\hat{\mb u }_{\mb R}(s,\omega)$ and cross-correlation data matrix $\hat{\mb C}(s,\omega)$
\begin{equation}
\hat{\mb u }_{\mb R}(s,\omega)=\xi(s,\omega)\mb A(s,\omega)\pmb \rho,
\label{eq:forward_field}
\end{equation}
\begin{equation}
\hat{\mb C}(s,\omega)=\hat{\mb u}_\mb R(s,\omega)\overline{\hat{\mb u}_\mb R(s,\omega)}^T=|\textcolor{black}{\xi(s,\omega)}|^2(\mb A (s,\omega)\pmb \rho)\overline{(\mb A (s,\omega)\pmb \rho)}^T=|\textcolor{black}{\xi(s,\omega)}|^2\mb A (s,\omega)\pmb \rho \pmb \rho ^T \overline{\mb A (s,\omega)}^T.
\label{eq:C_mat_exp}
\end{equation}
Denoting by $\mb X=\pmb \rho \pmb \rho ^T, X_{kk'}=\rho _k \rho_{k'}$, the outer product of reflectivities, then our model for the cross-correlation data in matrix form is
\begin{equation}
\hat{\mb C}(s,\omega)=|\textcolor{black}{\xi(s,\omega)}|^2\mb A (s,\omega) \mb X \overline{\mb A(s,\omega)}^T.
\label{eq:C_form}
\end{equation}
The cross-correlations depend on the reflectivities $\pmb \rho$ through their outer product $\mb X=\pmb \rho \pmb \rho^T$. As a result there can be several different extensions of Kirchhoff migration to cross-correlations, which we investigate in the next section. 

The model for the data of \eqref{eq:forward_field},\eqref{eq:C_mat_exp}, requires resolution of the rotation parameters, since $\mb A(s,\omega)$ is dependent on $\mathcal{R}(s)$ via \eqref{eq:t_rot}. As a result, the rotation parameters need to be resolved prior to the migration step, as explained in Section~\ref{sec:rot_estimate}.

\subsection{Imaging functions for cross-correlation data}
\label{sec:twopt_scheme}
Given the data $\hat{\mb u}_{\mb R}(s,\omega)$ and the model $\hat{\mb u}_{\mb R}(s,\omega)=\mb A(s,\omega )\pmb \rho$, Kirchhoff migration of the data $\hat{\mb u}_\mb R(s,\omega) $ is given by
\begin{equation}
\tilde{\pmb{\rho}}=\sum\limits_{s,\omega}\overline{\mb A(s,\omega)}^T\hat{\mb u}_{\mb R}(s,\omega).
\label{eq:KM_freq}
\end{equation}
It was shown in \cite{leibovich2020generalized} that a natural extension of \eqref{eq:KM_freq}  for migrating cross-correlations is the matrix $\tilde{\mb X}$
\begin{equation}
\tilde{\mb X}=\sum\limits_{s,\omega}\overline{\mb A(s,\omega)}^T\hat{\mb C}(s,\omega)\mb A (s,\omega) ,
\end{equation}
with elements
\begin{equation}
\tilde{X}_{kk'}=\sum\limits_{s,\omega,\mb R,\mb R'}\overline{ A}_{\mb R,k}(s,\omega)\hat{C}_{\mb R,\mb R'}(s,\omega)A_{\mb R',k'} (s,\omega).
\label{eq:M_tilde}
\end{equation}

The result of this two-point migration is an estimation of $\mb {X}$ rather than $\pmb \rho$, as our model is quadratic with respect to the reflectivities.
$\tilde{\mb X}$ is a square matrix with dimensions $K\times K$, where $K$ is the number of search points in the imaging domain. If points $\mb y_k,\mb y_{k '}$ are associated with reflectivities $\rho_k, \rho_{k'}$, we can think of $\tilde{\mb X}$ as a two-variable generalized cross-correlation imaging function
\begin{equation}
\mathcal{I}^{GCC}(\mb y_k,\mb y_{k'})=\tilde{X}_{kk'}.
\label{eq:2_point_mat_imag}
\end{equation}
Note that $\tilde{\mb X}\in \mathbb{C}^{K\times K}$  is Hermitian positive definite by definition. We will refer to it as the matrix interference pattern from which the image will be obtained. We next consider how in fact  an image of the reflectivity can be extracted from $\mathcal{I}^{GCC}$.
The functional $\mathcal{I}^{GCC}$ defined in \eqref{eq:2_point_mat_imag} lacks a direct physical interpretation and we refer to it as an interference pattern. It evaluates the outer product of reflectivities rather than the reflectivities themselves. We examine next two ways to extract an image from $\tilde{\mb X}$:  

\begin{enumerate}
	\item Reconstruct an image of $|\rho_k|^2=\tilde{X}_{kk}$. This is equivalent to the single point migration functional proposed and analyzed in \cite{fournier2017matched} since the diagonal terms of $\mathcal{I}^{GCC}(\mb y_k,\mb y_{k'})$ recreate the image generated by migrating the data to the same point $\mb y_{k}=\mb y_{k'}$.
	In terms of \eqref{eq:2_point_mat_imag} the image is evaluated by plugging in the same search point in both variables
	$$\mathcal{I}^{CC}(\mb y_k)=\mathcal{I}^{GCC}(\mb y_k,\mb y_k).$$
	\item Reconstruct an image of $|\rho_k|^2=|\mb v_1(\tilde{\mb X})|^2_k$, i.e., calculate the top eigenvector of $\tilde{\mb X}$.  In terms of \eqref{eq:2_point_mat_imag} the image is evaluated by taking, $\mathcal{V}(\mb y_k)$, the first eigenvector of $\mathcal{I}^{GCC}(\mb y_k,\mb y_{k'})$, thought of as a matrix
	$$\mathcal{I}^{R1CC}(\mb y_k)=\mathcal V(\mb y_k).$$
	We call this the \textbf{rank-1 image}.
\end{enumerate}
The single point migration and the rank-1 image are related by
\begin{equation}
\tilde{\mb X}=\mb V \mb \Lambda \overline{\mb V}^T \Rightarrow \tilde X_{kk}=\sum\limits_{i=1}^r\lambda_i |\mb v_i(\tilde{\mb X})\hspace{0.01em}_k|^2,
\label{eq:eig_to_diag}
\end{equation}
with $r$ being the rank of $\tilde{\mb X}$. i.e., the single point migration is a weighted sum of all the eigenvectors, squared, by their respective eigenvalues. When there is a rapid decay in the eigenvalues $\lambda_i$, we expect the two methods to give similar results. However, when $\mathcal{I}^{GCC}(\mb y_k,\mb y_{k'})$ is not close to rank one, we expect the two methods to provide different results. 
\subsection{Performance of the rank-1 image}
\label{sec:perf_rank_1}
In Sections~\ref{sec:numerical_simulations} and \ref{sec:prop_filt} we compare through simulations the performance of the single-point migration and the rank-1 image. We consider imaging in a plane, i.e., the $z$ coordinate (height) is fixed, so that the image coordinate $\mb y_k\in \mathbb{R}^2$. We show that for a large enough synthetic aperture, the rotation dramatically improves the resolution of the rank-1 image compared to the single point migration image.
This is the result of two combined effects: (i) rotation induces an anisotropy in the resolution and (ii) the rank-1 image benefits from this anisotropy because it provides a resolution that is an effective average of the resolution obtained along different directions. On the other hand, the single point migration corresponds to imaging along the diagonal direction $\mb y_k= \mb y_{k'}$ for which the resolution is determined by the physical aperture of the imaging system and is not affected by the inverse synthetic aperture and the rotation of the object.  This is similar to the imaging resolution analysis in \cite{leibovich2020generalized} where resolution improvement was observed through linear motion, i.e,  an inverse synthetic aperture effect. However, the rotation enhances more dramatically the resolution than linear motion and allows optimal imaging resolution to be achieved, i.e., resolution of the order of the wavelength. 
To explain this we analyze the peaks or stationary points of the imaging functional.  

\begin{figure}[htbp]
	\centering
	\begin{subfigure}[t]{0.3  \textwidth}
		\includegraphics[width=\textwidth]{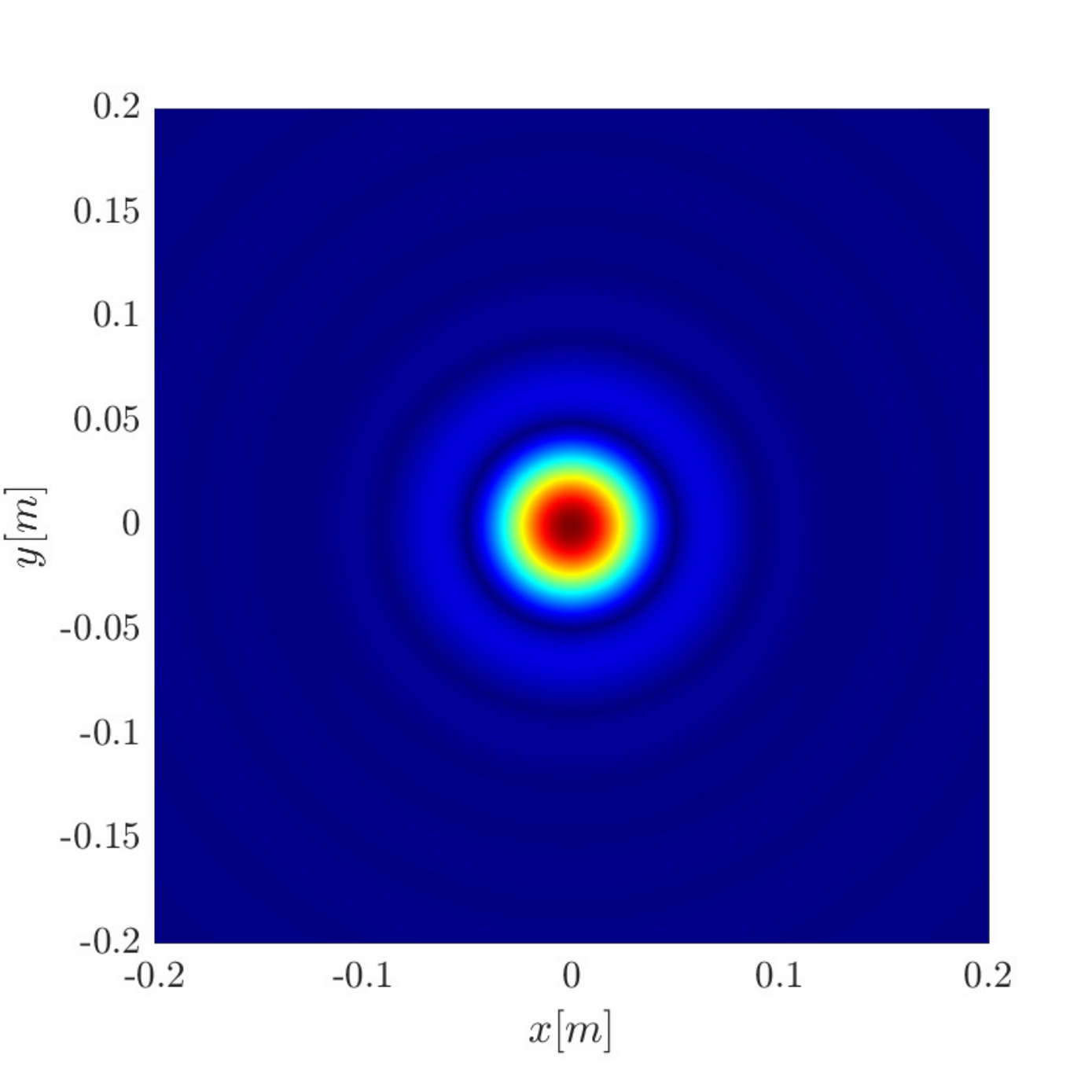}
		\caption{}
	\end{subfigure}
	\begin{subfigure}[t]{0.3  \textwidth}
		\includegraphics[width=\textwidth]{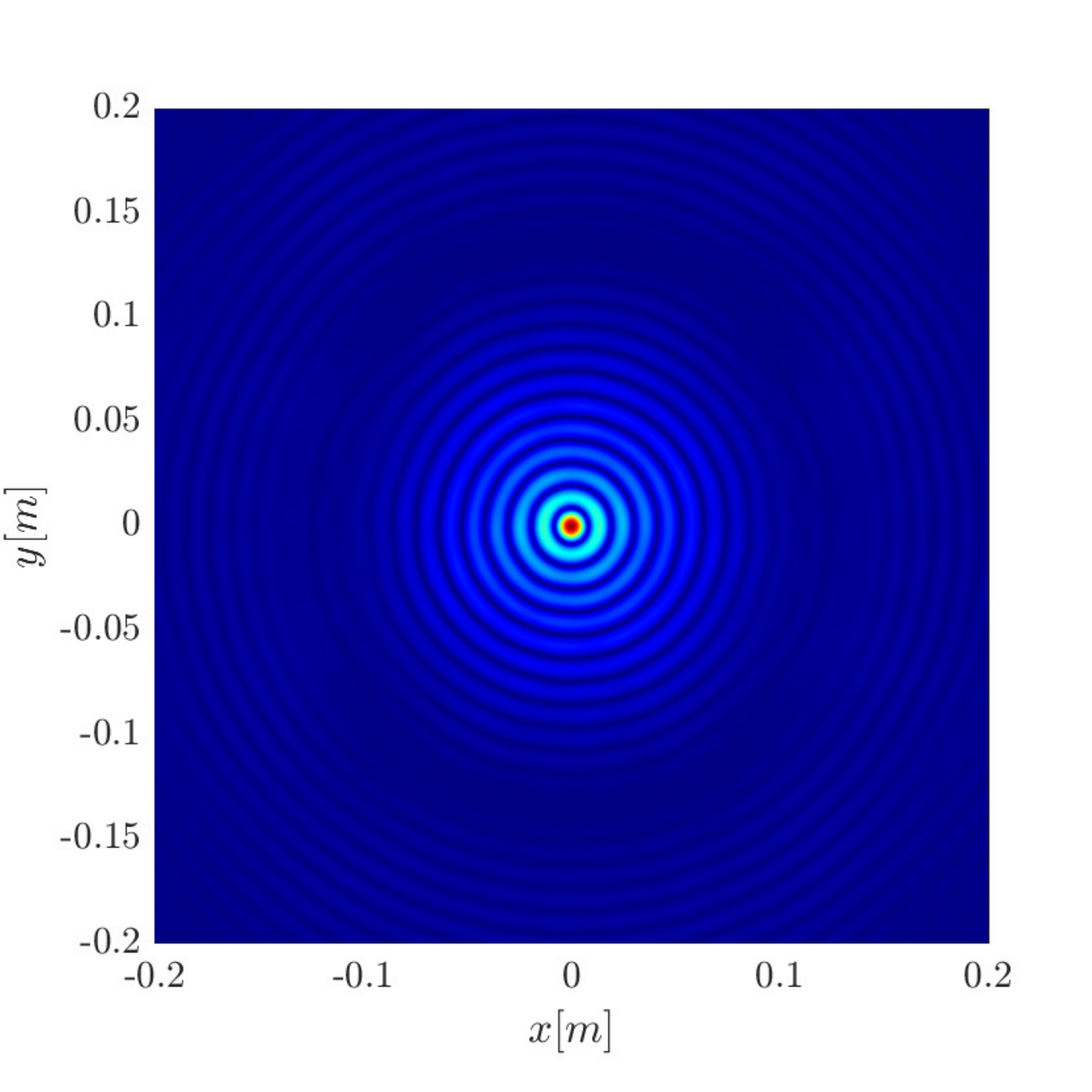}
		\caption{}
	\end{subfigure}
	\begin{subfigure}[t]{0.3\textwidth}
		\includegraphics[width=\textwidth]{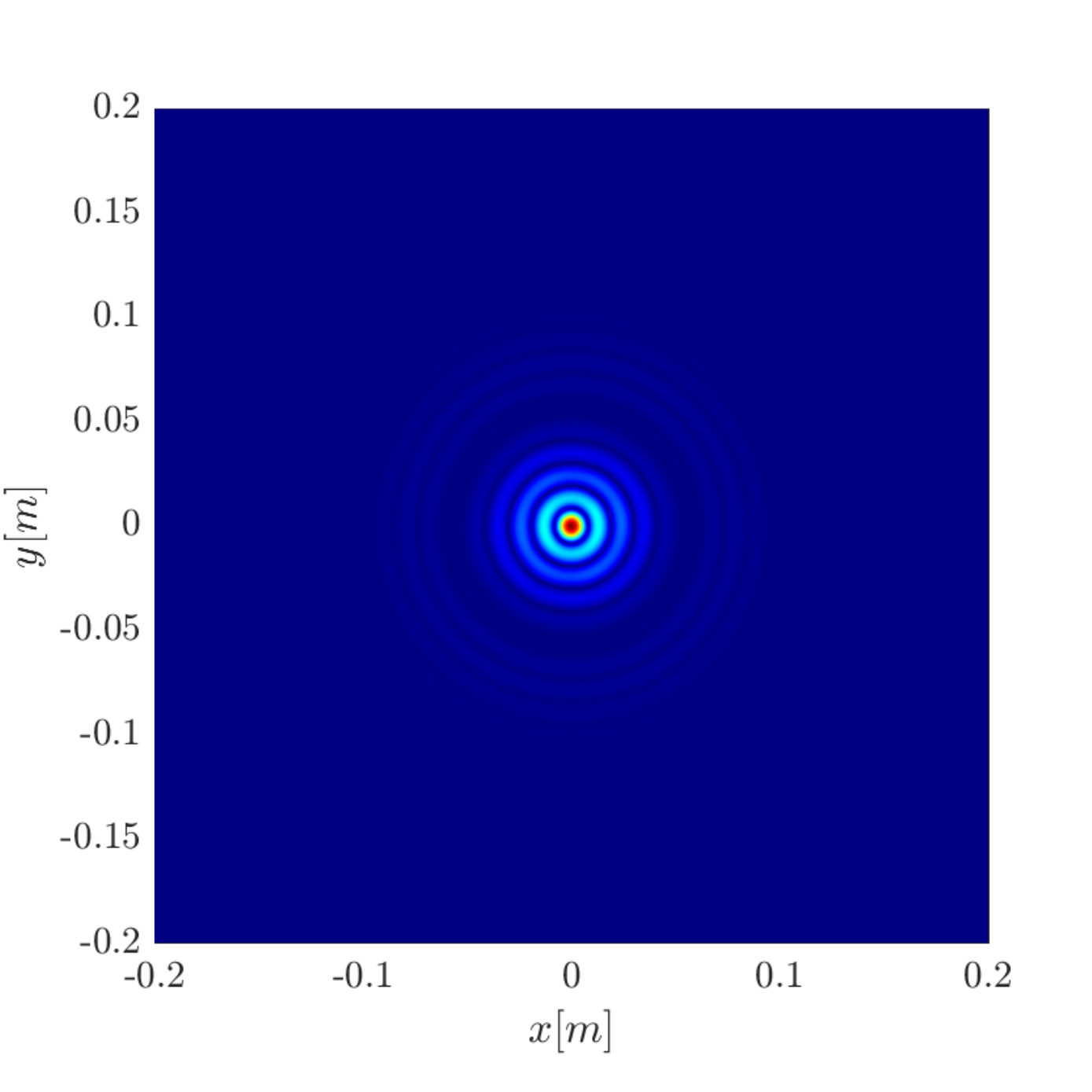}
		\caption{}
	\end{subfigure}
	\caption{\textcolor{black}{Rotation and receiver array effects for the terms in \eqref{eq:B_eff_approx}}. $(a)$ Numerical evaluation of the \textcolor{black}{induced receiver array aperture} $\int ds \mathcal{B}_A(\mathcal{R}(s) \mb x)$, $\theta=\pi/2$. We can see that the effect of time integration does not increase the resolution and it is on the scale of $\lambda H_\mb T/a$ (7 cm for our parameters). $(b)$ Evaluation of \textcolor{black}{the rotation affected phase} $\int ds e^{i\frac{\omega}{c_o}\frac{2\mb x_\mb T(s)}{H_\mb T}\cdot \mathcal{R}(s)\mb x}$. The resolution is much smaller, on the order of $\lambda$. $(c)$ Evaluation of \textcolor{black}{the product of induced array aperture and rotation affected phase}, $\int ds e^{i\frac{\omega}{c_o}\frac{2\mb x_\mb T(s)}{H_\mb T}\cdot \mathcal{R}(s)\mb x}\mathcal{B}_A(\mathcal{R}(s) \mb x)$. The resolution is dominated by the phase integral, suggesting the that the approximation in \eqref{eq:B_eff_approx} is valid. Since in the two-point interference pattern the argument of the phase is $\mb x-\mb y_i-(\mb y-\mb y_j)$, we would see anisotropy in the spot size, as illustrated in Figure~\ref{fig:single_cross}.}
	\label{fig:phase_comp}
\end{figure}

We give an estimate of the peaks of the two point interference function or matrix \eqref{eq:M_tilde} in Appendix~\ref{app:stat_phase}. The main result in this appendix is that for a large enough synthetic aperture, the rotation induces an anistoropy in the resolution of  $\mathcal{I}^{GCC}(\mb y_k,\mb y_{k'})$ in the space $\mb y_k\times \mb y_{k'}\in \mathbb{R}^4$. For a small synthetic aperture, the resolution is only determined by the size of the receiver array and is $\frac{\lambda H_\mb T}{a}$ in all directions, where $\lambda=c_0/f_0$ is the wavelength at the carrier frequency, $H_\mb T$ is the height of the target (the average distance from the receivers), and $a$ is the diameter of the imaging array spanned by the receivers. However, when the synthetic aperture grows large enough, the effect of rotation is to induce different resolution in different directions. We show in Appendix~\ref{app:stat_phase}, that  $\mathcal{I}^{GCC}(\mb y_k,\mb y_{k'})$ can be approximated as
\begin{equation}
\begin{split}
&\int d\omega |\xi(\omega)|^2 \int \limits_{-S/2}^{S/2}ds e^{i\textcolor{black}{\Phi\left((\mb x-\mb y_i)-(\mb y-\mb  y_j)\right)}}\mathcal{B}_{A}(\mathcal{R}(s)(\mb x-\mb y_i))\mathcal{B}_{A}^*(\mathcal{R}(s)(\mb y-\mb y_j)),\\
\propto&\int d\omega |\xi(\omega)|^2 \mathcal{B}_{\text{eff}}(\mb y_k-\mb y_i)\mathcal{B}_{\text{eff}}^*(\mb y_{k'}-\mb y_j)J_0\left( \frac{\omega }{c_0}2 \sin \theta_\text{rot} |(\mb y_k-\mb y_i)-(\mb y_{k'}-\mb  y_j)|\right)
\end{split}
\label{eq:B_eff_approx}
\end{equation}
with 
$$ \textcolor{black}{\Phi\left((\mb x-\mb y_i)-(\mb y-\mb  y_j)\right)={ \frac{\omega }{c_0}\frac{2\mb x_\mb T(s)-\mb x_\mb E-H_\mb R \hat{z}}{H_\mb T}\cdot\mathcal{R}(s)\left((\mb x-\mb y_i)-(\mb y-\mb  y_j)\right)}}.$$
Here $\mathcal{B}_A(\mb x)$ is the array induced resolution, which  we approximate in the continuum limit as
\begin{equation}
\mathcal{B}_A(\mb x-\mb y_i)=\int d\mb x_\mb R e^{-i \frac{\omega}{c_0}\frac{\mb x_\mb R-H_\mb R\hat{z}}{H_\mb T}\cdot(\mb x-\mb y_i)}=a^2\text{sinc}\left(\frac{\omega}{c_0}\frac{a}{2H_\mb T}(x_1-y_{i,1})\right)\text{sinc}\left(\frac{\omega}{c_0}\frac{a}{2H_\mb T}(x_2-y_{i,2})\right).
\label{eq:B_A}
\end{equation}
We see from \eqref{eq:B_A} that $\mathcal{B}_A(\mb x)$ has an effective resolution of $\frac{\lambda H_\mb T}{a}$. In \eqref{eq:B_eff_approx}, 
 $\mathcal{B}_\text{eff}(\mb x)$ is the effective array induced resolution, averaged over the synthetic aperture which, as we show in Appendix~\ref{app:stat_phase}, can be approximated as
\begin{equation}
\mathcal{B}_{\text{eff}}(\mb x)=\int ds \mathcal{B}_A(\mathcal{R}(s) \mb x).
\end{equation}
As illustrated in  Figure~\ref{fig:phase_comp}, $\mathcal{B}_\text{A}(\mathcal{R}(s)\mb x)$ is slowly varying and $\mathcal{B}_{\text{eff}}(\mb x)$ has the same resolution as $\mathcal{B}_A(\mb x)$. In \eqref{eq:B_eff_approx}, 
$J_0(x)$ is the zeroth order Bessel function. It is induced by the rotation phase $$e^{i \frac{\omega }{c_0}\frac{2\mb x_\mb T(s)-\mb x_\mb E-H_\mb R \hat{z}}{H_\mb T}\cdot\mathcal{R}(s)\left((\mb x-\mb y_i)-(\mb y-\mb  y_j)\right)},$$ whose first zero is at $\approx 2.408$.
Thus, the peak becomes anisotropic. The narrowest spot width is approximately $\frac{\lambda} {2\sin\theta_{\text{rot}}}$ and is in the direction $\hat{\mb y}_k=-\hat{\mb y}_{k'}$. On the other hand, the spot width remains unchanged, $\lambda H_\mb T/a$, in $\hat{\mb y}_k=\hat{\mb y}_{k'}$, which is the direction corresponding to the diagonal/single point migration $\mathcal{I}^{GCC}(\mb y_k,\mb y_k)$.

\begin{figure}[htbp]
	\begin{subfigure}[t]{\textwidth}
		\centering
		\begin{subfigure}[t]{0.3\textwidth}
			\includegraphics[width=\textwidth]{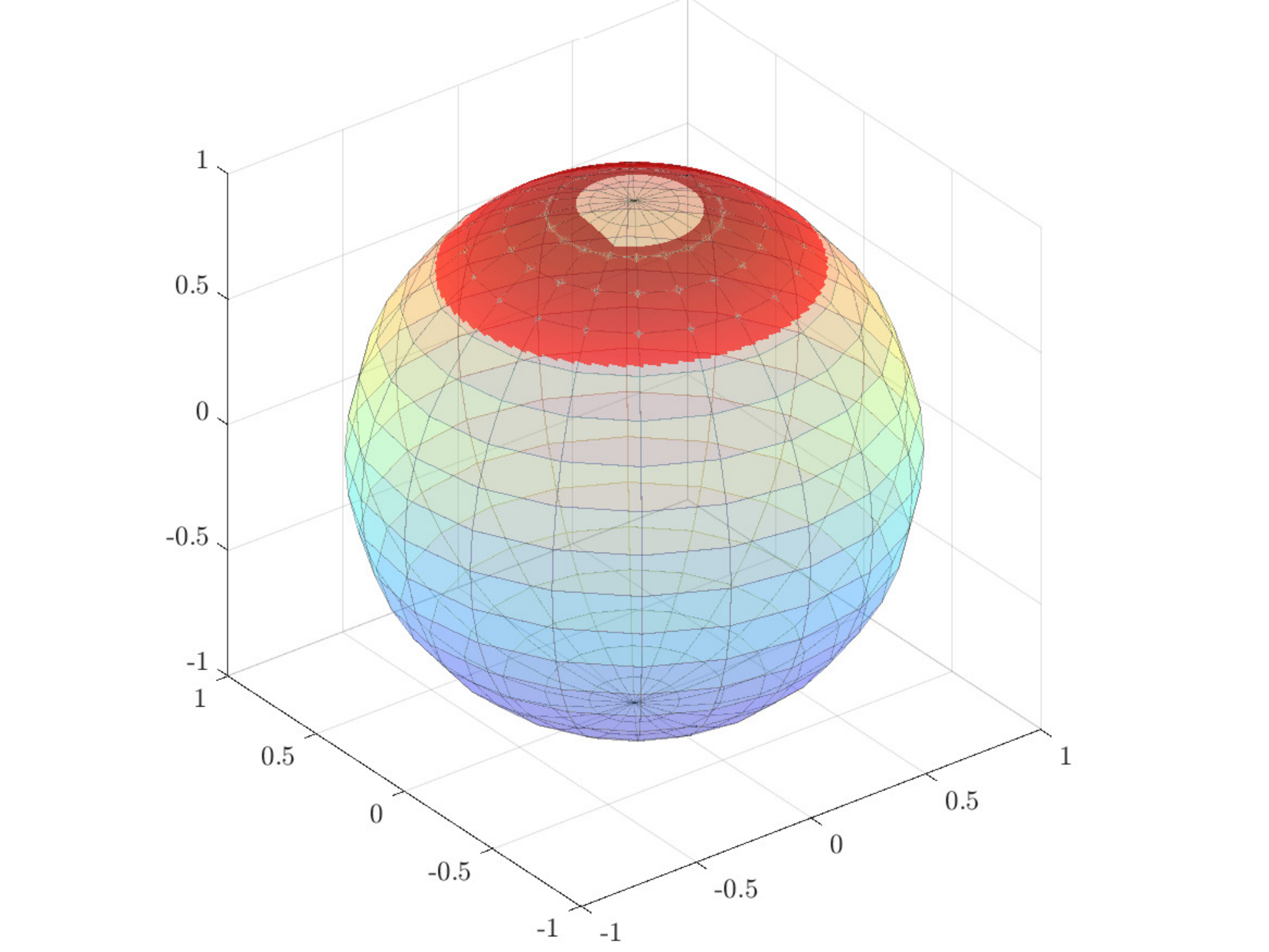}
		\end{subfigure}
		\begin{subfigure}[t]{0.3\textwidth}
			\includegraphics[width=\textwidth]{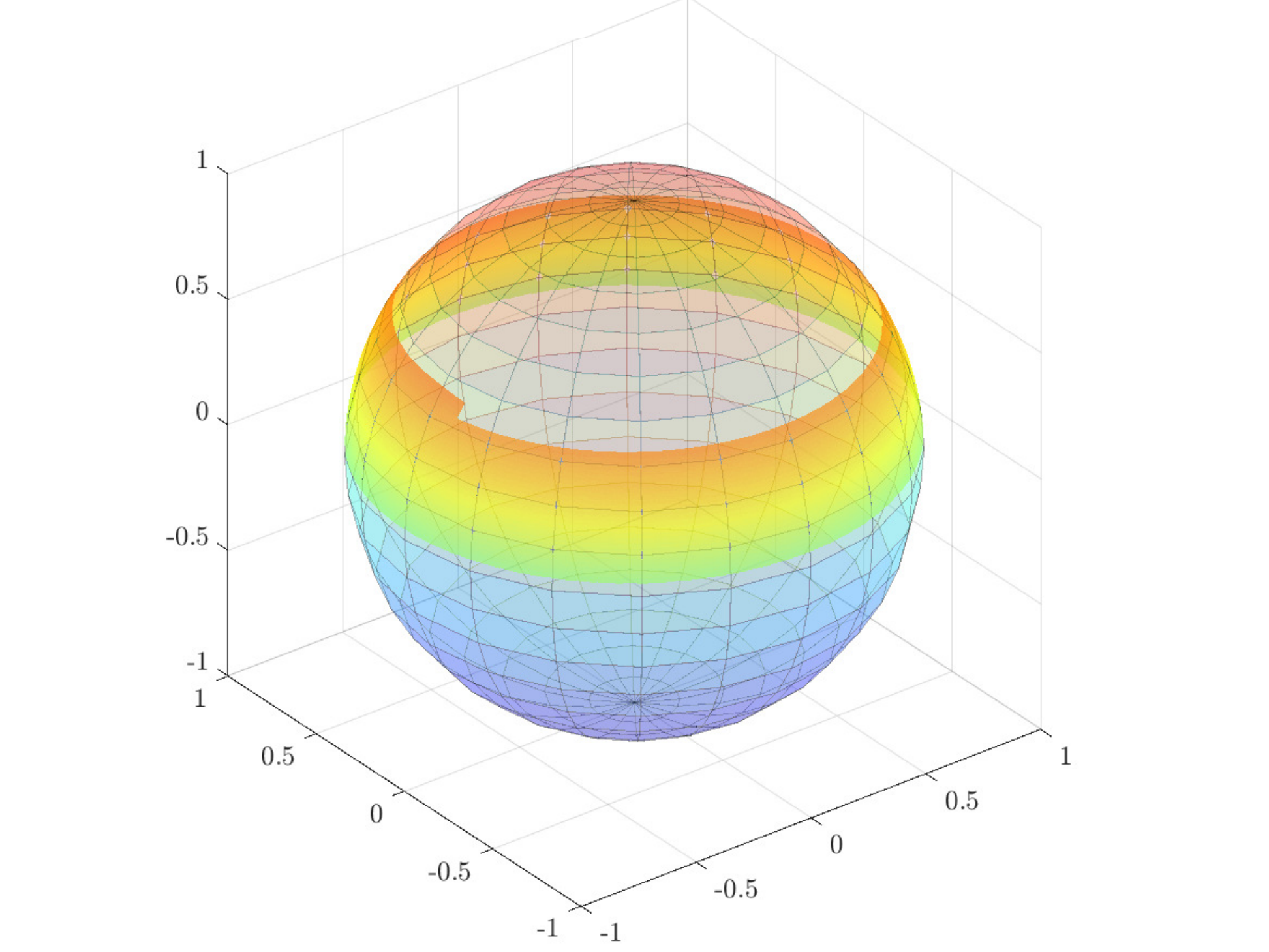}
		\end{subfigure}
	\begin{subfigure}[t]{0.3\textwidth}
		\includegraphics[width=\textwidth]{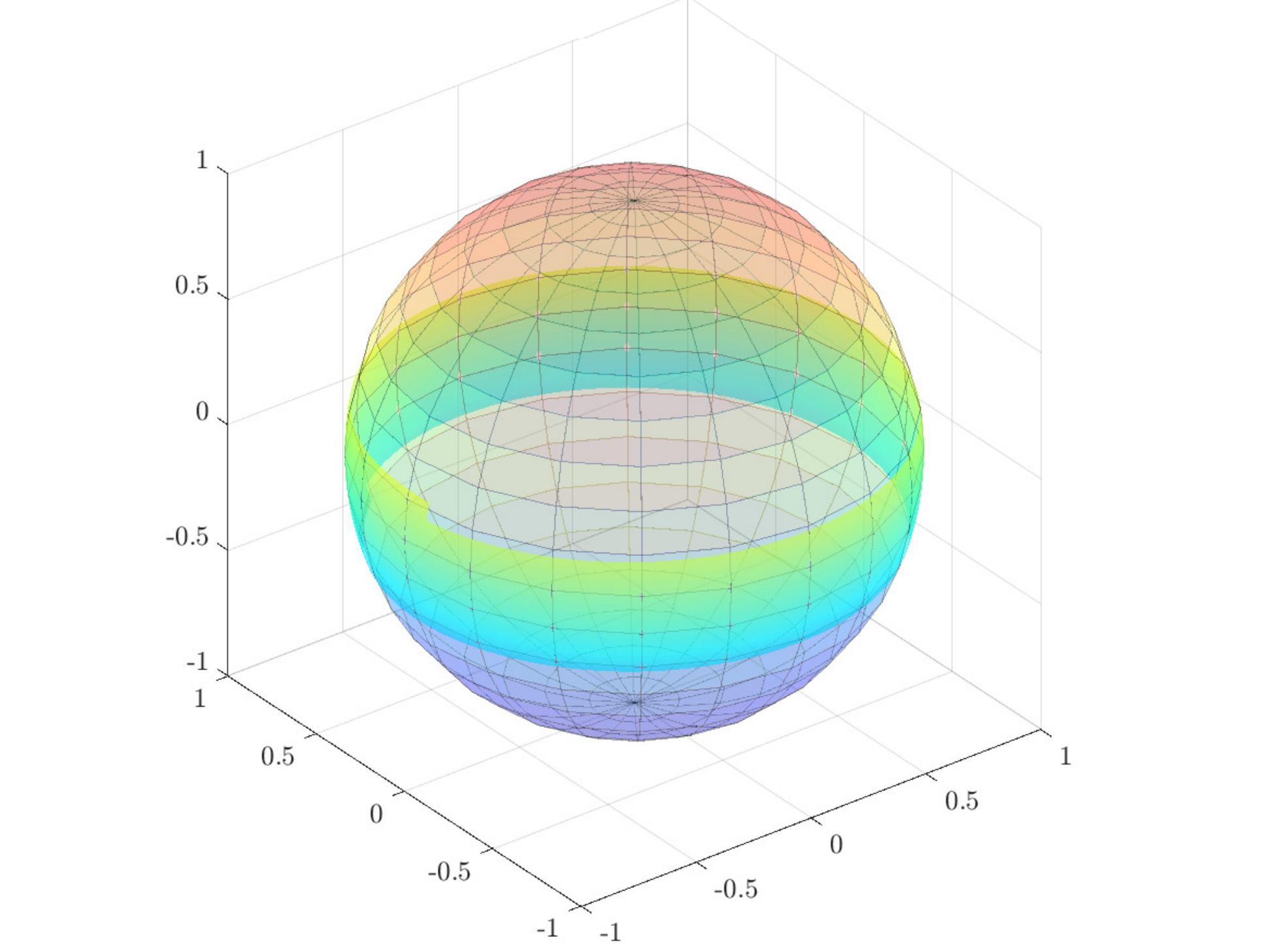}
	\end{subfigure}
	\end{subfigure}
	\begin{subfigure}[t]{\textwidth}
	\centering
	\begin{subfigure}[t]{0.3\textwidth}
		\includegraphics[width=\textwidth]{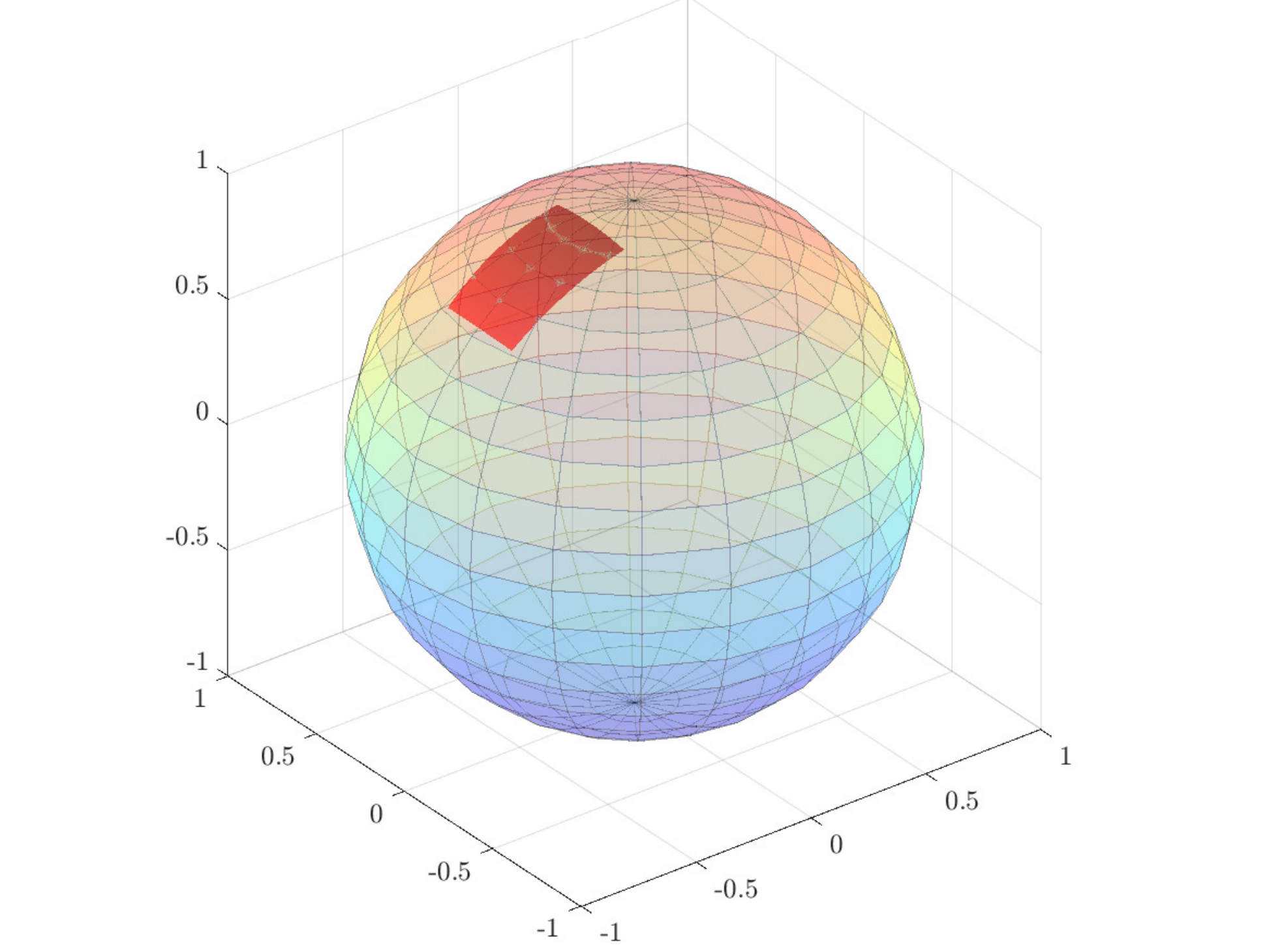}\caption{}
	\end{subfigure}
	\begin{subfigure}[t]{0.3\textwidth}
		\includegraphics[width=\textwidth]{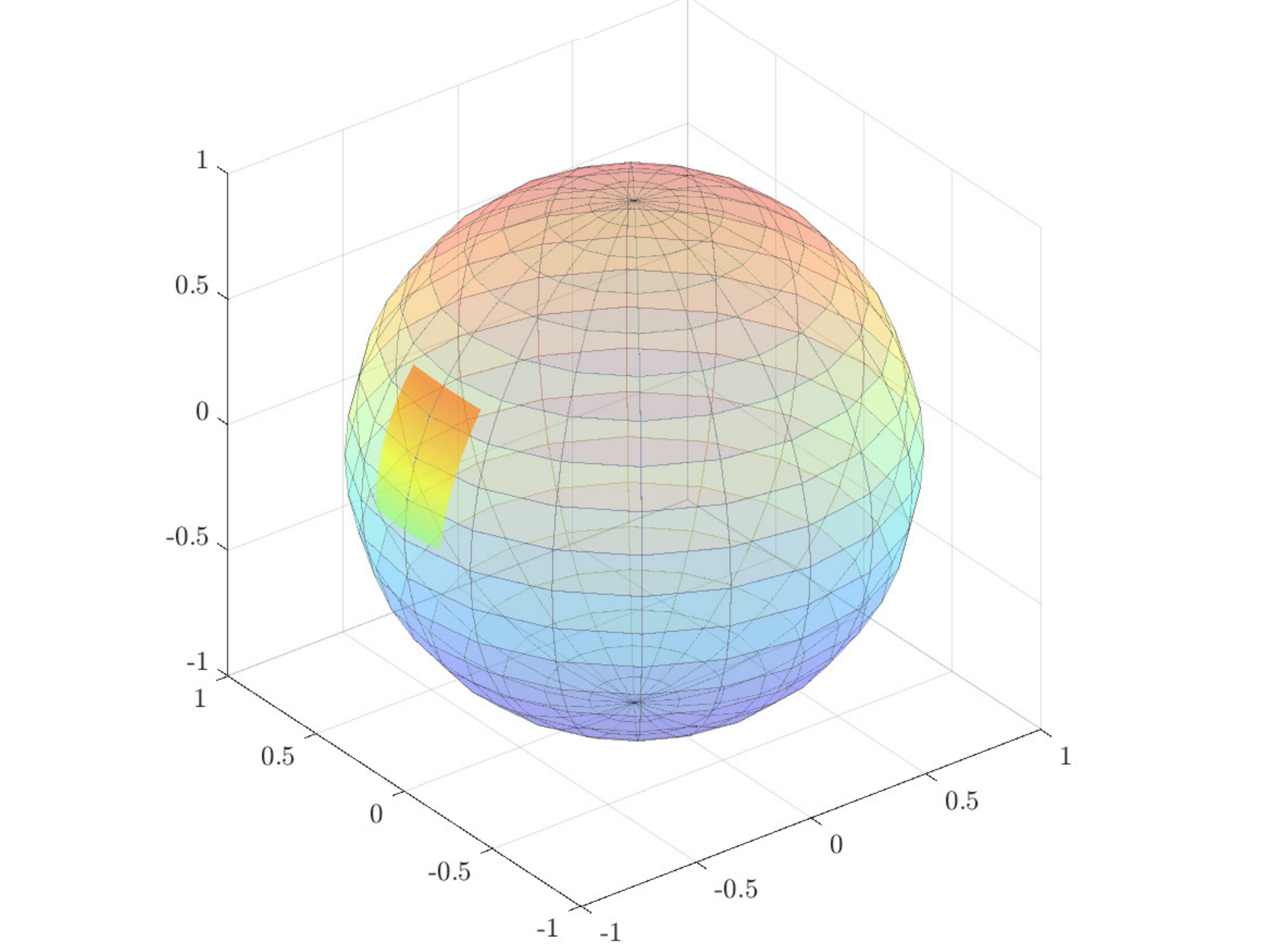}\caption{}
	\end{subfigure}
	\begin{subfigure}[t]{0.3\textwidth}
		\includegraphics[width=\textwidth]{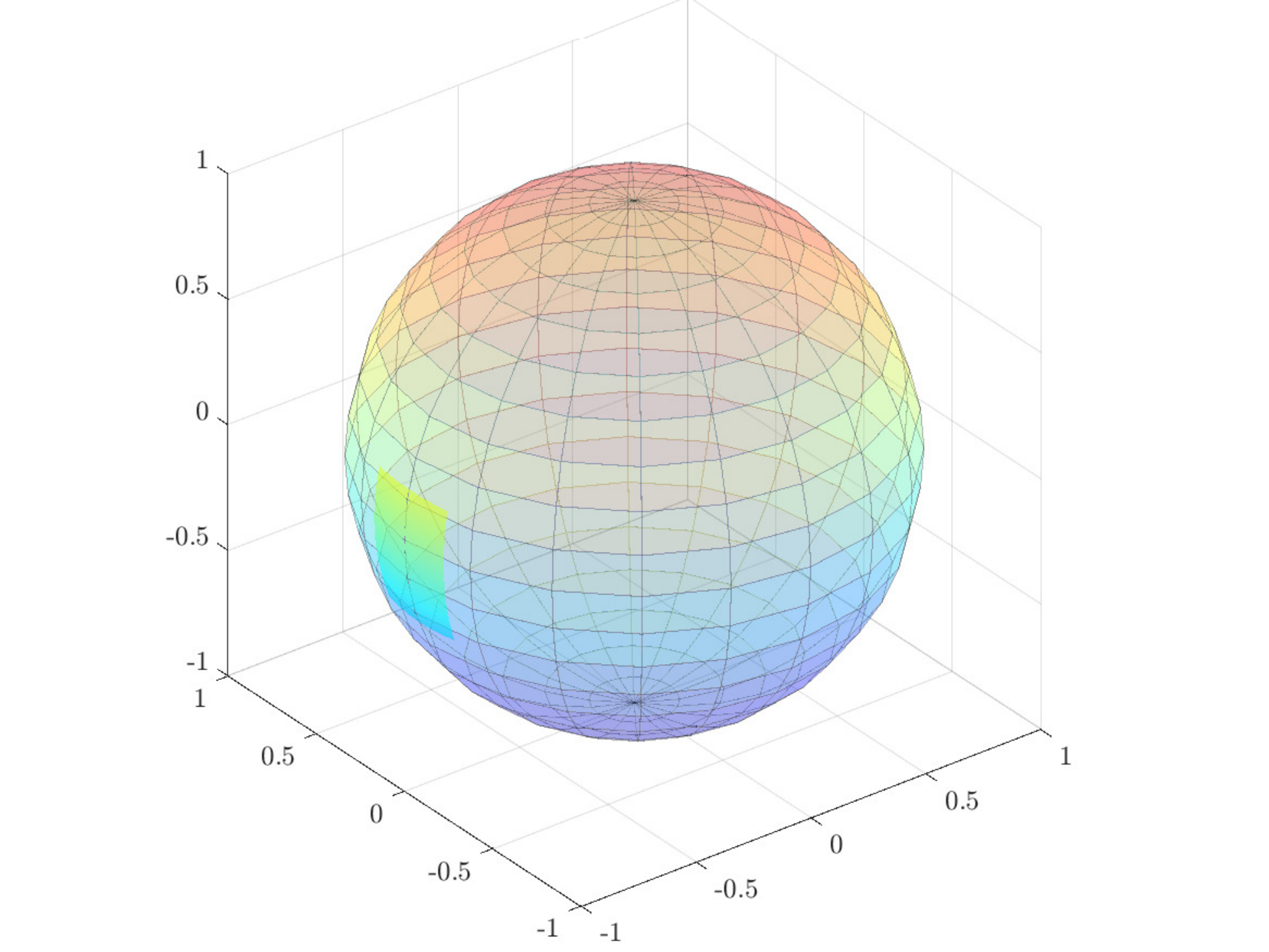}\caption{}
	\end{subfigure}
\end{subfigure}

	\caption{Illustration of the effective aperture size for different values of $\theta_{\text{rot}}$ with rotation (top) and without rotation (bottom). $(a)$ $\theta_{\text{rot}}=7\pi/8$, $(b)$ $\theta_{\text{rot}}=5\pi/8$, $(c)$ $\theta_{\text{rot}}=\pi/2$. Rotation greatly increases the effective aperture size, and thus improves the resolution. For a detailed resolution analysis see Appendix~\ref{app:stat_phase}.}
	\label{fig:k_w_1}
\end{figure}

	The effect of rotation is also illustrated by considering the effective array size induced by rotation, as shown in Figure~\ref{fig:k_w_1}. We observe that rotation greatly increases the effective aperture size which significantly improves the resolution. The effect is depended on $\theta_{\text{rot}}$, as suggested by numerical simulations in Section~\ref{sec:numerical_simulations}, and the analysis in Appendix~\ref{app:stat_phase}. 

In \cite{leibovich2020generalized} we analyzed the effect of anisotropy on resolution by considering the eigenfunction of one dimensional continuous kernels that have an anisotropic form.  We show there that the top eigenfunction of anisotropic kernels can be narrower than the maximal width of the kernel, especially for functions like sinc, which show up in our analysis. This is also true when considering a superposition of such kernels in the form that arises in the two point interference pattern, in which the kernels are localized around all possible pairs of target locations.
In the continuum limit, if a kernel $\mathcal{K}(x,y)$ is a linear combination of translations of the localized kernel $K(x,y)$
\begin{equation}
\mathcal{K}{}(x,y)=\sum\limits_{i,j} c_{ij}K(x-a_i,y-a_j),
\end{equation}
 then as long as the translations are far enough apart such that
\begin{equation}
\int dyK(x-a_i,y-a_j)u(y-a_k)= c_{ij} u(x-a_i)\delta_{jk},
\label{eq:loc_cond}
\end{equation}
the top eigenfunction of $\mathcal{K}$ is a linear combination of the translated local top eigenfunctions
$$
\mathcal{U}(x)=\sum\limits_i \alpha_i u(x-a_i)$$
with the restriction
$$
\sum\limits_{i,j} c_{ij} \alpha_j=\lambda\alpha_i$$
i.e, the eigenfunction's coefficient vector is an eigenvector of the coefficient matrix $c_{ij}$.

The combined effect of these observations is that the rank-1 image provides better resolution than the single point migration image. Since in the case of a rotating object the main cause for the anisotropy is the object rotation, it leads to higher resolution on a time scale of a period of rotation, and the resolution itself can be up to the order of a wavelength. 

\begin{figure}
	\centering
\begin{subfigure}[t]{0.25\textwidth}
	\includegraphics[width=\textwidth]{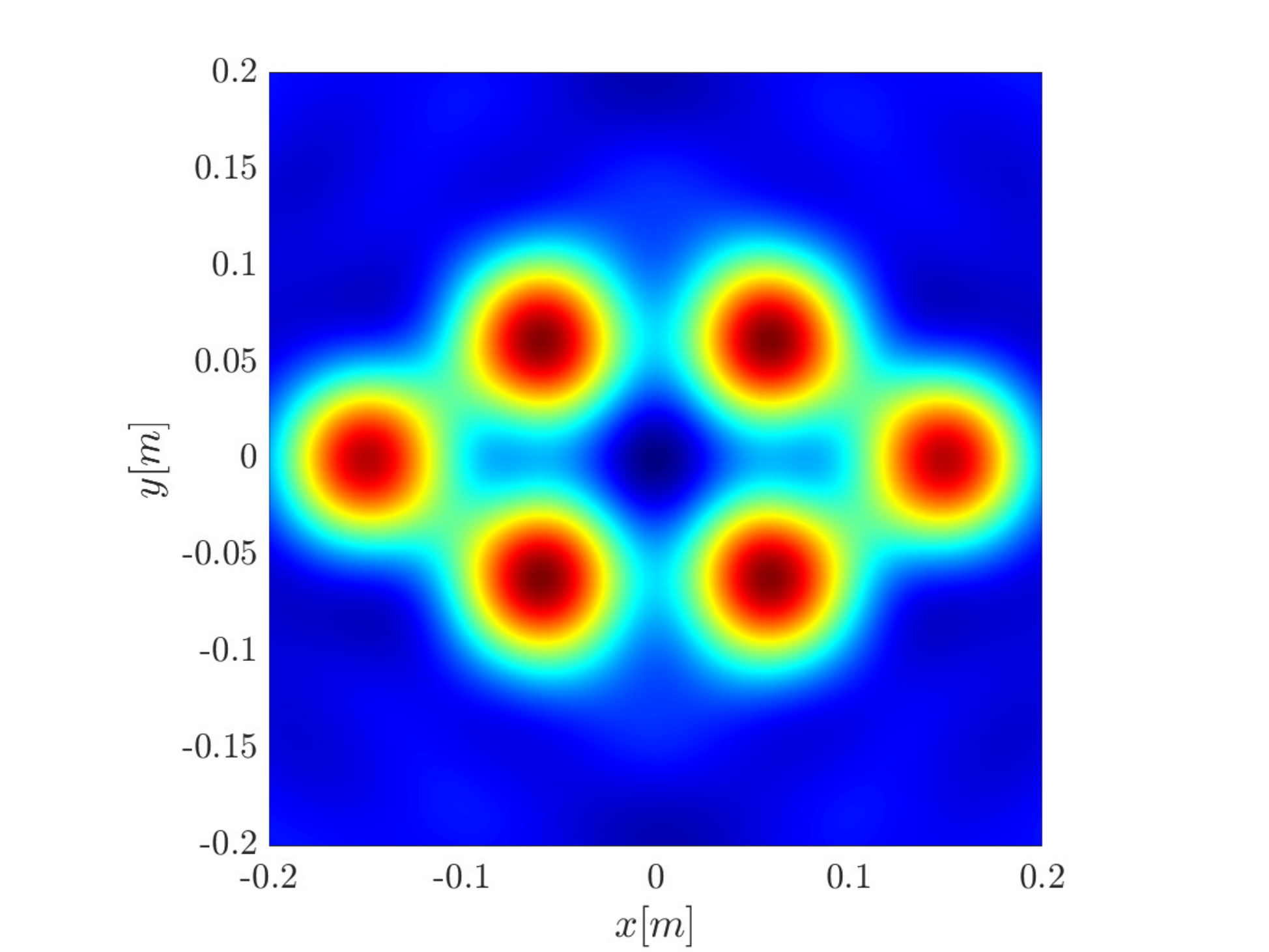}
	\caption{}
\end{subfigure}
\begin{subfigure}[t]{0.25\textwidth}
	\includegraphics[width=\textwidth]{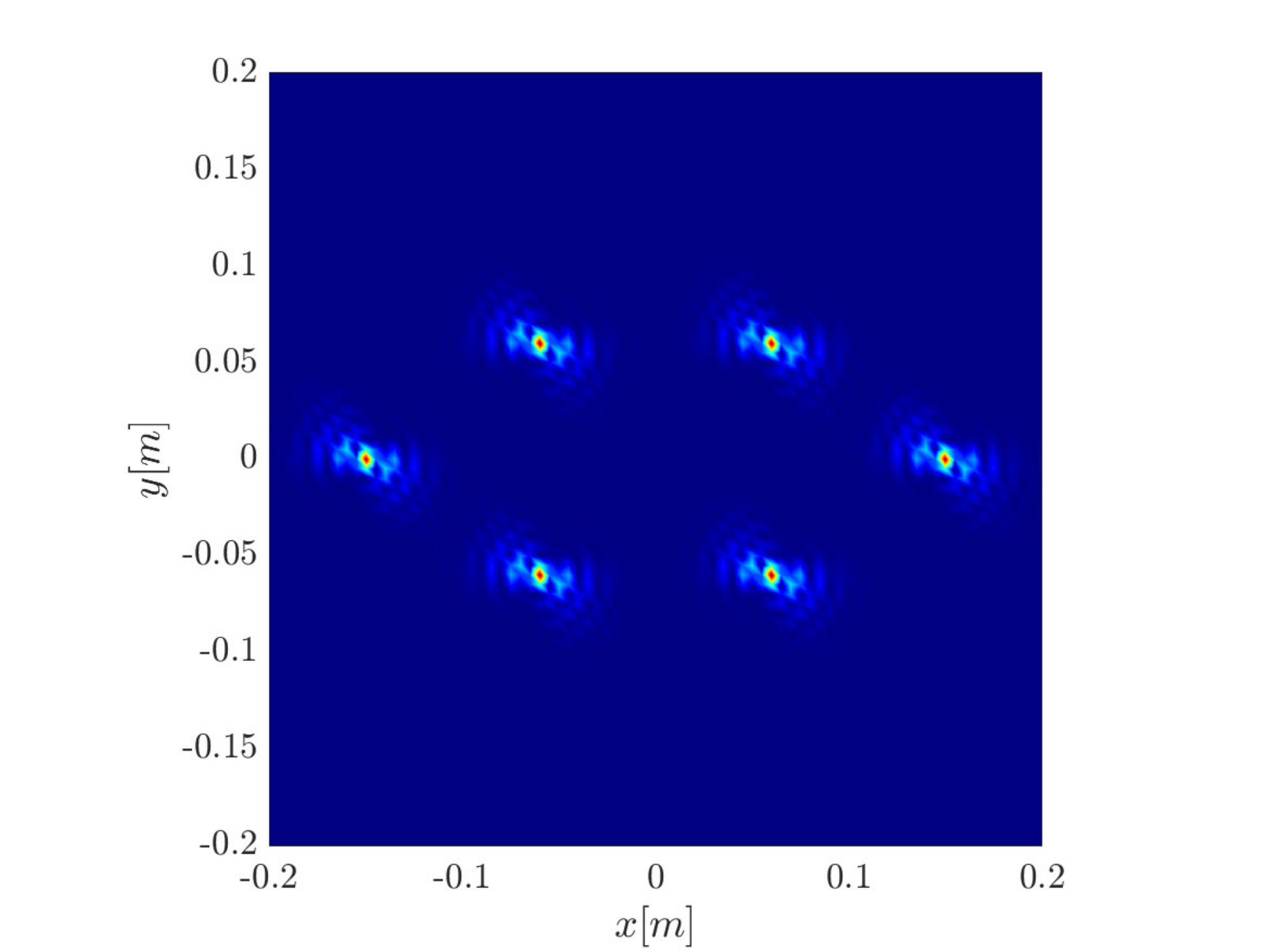}
	\caption{}
\end{subfigure}
\begin{subfigure}[t]{0.25\textwidth}
	\includegraphics[width=\textwidth]{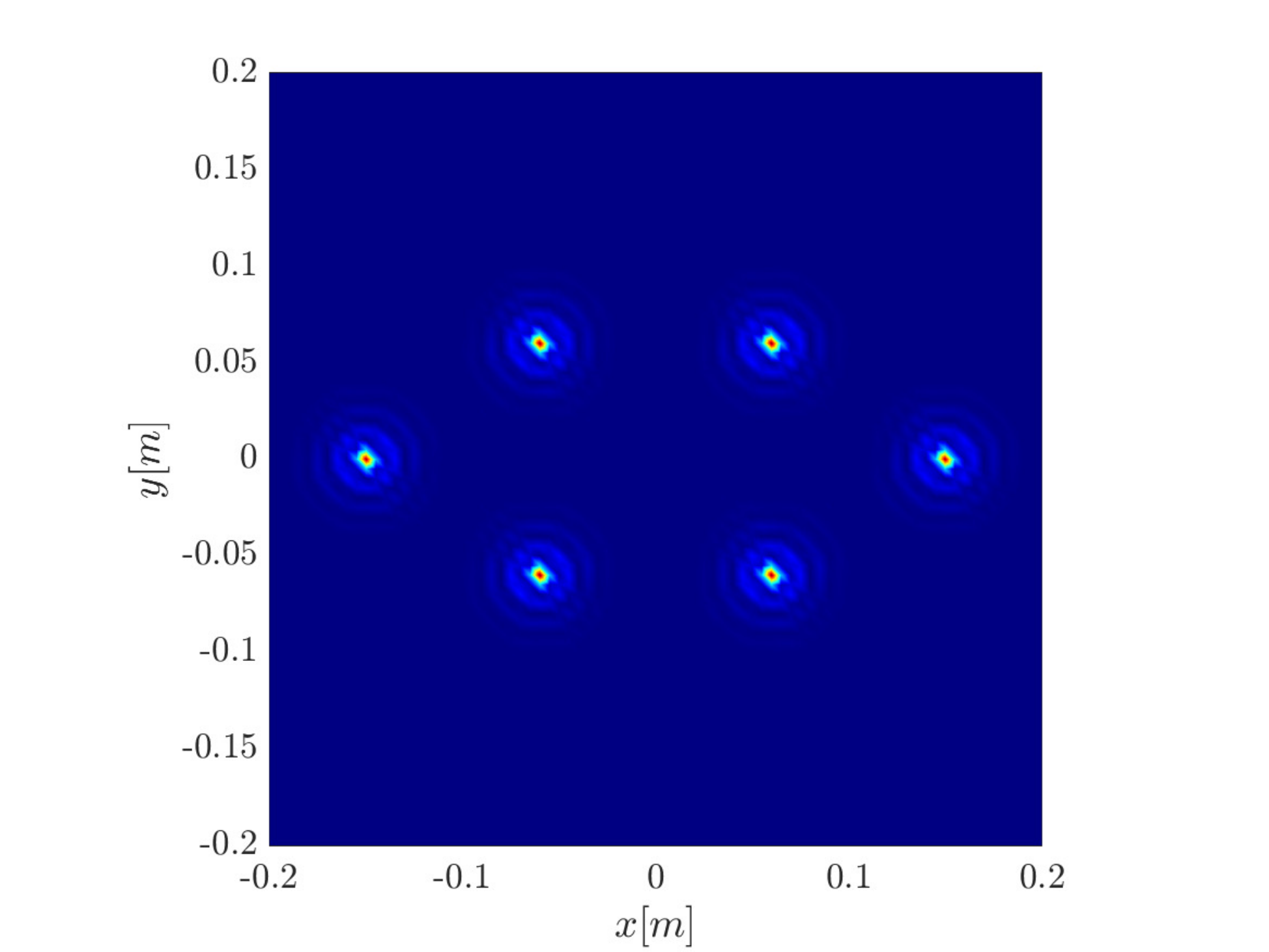}
	\caption{}
\end{subfigure}
\caption{Example of the performance of the rank-1 image for rotating objects. $\theta_{\text{rot}}=3\pi/4$. Aperture size is $1500\Delta s$. $(a)$ Single point migration. $(b)$ Rank-1 image, $(c)$ Kirchhoff migration. The rank-1 image achieves a resolution on the scale of $\lambda$, comparable to the linear case.}
\label{fig:rot_example}
\end{figure}

An illustration of the performance of the rank-1 image is shown in Figure \ref{fig:rot_example} where we see that the rank-1 image (see  Figure~\ref{fig:rot_example}-(b)) provides a significant improvement in resolution compared to the single point migration (see Figure~\ref{fig:rot_example}-(a)). The true object to be reconstructed is shown in Figure \ref{fig:sat_demo_layout} and the parameters used in the numerical simulations are given in Section \ref{sec:numerical_simulations}.
In Section~\ref{sec:numerical_simulations} we present more numerical simulations and explore the range of parameters for which the rank-1 image performs well. We further investigate the performance of the rank-1 image in Section~\ref{sec:prop_filt} by considering the case of a single target. The numerical results are in accordance with the analysis of Appendix~\ref{app:stat_phase} that we summarized above. 

\section{Numerical simulations}
\label{sec:numerical_simulations}
In this section we compare with numerical simulations the performance of the different migration schemes for a fast moving and rotating target.  The object to be imaged is illustrated in Figure~\ref{fig:sat_demo_layout}. It is a cluster composed of 6 point scatterers. The scatterers are at $z=500$km height and they all move with the same linear speed $7.6$ km/s, as well as rotating in plane with angular velocity $\omega_r=2\pi/5$[sec$^{-1}$]. The rotation axis has an elevation angle $\theta_{\text{rot}}$ varying between $5\pi/8$ and $\pi$. In the inertial frame of reference the scatterers are in the $x-y$ plane. The targets are located at $(0,\pm0.15)[m]$ and $(\pm 0.06,\pm 0.06)$ [m]. The total synthetic aperture size used is $22.5$s with $\Delta s=0.015$s. For the signal, the carrier frequency $f_o=9.6$GHz, and  $B=0.5\times622$MHz. There are 15 receivers uniformly distributed in an area 200km$\times$200km. \textcolor{black}{These parameters are representative of realistic scenarios, as mentioned in Section~\ref{sec:intro}. While the six point scatterer model is a simplification, it captures the geometry of the satellite with the reflection concentrated on the corners of the bulk and solar panels.}

\begin{figure}[htbp]
	\centering
	\begin{subfigure}[t]{0.24\textwidth}
		\includegraphics[width=\textwidth]{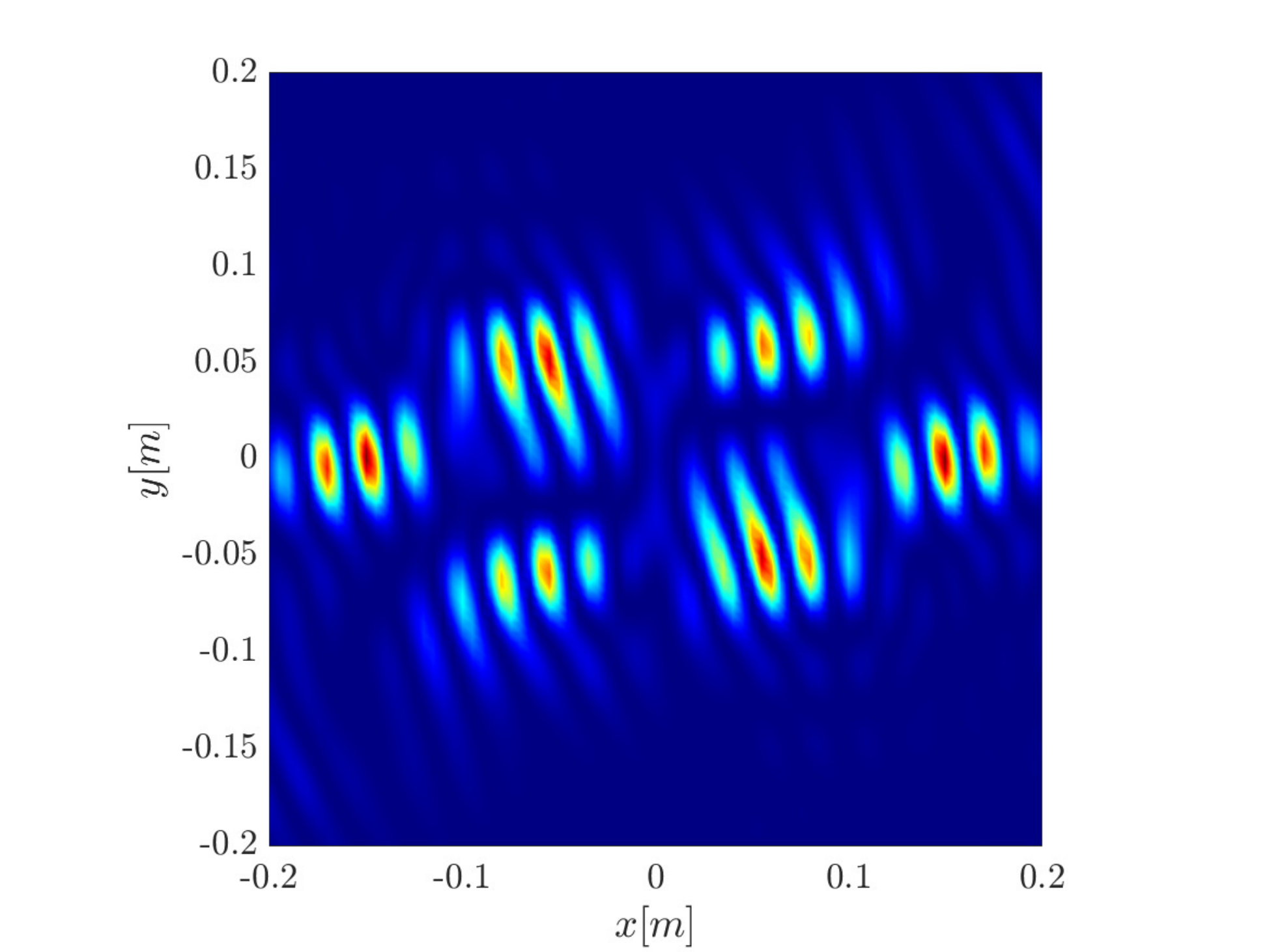}
		\caption{}
	\end{subfigure}
	\begin{subfigure}[t]{0.24\textwidth}
		\includegraphics[width=\textwidth]{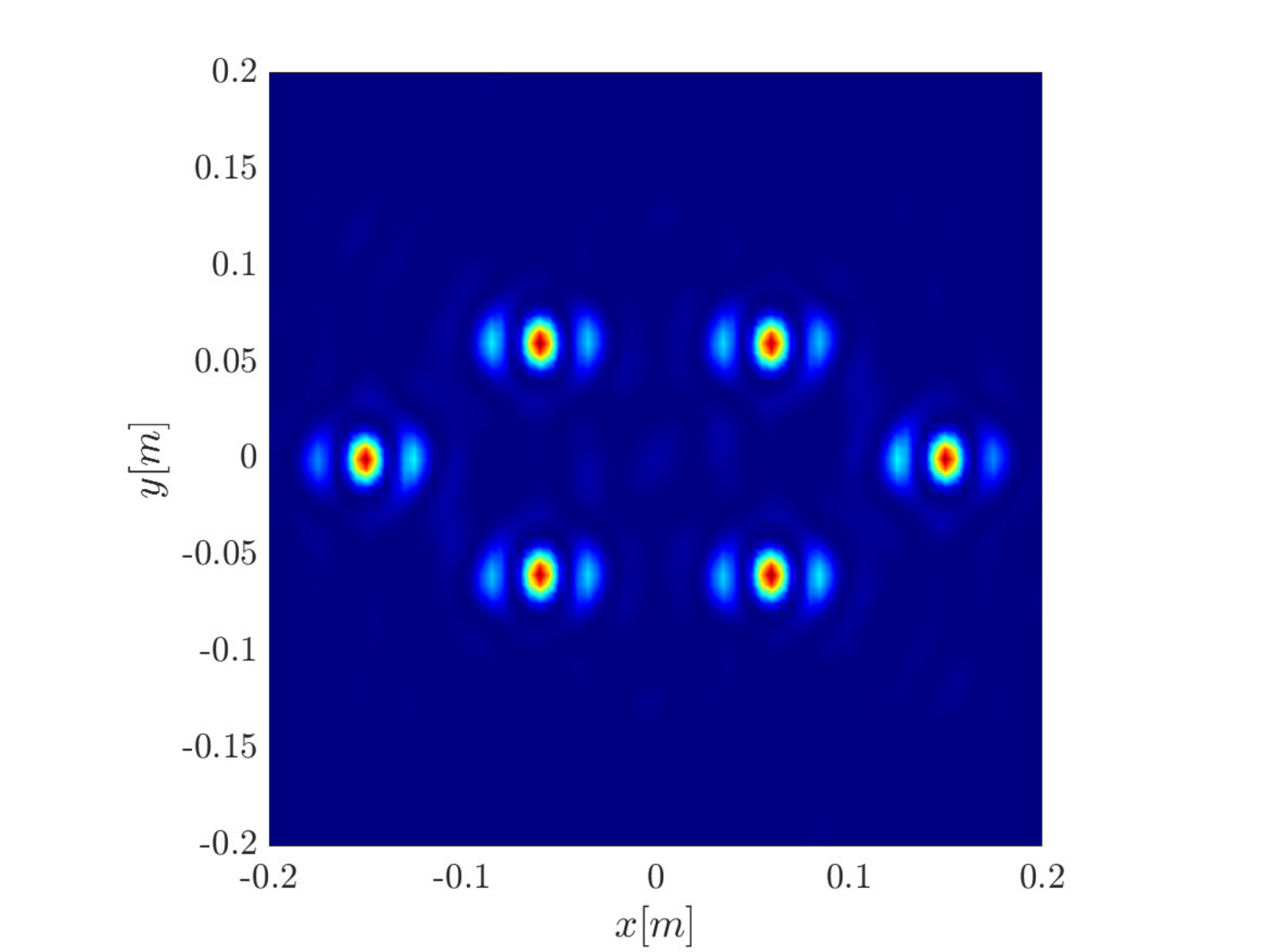}
		\caption{}
	\end{subfigure}
	\begin{subfigure}[t]{0.24\textwidth}
	\includegraphics[width=\textwidth]{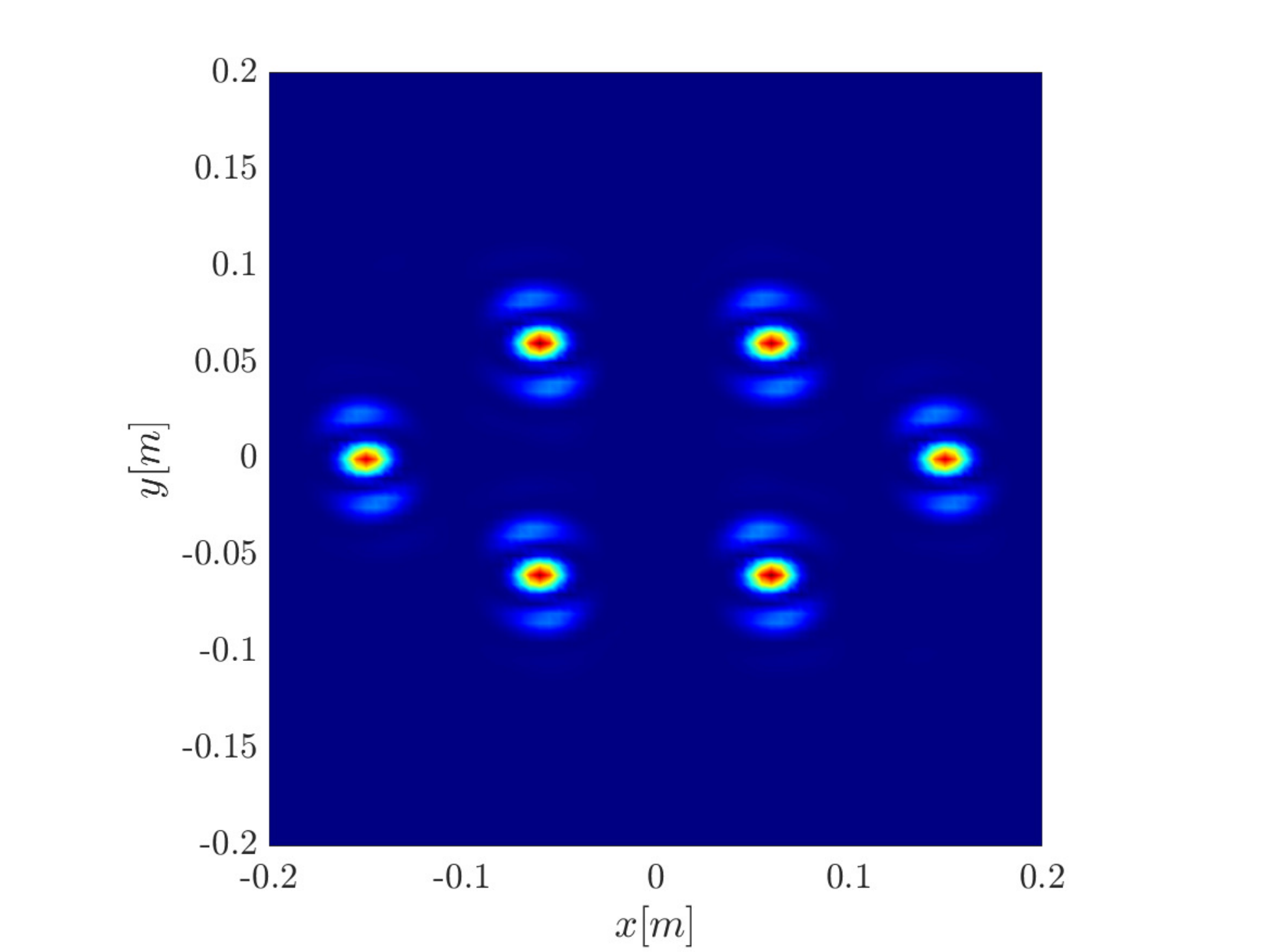}
	\caption{}
\end{subfigure}
	\begin{subfigure}[t]{0.24\textwidth}
		\includegraphics[width=\textwidth]{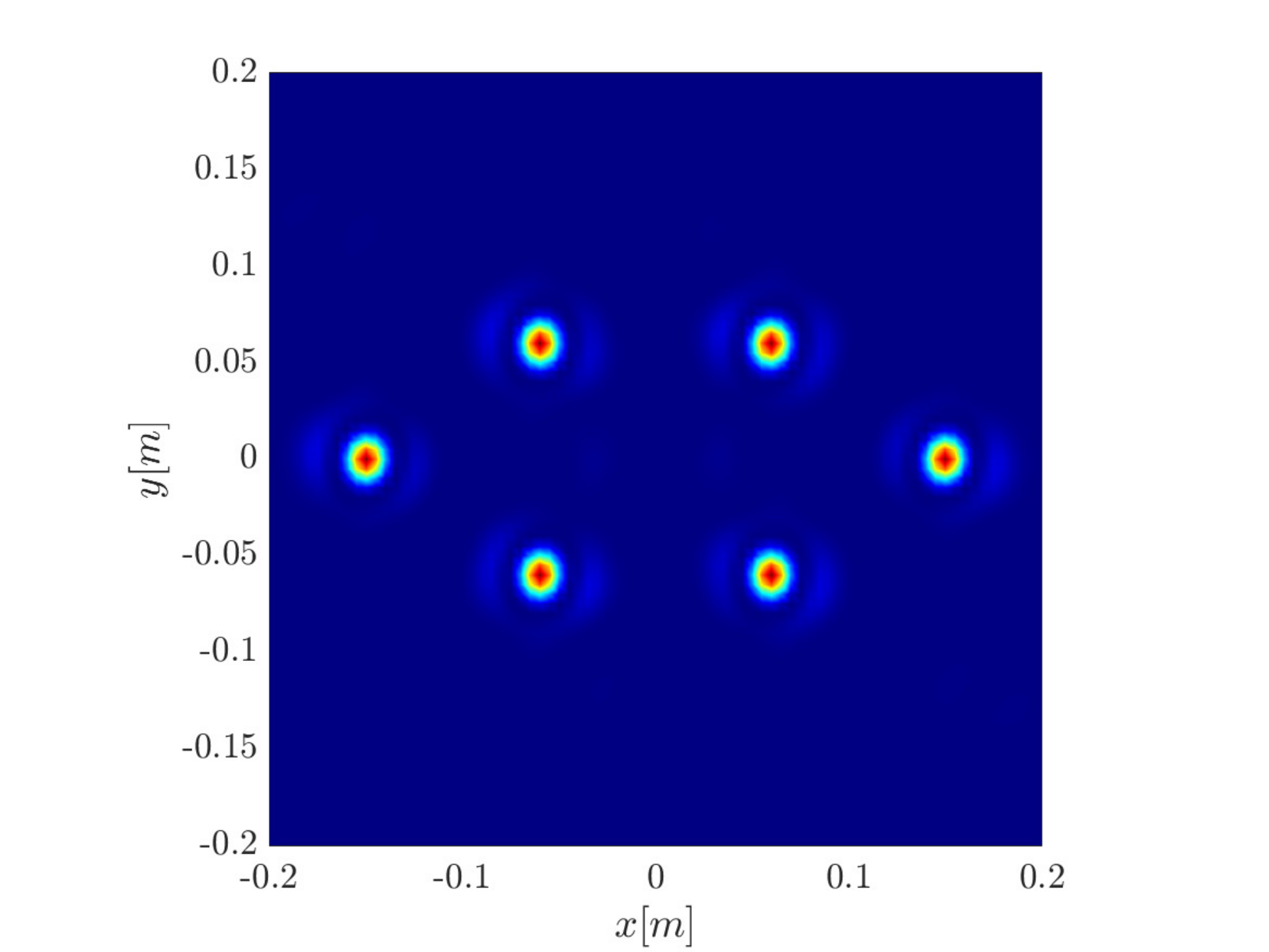}
		\caption{}
	\end{subfigure}
	\caption{Rank-1 image as a function of the synthetic aperture size $\theta_{\text{rot}}=7\pi/8$. $(a)$ $50\Delta s$,$(b)$ $500\Delta s$ and $(c)$ $1000\Delta s$,$(d)$ $1500\Delta s$. We can see that as the synthetic aperture increases the image converges to a finite resolution. }
	\label{fig:rot_converge}
\end{figure}

In Figure~\ref{fig:rot_converge} we show results for the rank-1 image for $\theta_{\text{rot}}=7\pi/8$ and different synthetic aperture sizes ranging from $50\Delta s$ to $1500\Delta s$. Resolution is improved as the synthetic aperture increases. However, as the reason for resolution enhancement is the rotation of the object, the maximal resolution is attained after a full object rotation which is of the order of 300$\Delta s$. This is in accordance with the analysis in Appendix \ref{app:stat_phase}, which suggests that the resolution for the two point interference functional converges to a finite size after a period of the revolution. 

\begin{figure}[htbp]
	\centering
	\begin{subfigure}[t]{0.25\textwidth}
		\includegraphics[width=\textwidth]{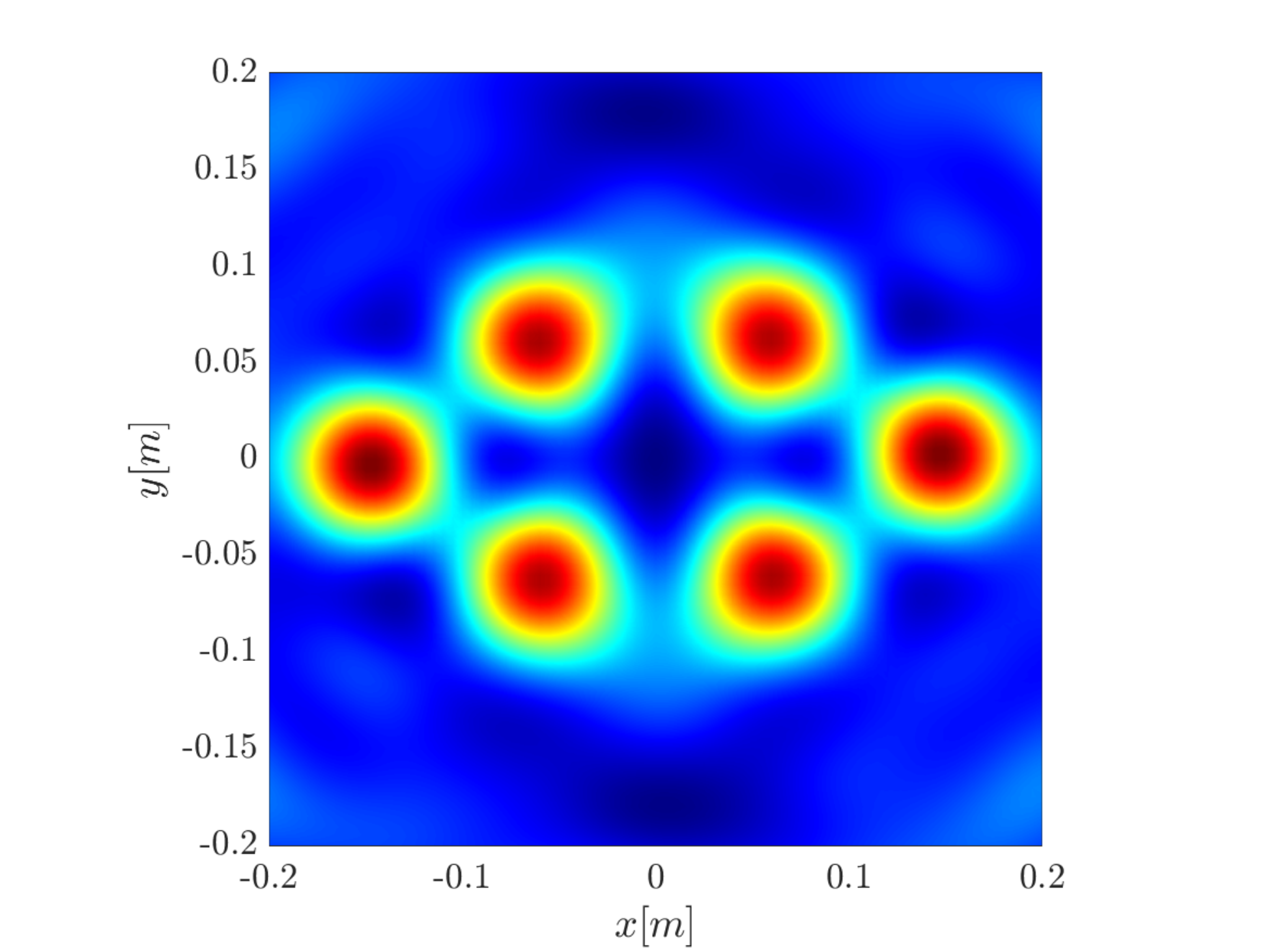}
	\end{subfigure}
	\begin{subfigure}[t]{0.25\textwidth}
		\includegraphics[width=\textwidth]{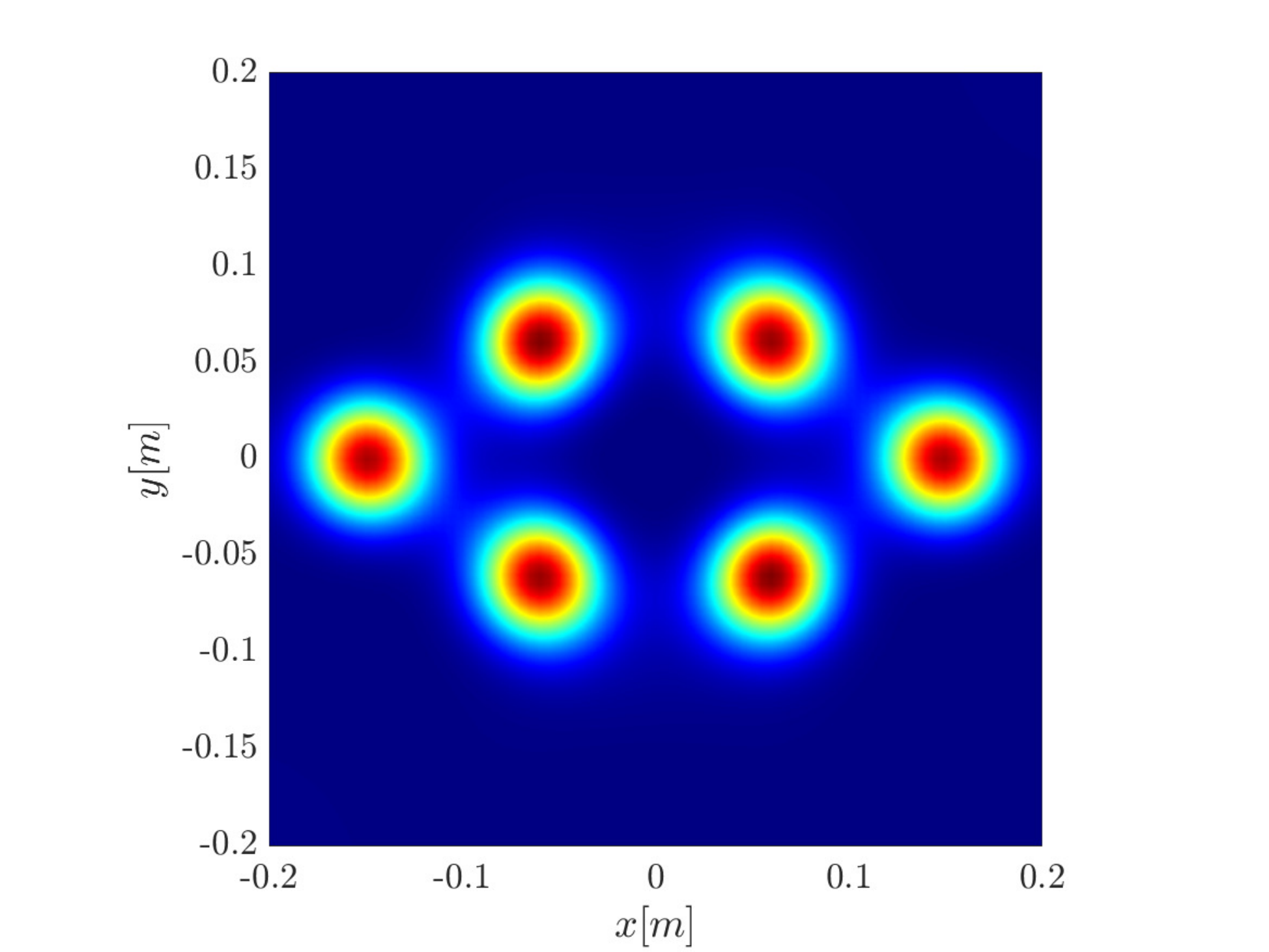}
	\end{subfigure}
	\begin{subfigure}[t]{0.25\textwidth}
		\includegraphics[width=\textwidth]{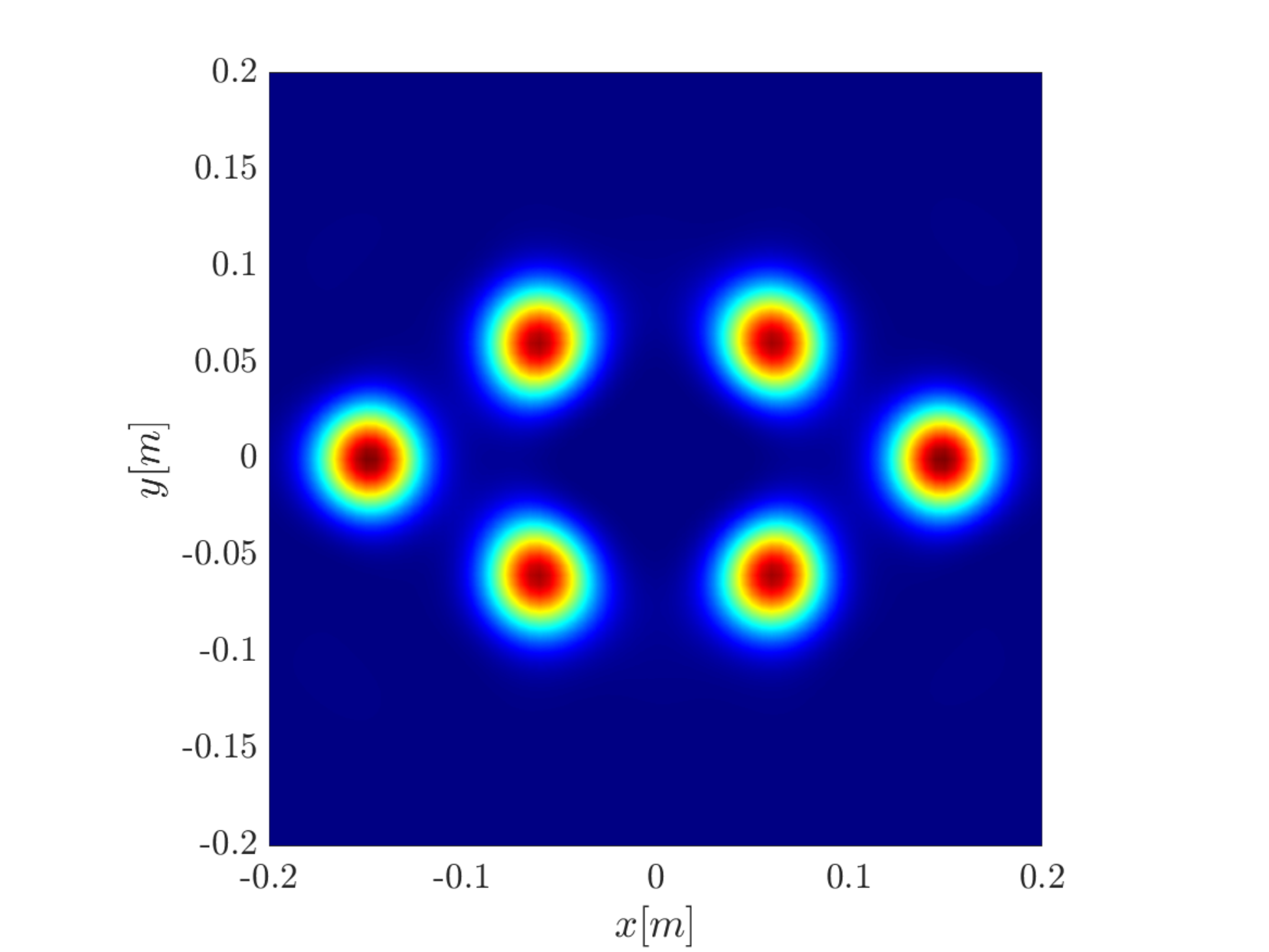}
		\end{subfigure}
\begin{subfigure}[t]{0.25\textwidth}
	\includegraphics[width=\textwidth]{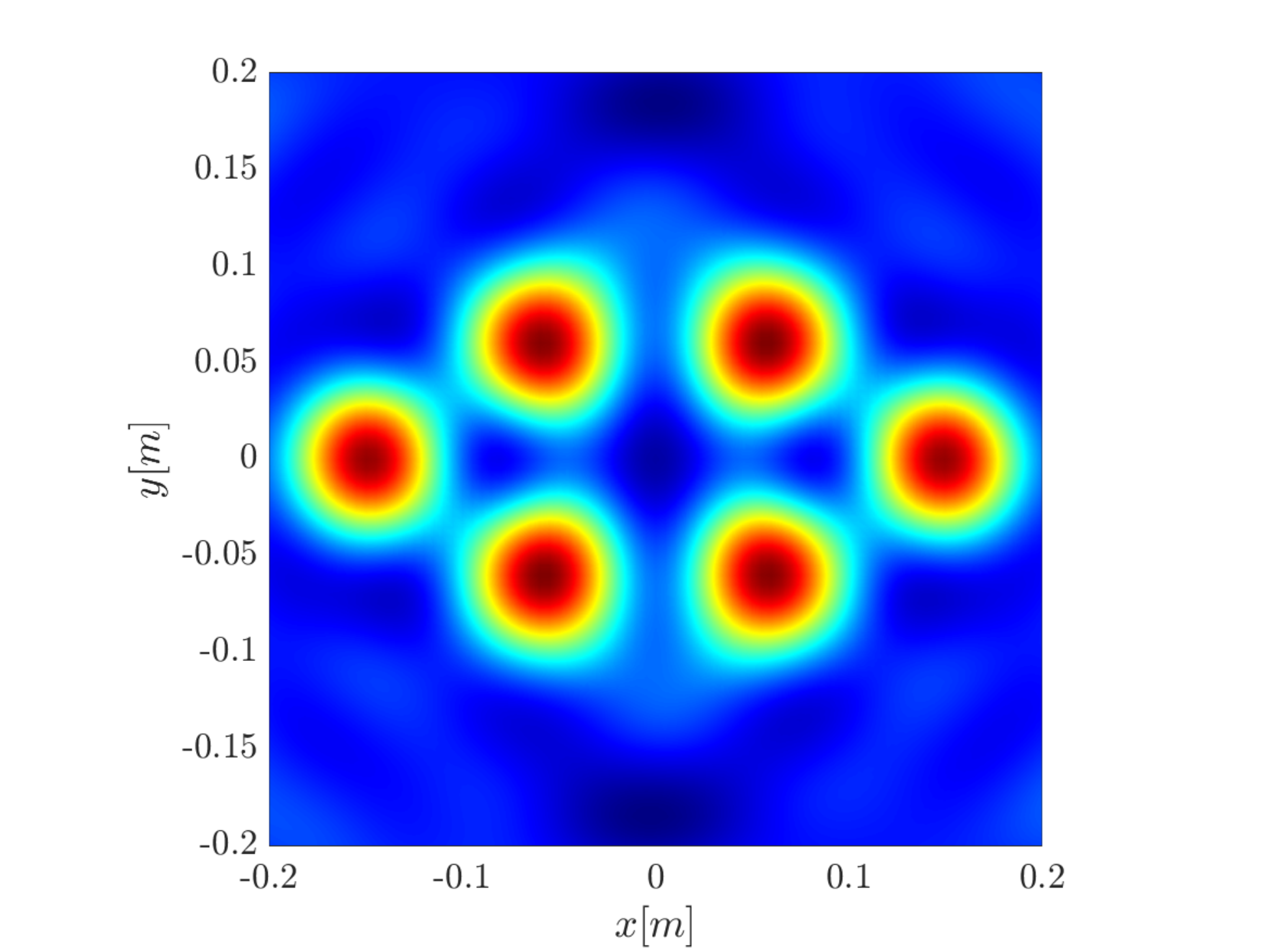}
\end{subfigure}
\begin{subfigure}[t]{0.25\textwidth}
	\includegraphics[width=\textwidth]{LPT_Figures_accepted/1st_mode_1501ds_theta_3-eps-converted-to.pdf}
\end{subfigure}
\begin{subfigure}[t]{0.25\textwidth}
	\includegraphics[width=\textwidth]{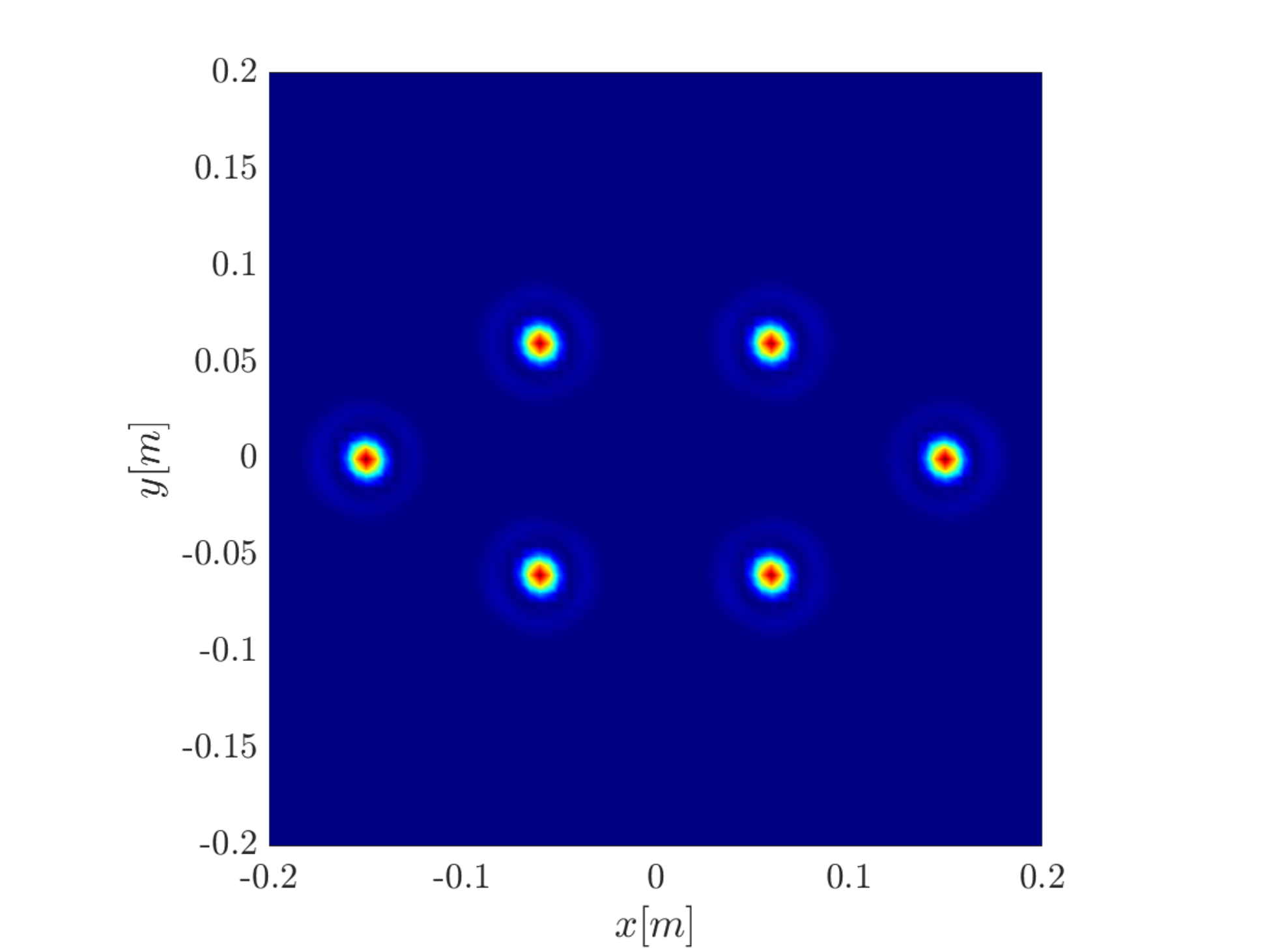}
\end{subfigure}

\begin{subfigure}[t]{0.25\textwidth}
	\includegraphics[width=\textwidth]{LPT_Figures_accepted/diag_1501ds_theta_4-eps-converted-to.pdf}
	\caption{}
\end{subfigure}
\begin{subfigure}[t]{0.25\textwidth}
	\includegraphics[width=\textwidth]{LPT_Figures_accepted/1st_mode_1501ds_theta_4-eps-converted-to.pdf}
	\caption{}
\end{subfigure}
\begin{subfigure}[t]{0.25\textwidth}
	\includegraphics[width=\textwidth]{LPT_Figures_accepted/linear_1501ds_theta_4-eps-converted-to.pdf}
	\caption{}
\end{subfigure}
	\caption{Image as a function of $\theta_{\text{rot}}$. Top: $\theta_{\text{rot}}=\pi$; Middle: $\theta_{\text{rot}}=7\pi/8 $; Bottom: $\theta_{\text{rot}}=3\pi/4$. Aperture size is $1500\Delta s$. $(a)$ Single point migration. $(b)$ Rank-1 image, $(c)$ Kirchhoff migration. We observe that as $\theta_{\text{rot}}$ decreases and the axis is not aligned with the $z$ axis, the rotation of the object has a bigger effect on resolution. This is true for the rank-1 image (b) and the Kirchhoff migration (c).}
	\label{fig:eig_v_diag_quad_rot_2}
\end{figure}

We next compare in Figure~\ref{fig:eig_v_diag_quad_rot_2} the results of single point migration, rank-1 image and Kirchhoff migration for different elevation angles $\theta_{\text{rot}}=\pi,7\pi/8$ and $3\pi/4$. We observe that as $\theta_{\text{rot}}$ becomes smaller an improvement in resolution for the rank-1 image is obtained, which is comparable to the one achieved by Kirchhoff migration. This is in accordance with the analysis in Appendix~\ref{app:stat_phase}, which suggests that the resolution has a $\sin \theta_{\text{rot}}$ dependence. The resolution for the single point migration remains the same as resolution along the diagonal of the two-point interference pattern is not affected by the rotation. This also agrees with our analysis in Appendix~\ref{app:stat_phase}. The results in Figure~\ref{fig:eig_v_diag_quad_rot_2} are for the largest synthetic aperture $1500\Delta s$. Note however, that as mentioned already the image converges to a finite resolution much faster, on a length scale comparable to the rotation period, since the main effect on resolution is the rotation. 

\begin{figure}[htbp]
	\centering	
	\begin{subfigure}[t]{0.25\textwidth}
		\includegraphics[width=\textwidth]{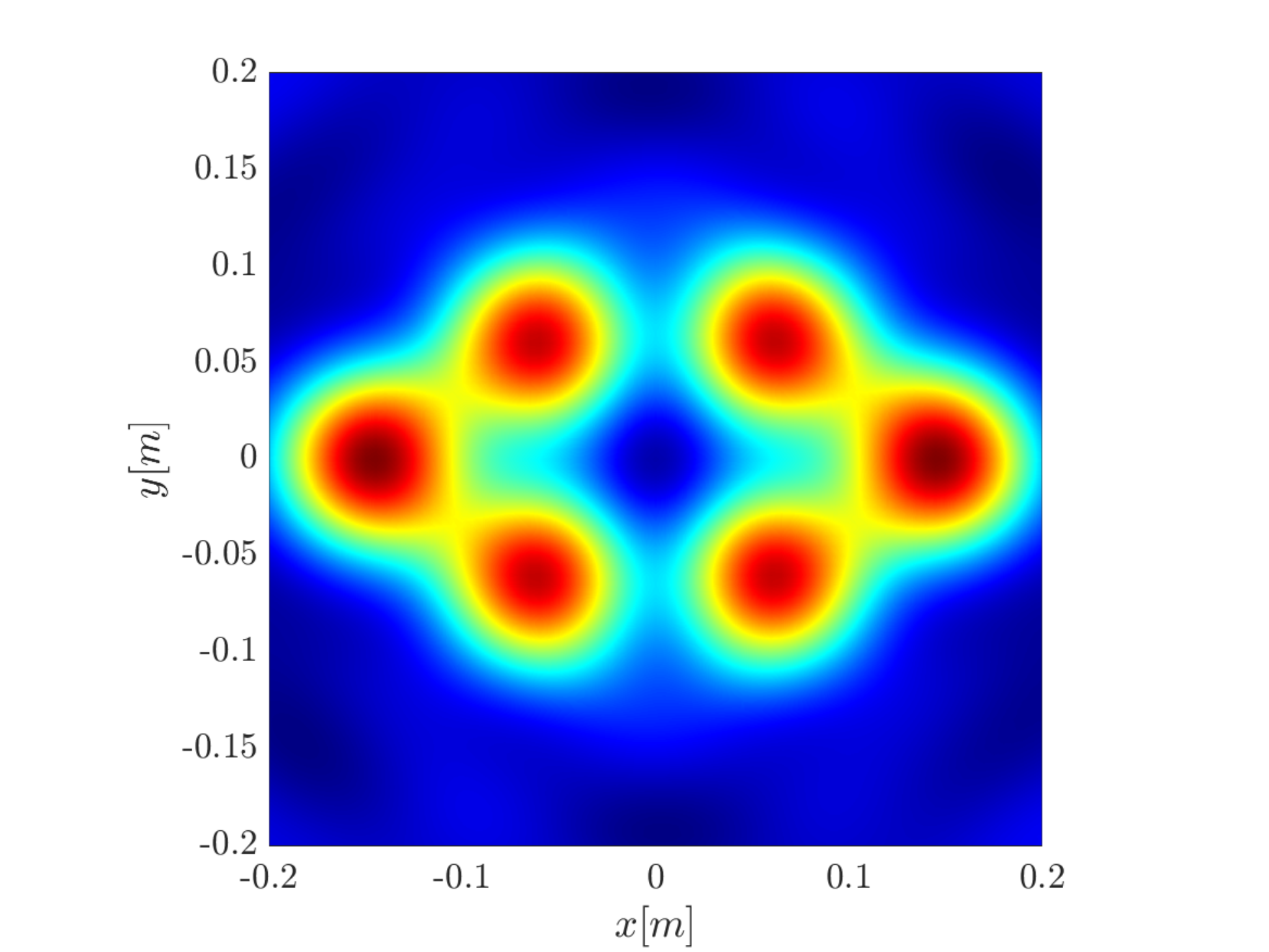}
	\end{subfigure}
	\begin{subfigure}[t]{0.25\textwidth}
		\includegraphics[width=\textwidth]{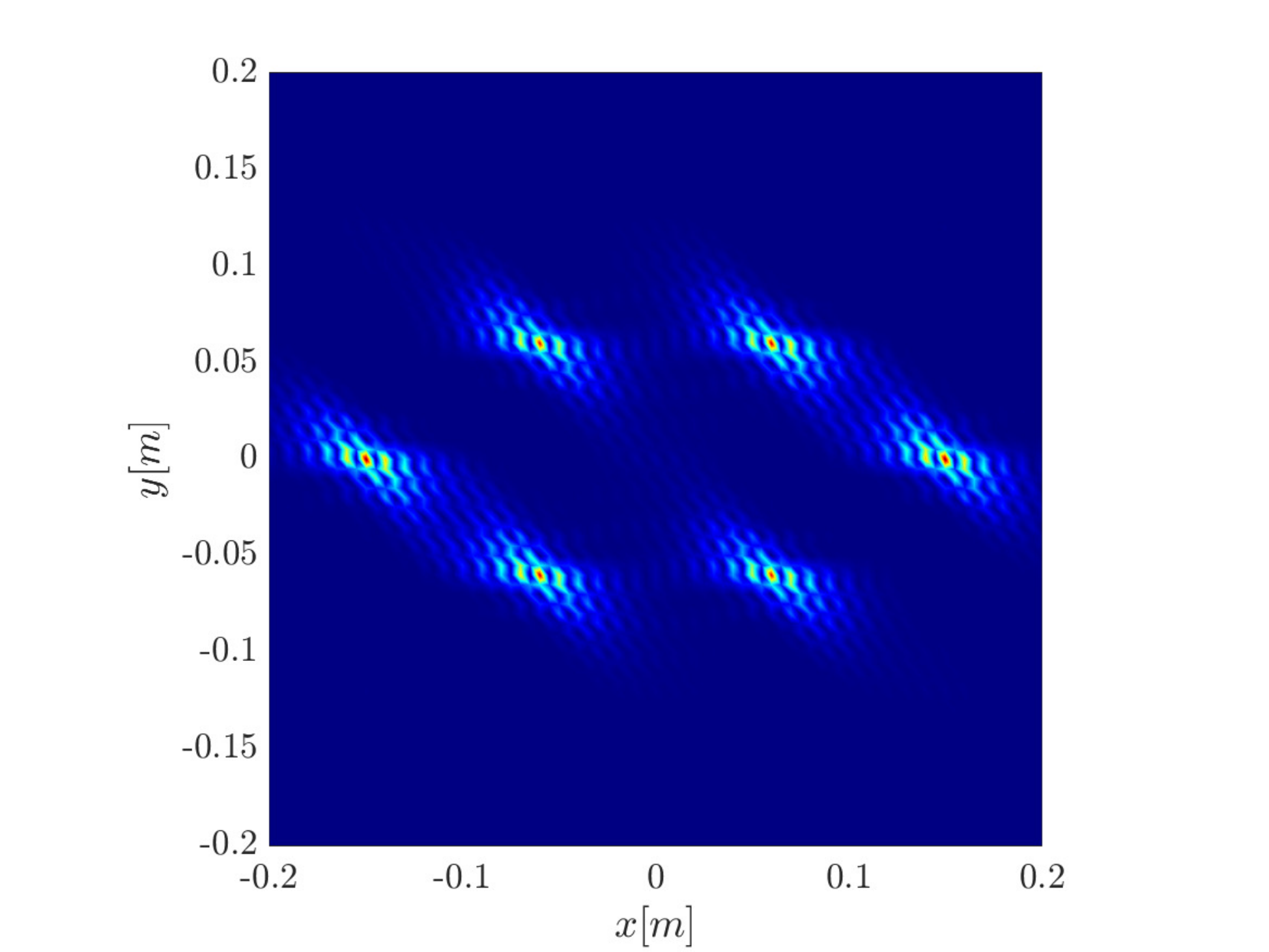}
	\end{subfigure}
	\begin{subfigure}[t]{0.25\textwidth}
		\includegraphics[width=\textwidth]{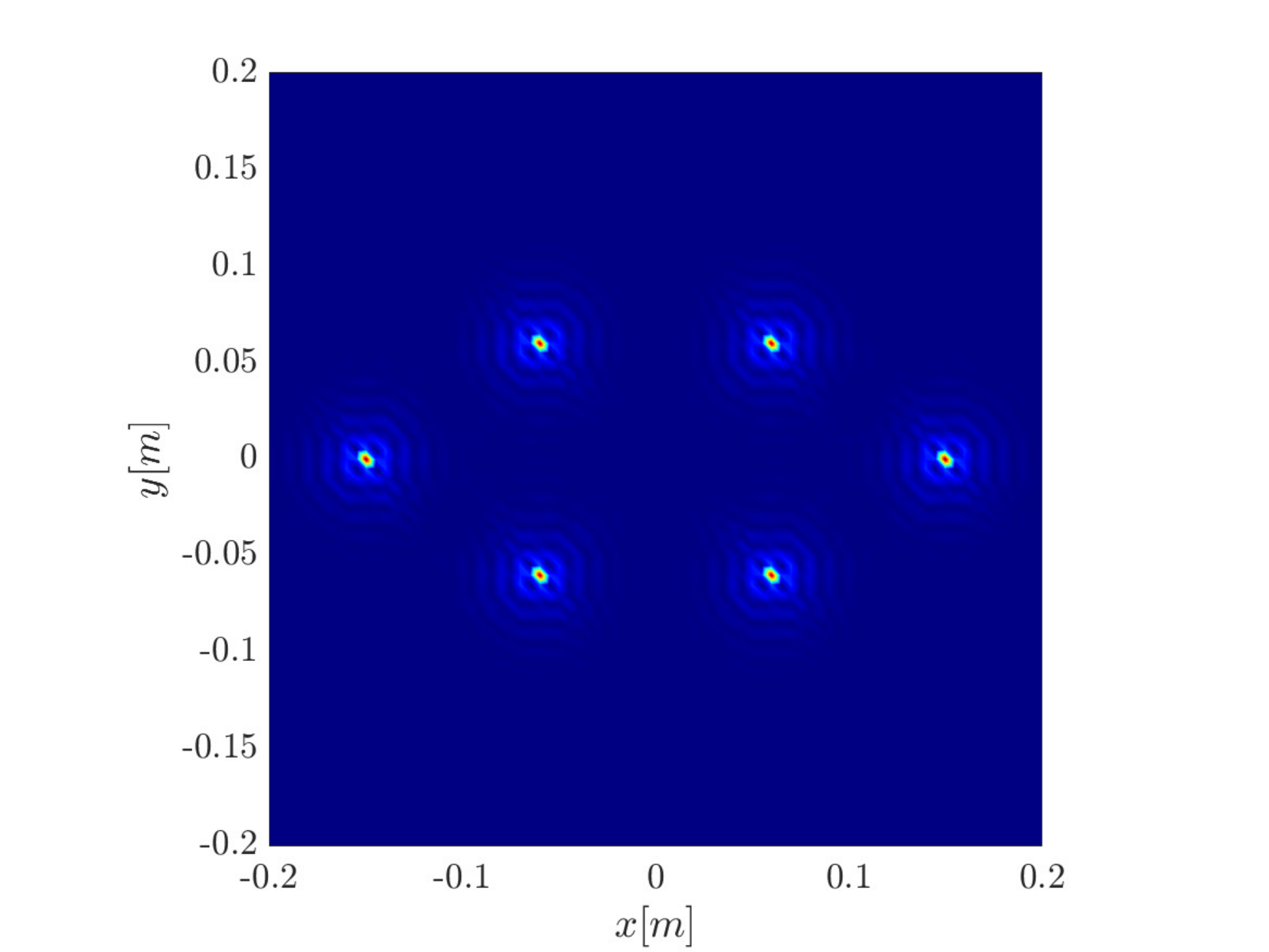}
	\end{subfigure}
	
	\begin{subfigure}[t]{0.25\textwidth}
		\includegraphics[width=\textwidth]{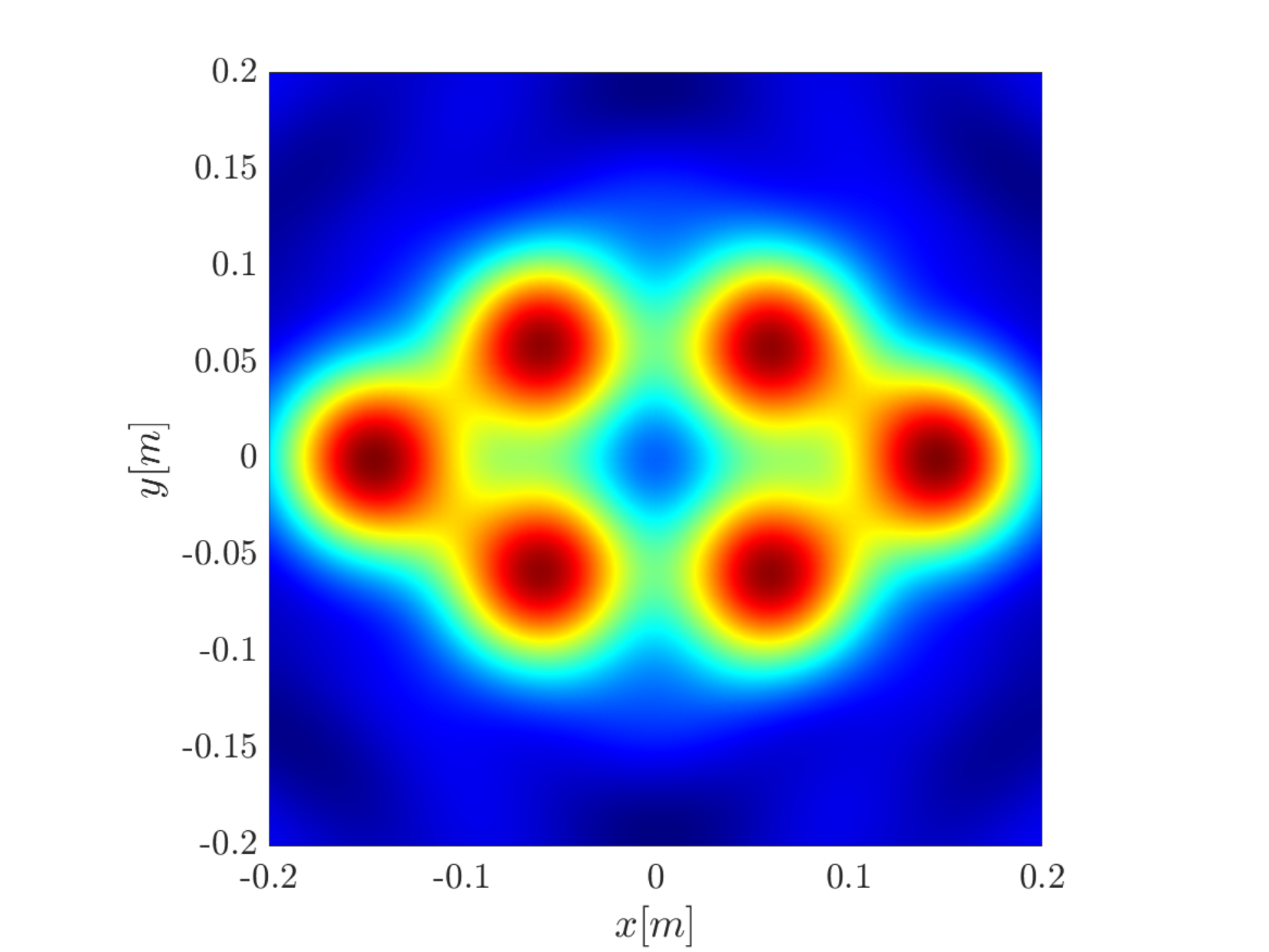}
	\end{subfigure}
	\begin{subfigure}[t]{0.25\textwidth}
		\includegraphics[width=\textwidth]{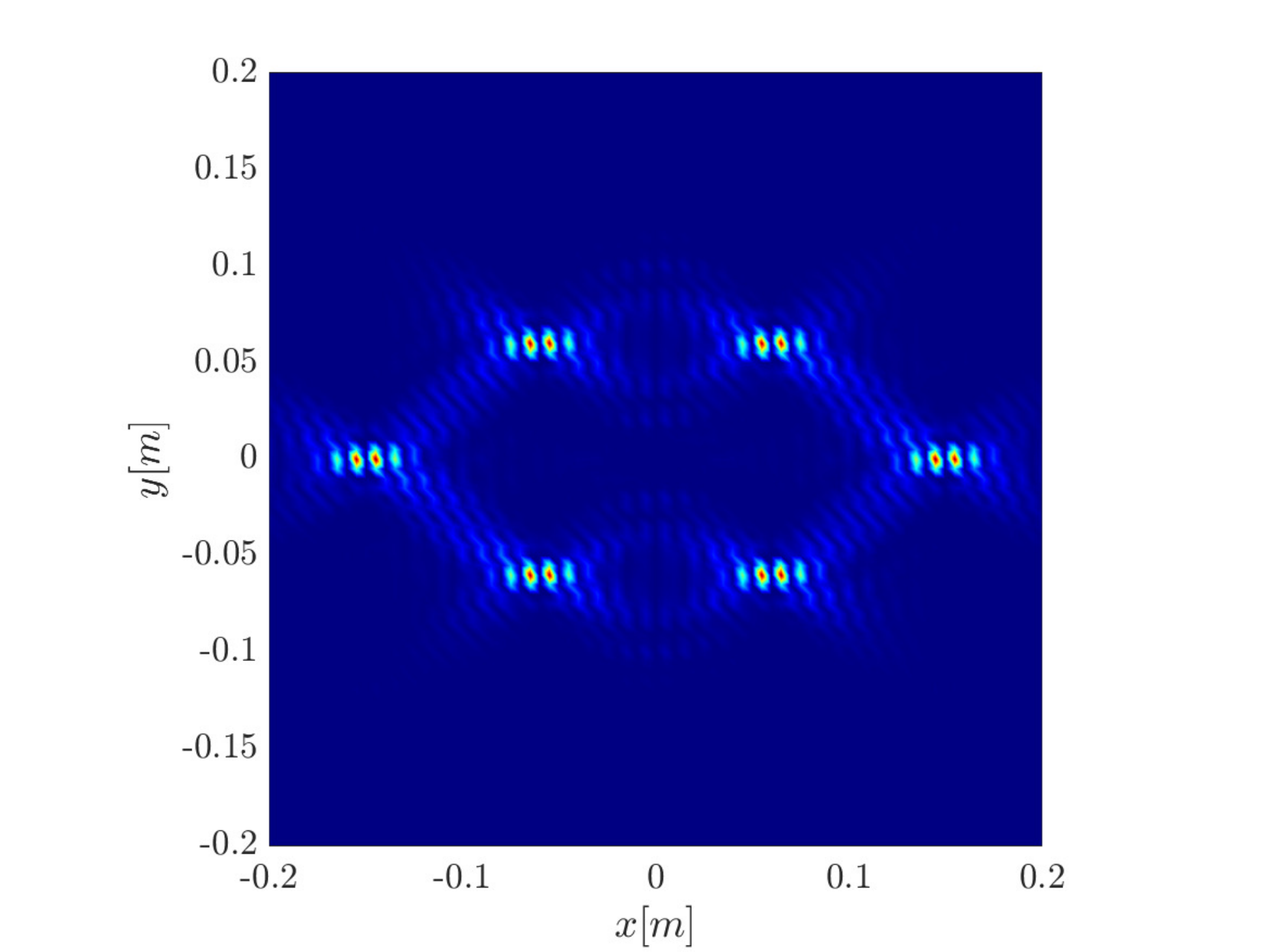}
	\end{subfigure}
	\begin{subfigure}[t]{0.25\textwidth}
		\includegraphics[width=\textwidth]{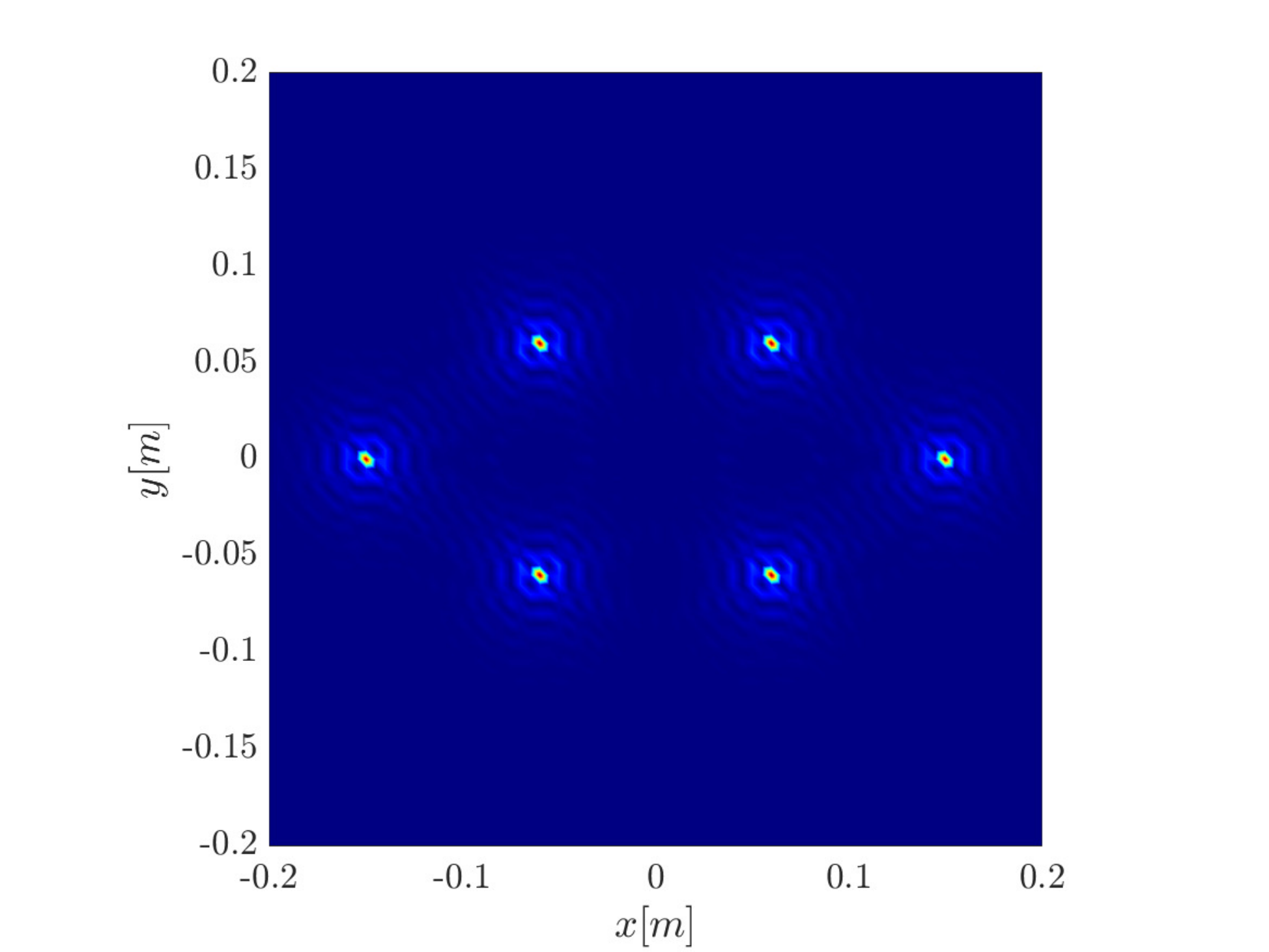}
	\end{subfigure}

	\begin{subfigure}[t]{0.25\textwidth}
	\includegraphics[width=\textwidth]{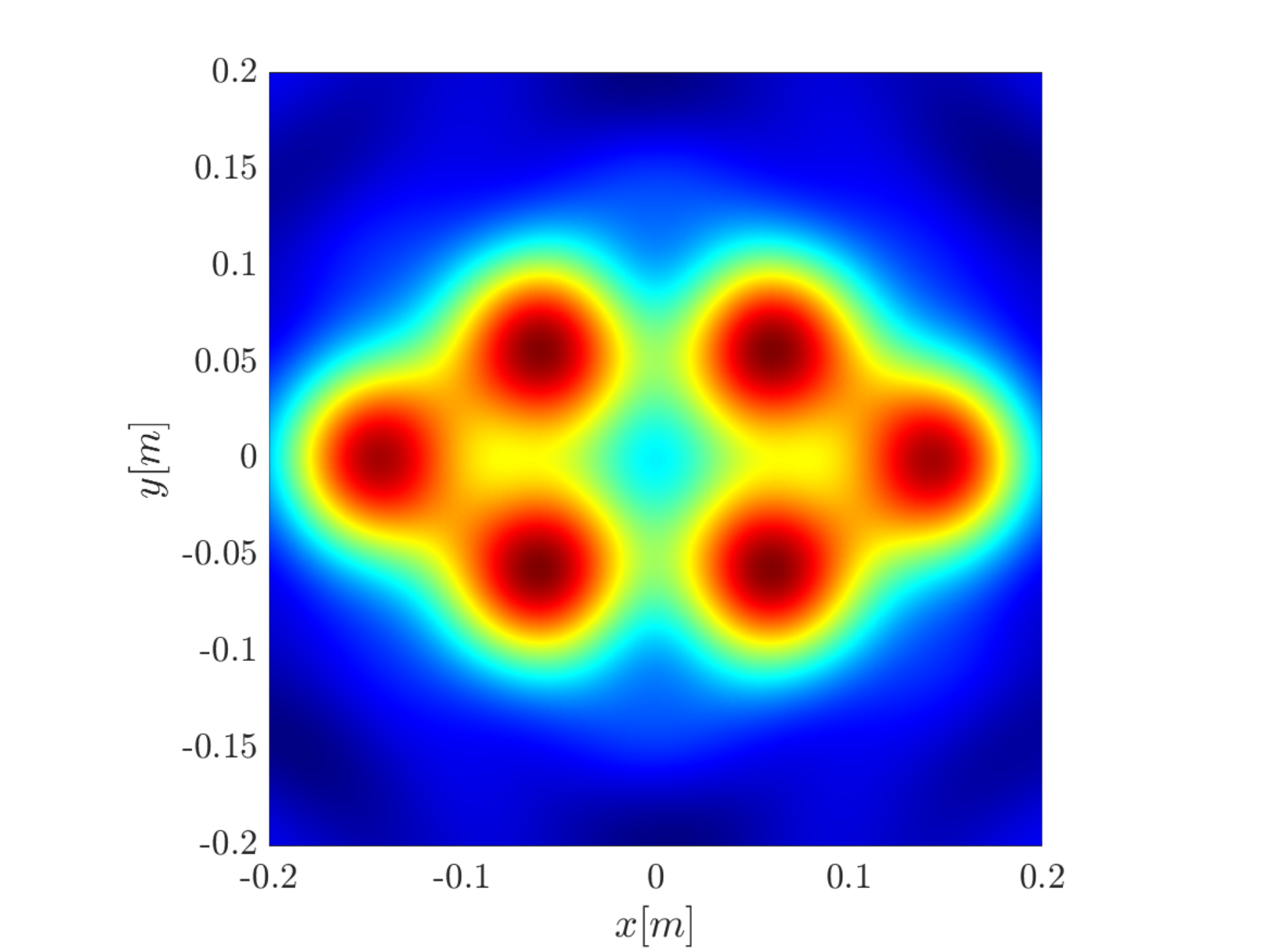}
	\caption{}
\end{subfigure}
\begin{subfigure}[t]{0.25\textwidth}
	\includegraphics[width=\textwidth]{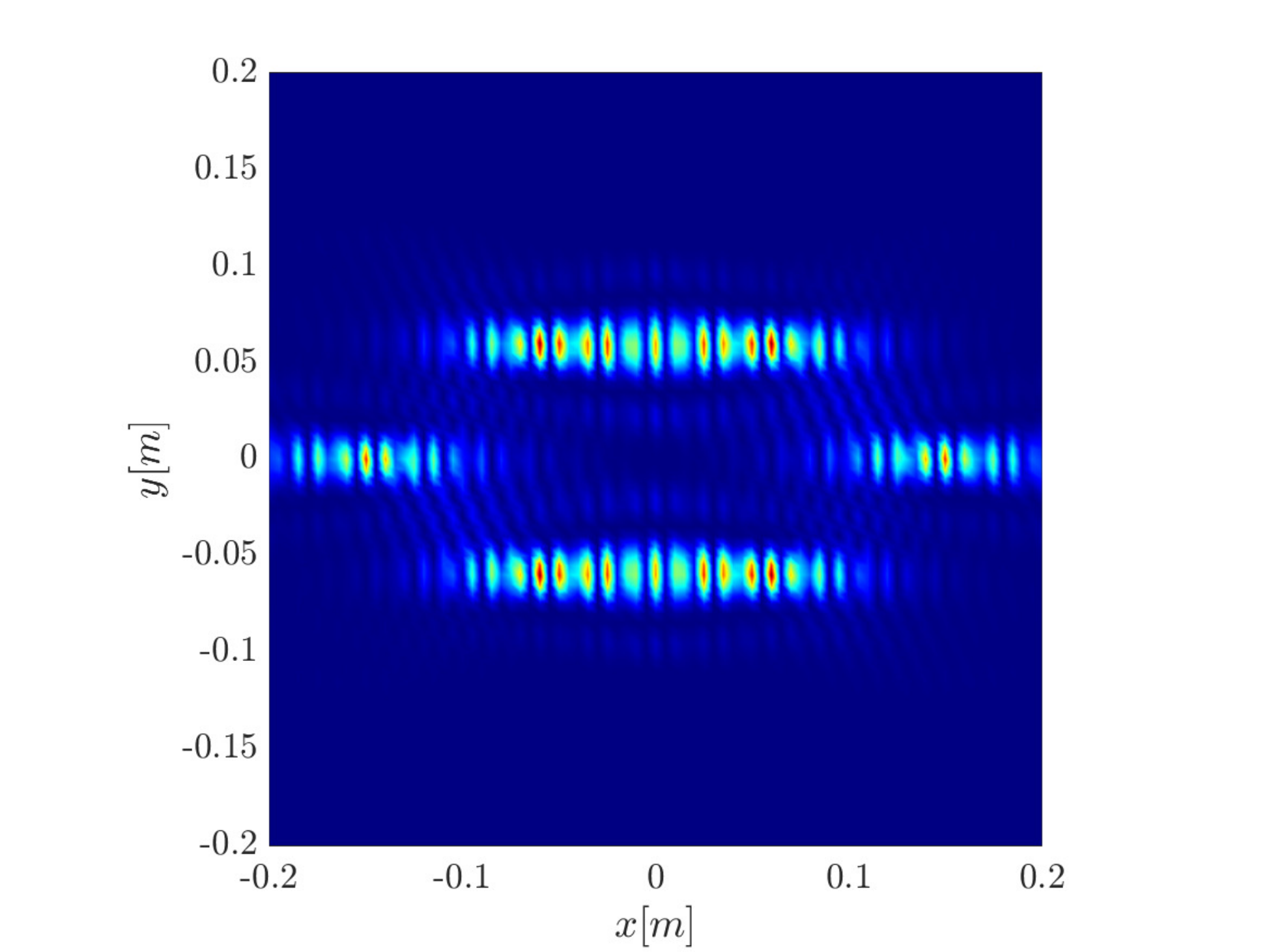}
	\caption{}
\end{subfigure}
\begin{subfigure}[t]{0.25\textwidth}
	\includegraphics[width=\textwidth]{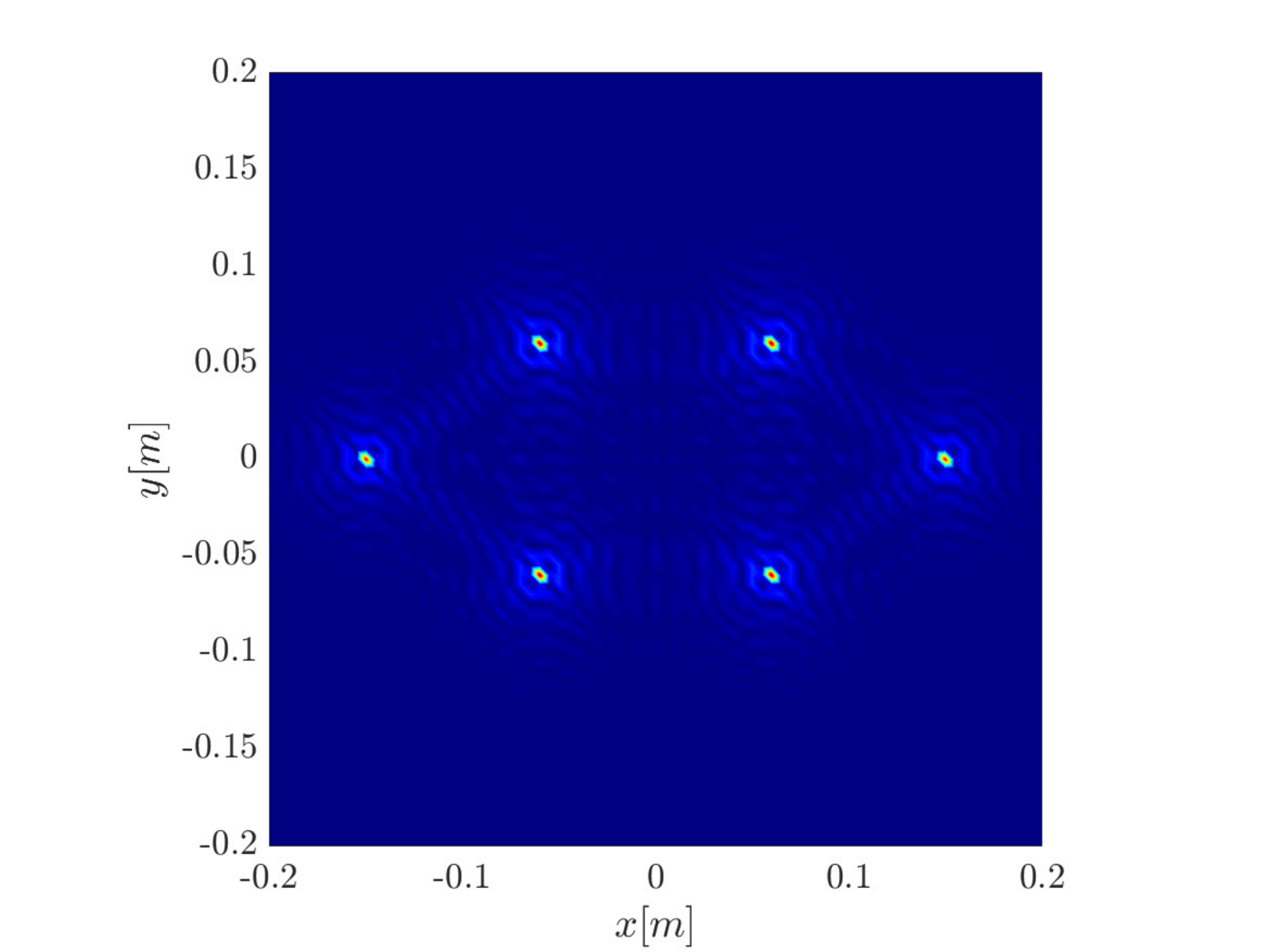}
	\caption{}
\end{subfigure}
	\caption{Image as a function of $\theta_{\text{rot}}$. Top: $\theta_{\text{rot}}=23\pi/32$; Middle: $\theta_{\text{rot}}=11\pi/16 $; Bottom: $\theta_{\text{rot}}=21\pi/32$. Aperture size is $1500\Delta s$. $(a)$ Single point migration. $(b)$ Rank-1 image, $(c)$ Kirchhoff migration. We can see that as $\theta_{\text{rot}}$ further decreases the rank-1 image resolution deteriorates while the Kirchhoff migration image remains good.}
	\label{fig:theta_deter}
\end{figure}

If we continue to decrease the value of $\theta_{\text{rot}}$ beyond $3\pi/4$ we begin to observe a deterioration in the performance of the rank-1 image as illustrated in Figure \ref{fig:theta_deter}. This observation is explained by simulations in Section~\ref{sec:prop_filt} and the analysis in Appendix~\ref{app:stat_phase}. 

\section{Properties of the rank-1 image}
\label{sec:prop_filt}
The numerical simulation results presented in Section~\ref{sec:numerical_simulations} indicate that the performance of the rank-1 image is superior to the single point migration and comparable to the linear Kirchhoff migration for a wide range of $\theta_{\text{rot}}$. In order to better understand properties of the rank-1 image we consider in this section a simplified model problem where there is only a single point scatterer, rotating around a fixed axis. The target is located at $(0.06,0.06)$[m], and all other parameters are as in the previous section. While this model is less realistic, the simulations provide insight to the properties of the rank-1 image.

\begin{figure}[htbp]
		\centering
	\begin{subfigure}[t]{0.32\textwidth}
	\includegraphics[width=\textwidth]{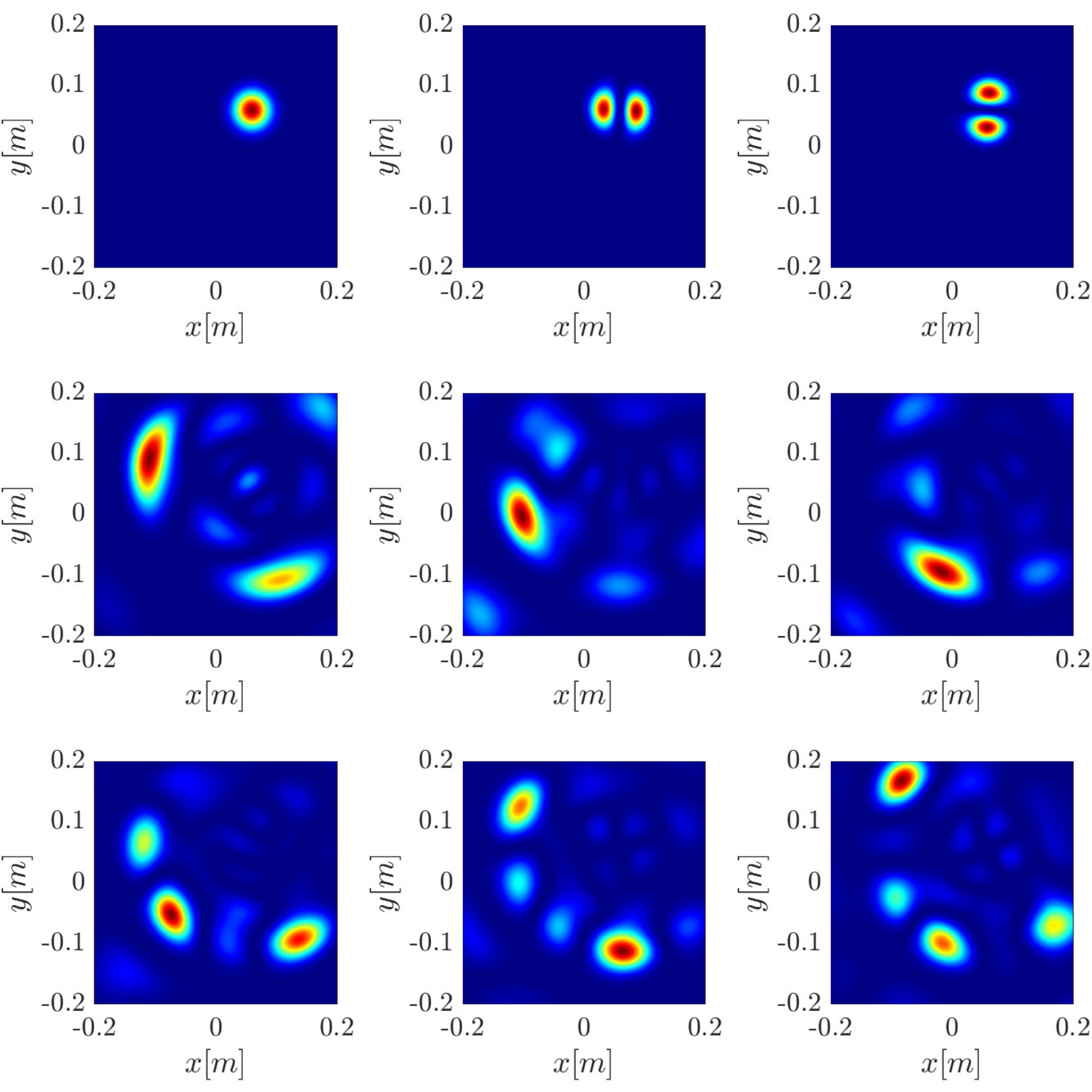}
\end{subfigure}
\hspace{3em}
\begin{subfigure}[t]{0.40\textwidth}
	\includegraphics[width=\textwidth]{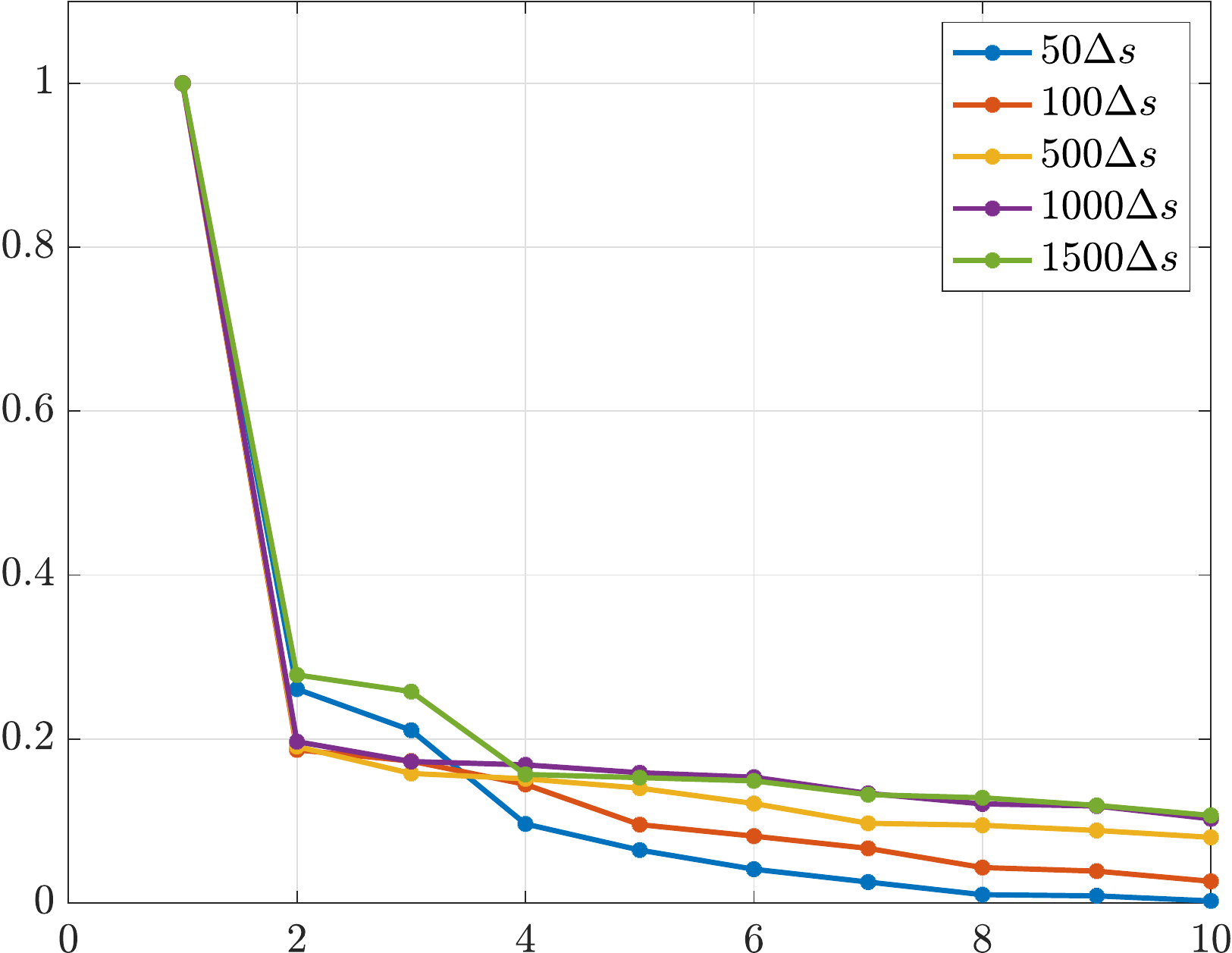}
\end{subfigure}
\\[4ex]
	\begin{subfigure}[t]{0.32\textwidth}
	\includegraphics[width=\textwidth]{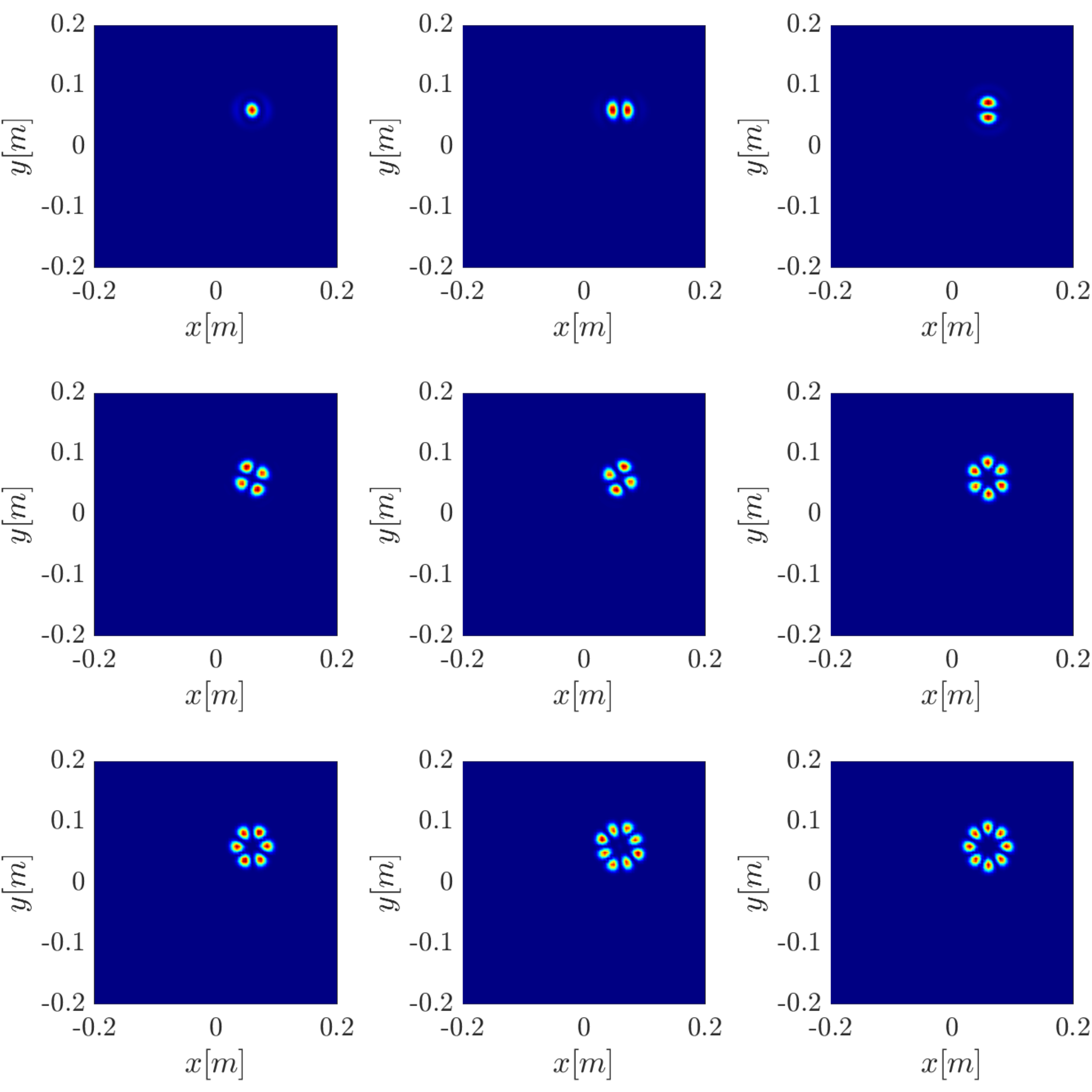}
\end{subfigure}
\hspace{3em}
\begin{subfigure}[t]{0.40\textwidth}
	\includegraphics[width=\textwidth]{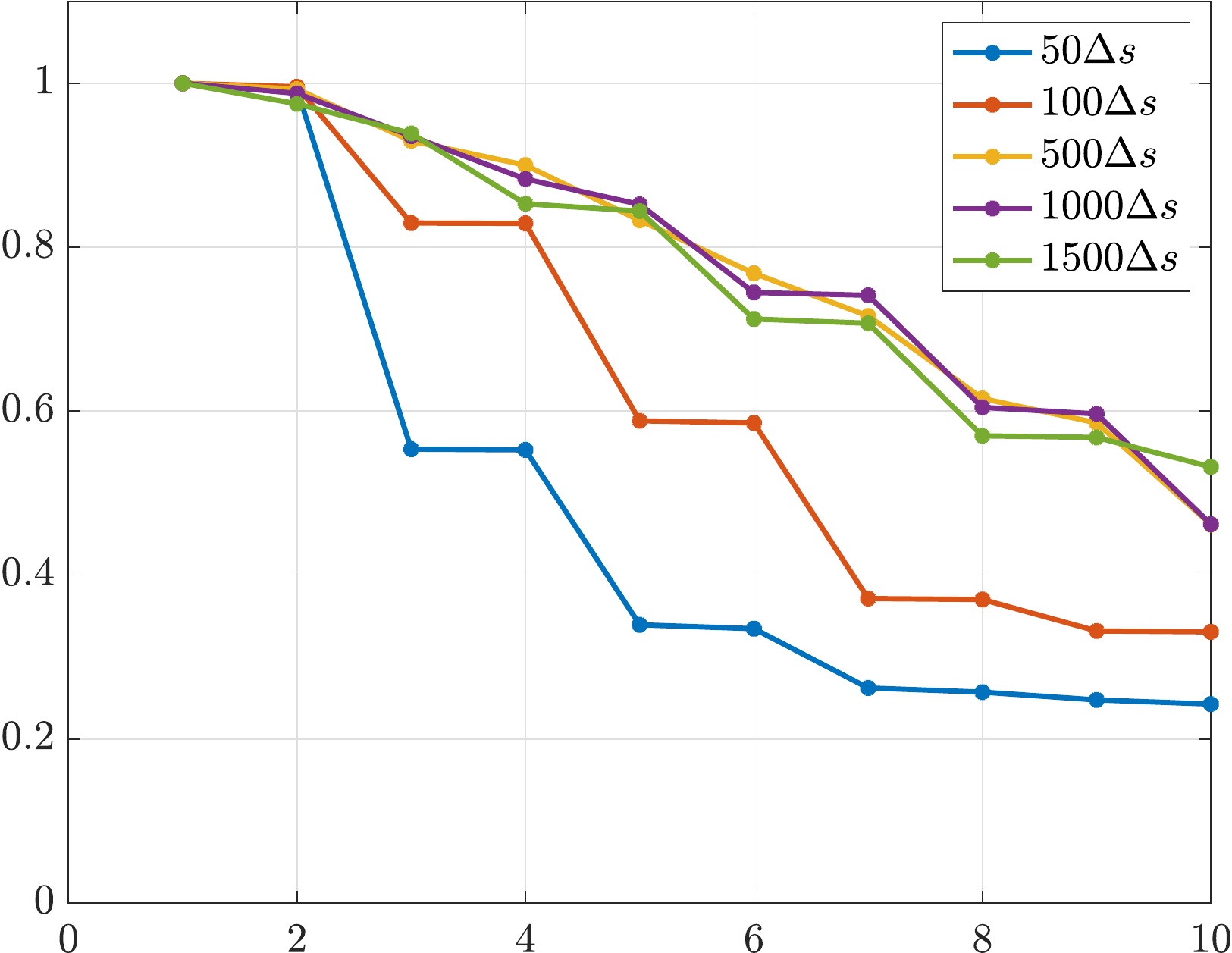}
\end{subfigure}
\\[4ex]
	\begin{subfigure}[t]{0.32\textwidth}
	\includegraphics[width=\textwidth]{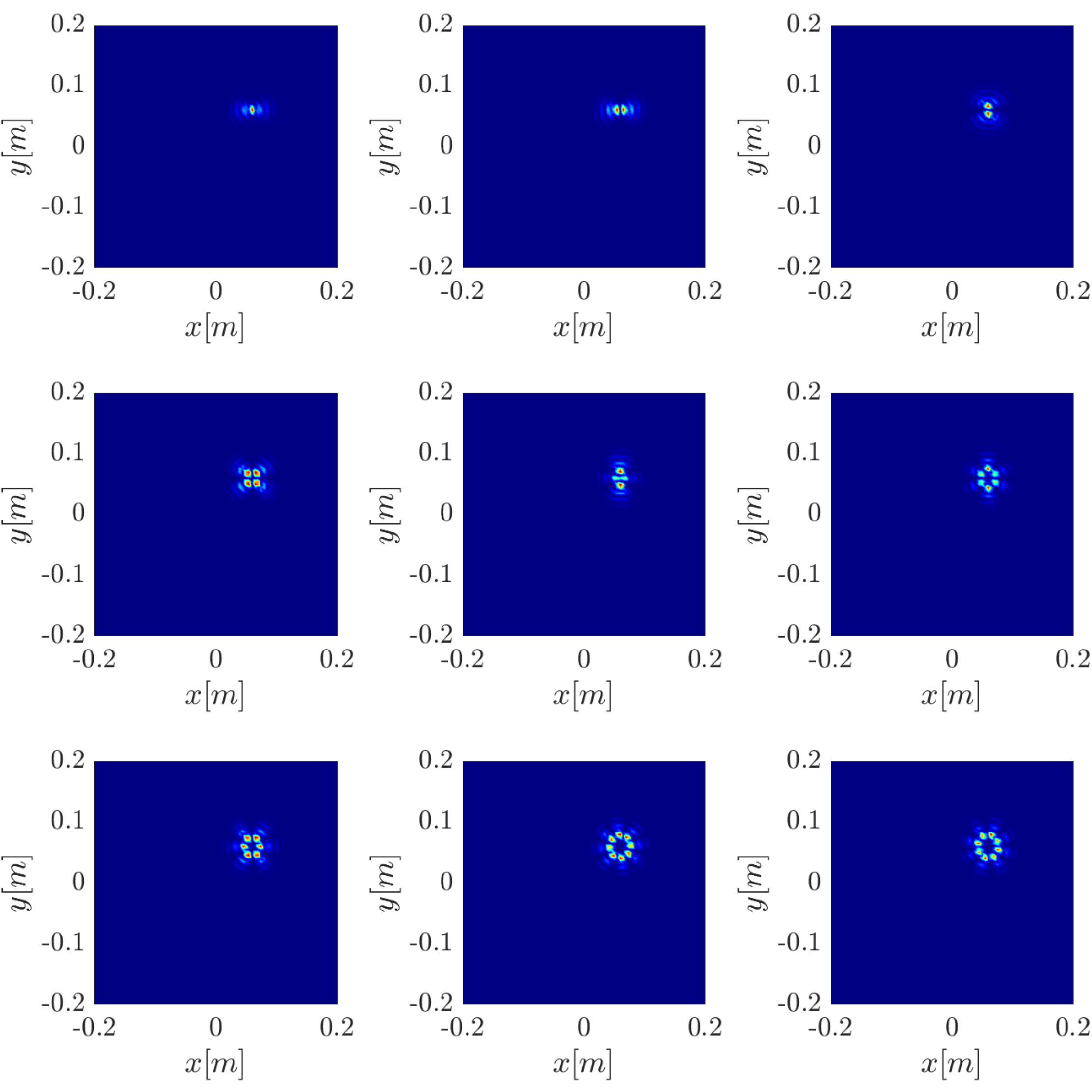}
	\caption{}
\end{subfigure}
\hspace{3em}
\begin{subfigure}[t]{0.40\textwidth}
	\includegraphics[width=\textwidth]{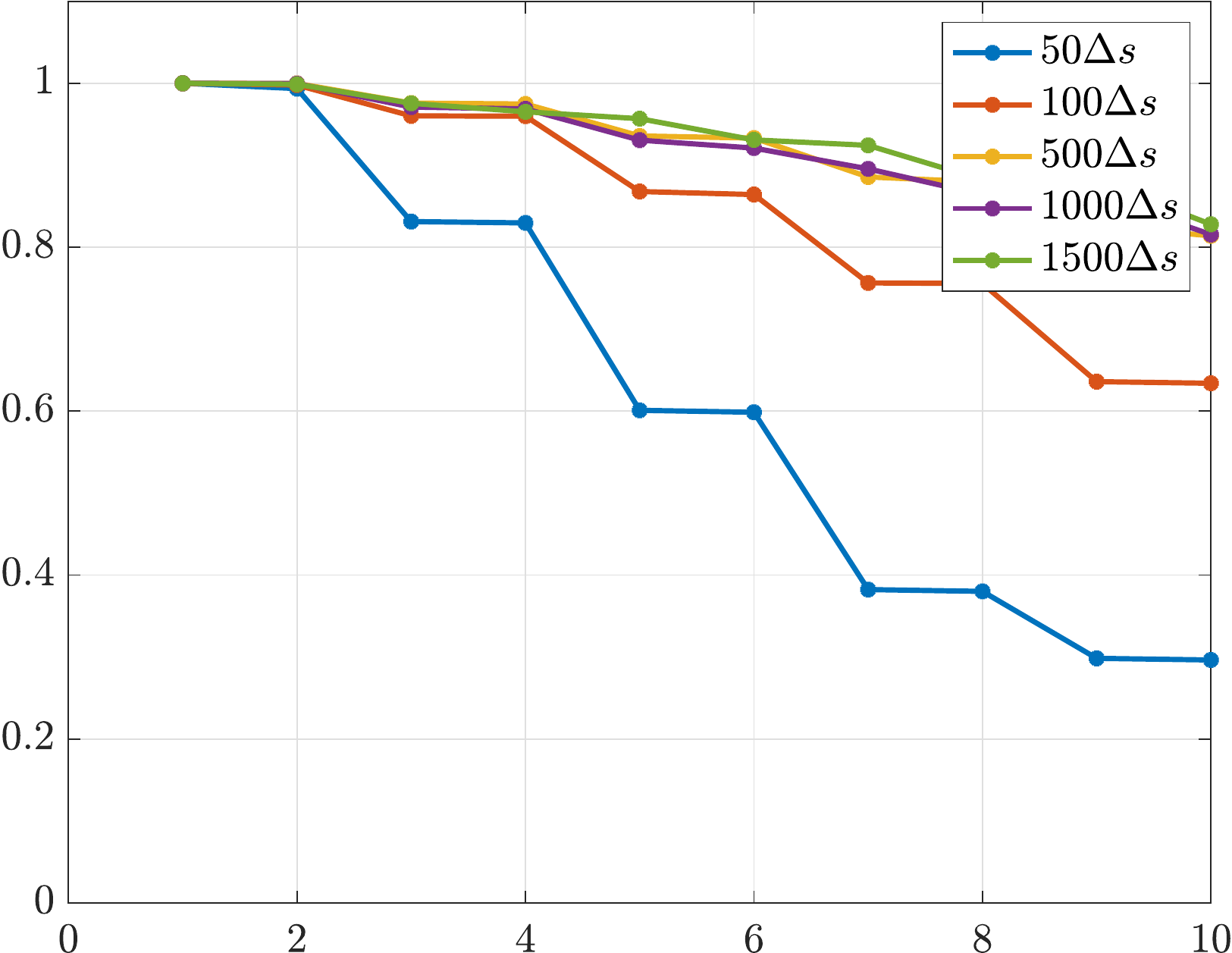}
	\caption{}
\end{subfigure}
\caption{ Eigendecomposition of $\tilde{\mb X}$ for a single rotating target, as a function of $\theta_{\text{rot}}$. $(a)$ Top 9 eigenvectors for aperture size $1500 \Delta s$. $(b)$ Top 10 Eigenvalues for varying synthetic aperture sizes (normalized with respect to the top eigenvalue). Top: $\theta_{\text{rot}}=\pi$; Middle: $\theta_{\text{rot}}=7\pi/8$; Bottom: $\theta_{\text{rot}}=3\pi/4$. As $\theta_{\text{rot}}$ deviates from $\pi$ we notice that the effective rank of $\tilde{\mb X}$ increases. We also see that the modes become more localized.  The combined effect of these two phenomena is in accordance with the observations made in Section \ref{sec:perf_rank_1} and the analysis of Appendix~\ref{app:stat_phase}. This demonstrates the advantage of the rank-1 image compared to the single point migration image.}
\label{fig:eig_mods}
\end{figure}

We first look at the eigenvectors of the two-point interference function $\mathcal{I}^{GCC}(\mb y_k,\mb y_{k'})$, as well as the evolution of its eigenvalues. As illustrated in Figure~\ref{fig:eig_mods}, as the synthetic aperture increases two effects take place. The eigenvalues decay at a slower rate, while the eigenfunctions become more localized. This explains the difference between the single point migration image and the rank-1 image, since the single point migration is the sum of all the eigenfunctions, weighted by the eigenvalues  \eqref{eq:eig_to_diag}. As the first mode gets more localized, higher modes, which have to be orthogonal to it, get spread out which results in a deterioration of the resolution for the single migration image compared to the rank-1 image. The extend of this effect depends on $\theta_{\text{rot}}$. For $\theta_\text{rot}=\pi$, no change is observed in resolution between the single point migration and the rank-1 image. In this case the spectrum indicated that $\mathcal{I}^{GCC}(\mb y_k,\mb y_{k'})$ remains close to being rank one  (see top row plots in Figure~\ref{fig:eig_mods}). As $\theta_{\text{rot}}$ decreases we observe a slower decay rate in the spectrum as the synthetic aperture increases. The limit behavior is obtained after a full rotation of the object corresponding to $300\Delta s$ (see middle and bottom row plots in Figure~\ref{fig:eig_mods}). 

\begin{figure}[htbp]
	\centering
	\begin{subfigure}[t]{0.75\textwidth}
		\includegraphics[width=\textwidth]{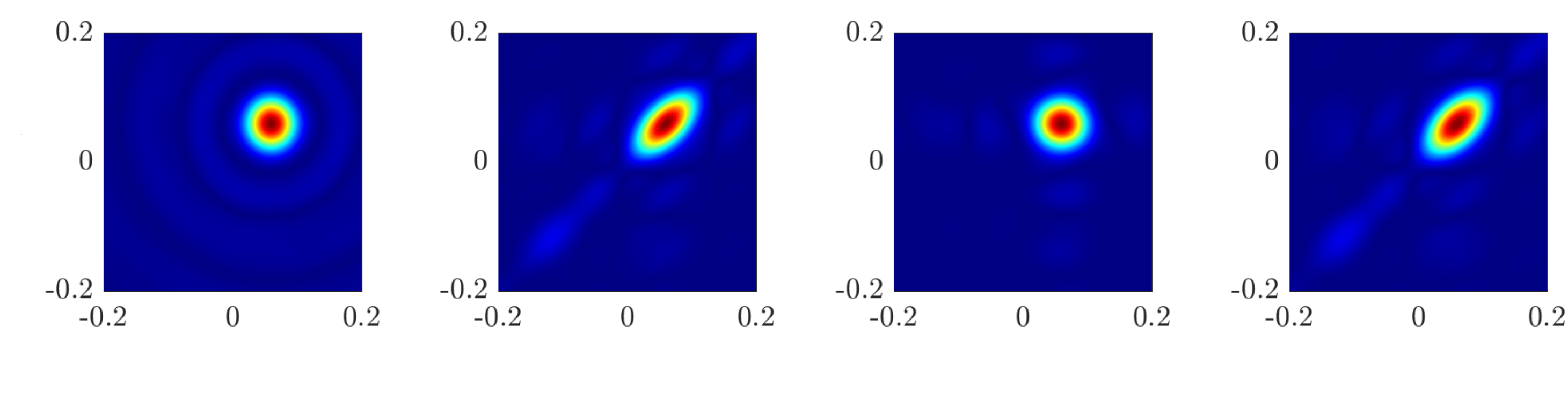}
		\caption{}
	\end{subfigure}
	\begin{subfigure}[t]{0.75\textwidth}
	\includegraphics[width=\textwidth]{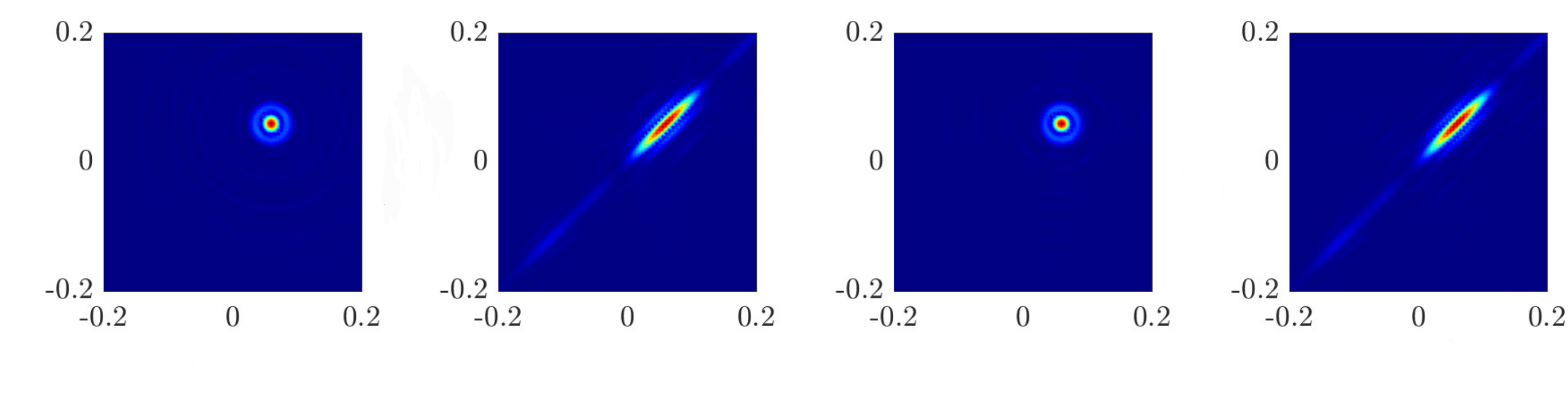}
	\caption{}
\end{subfigure}
	\begin{subfigure}[t]{0.75\textwidth}
	\includegraphics[width=\textwidth]{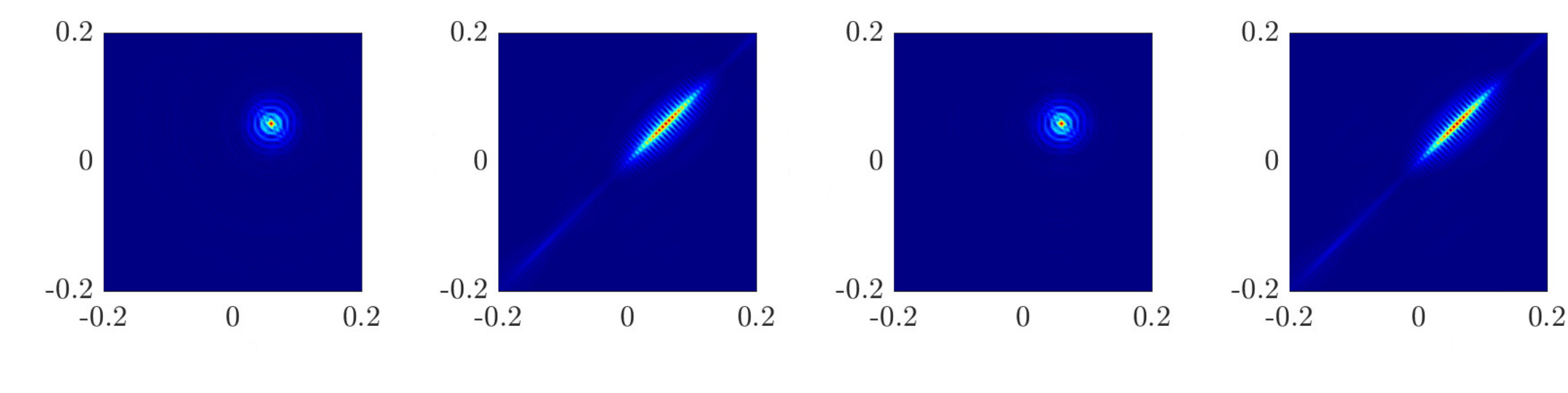}
	\caption{}
\end{subfigure}
\caption{Cross -sections of $\tilde{\mb X}$ around the peak $\mb y$ for different values of $\theta_{\text{rot}}$. We use a verbose notation $\mathcal{I}(y_{k,1},y_{k,2}, y_{k',1}, y_{k',2})$. They are ordered (left to right): $\mathcal{I}(y_1,y_2,\cdot,\cdot ),\mathcal{I}(\cdot,y_2,\cdot,y_2 ),\mathcal{I}(y_1,\cdot,\cdot,y_2 ),\mathcal{I}(y_1,\cdot,y_1,\cdot )$. The synthetic aperture size is $1500\Delta s$. Top: $\theta_{\text{rot}}=\pi$; Middle: $\theta_{\text{rot}}=7\pi/8$; Bottom: $\theta_{\text{rot}}=3\pi/4$. As the synthetic aperture size increases the peak becomes more anisotropic in the sense that the ratio of the widths of the main lobe with respect to different directions increases. Specifically, the directions with the weakest decay are the ones for which $\hat{\mb y}_k=\hat{\mb y}_{k'}$- see the diagonal of the second and third columns. This is the direction along which the single point migration is computed.}
\label{fig:single_cross}
\end{figure}

As we noted in Section \ref{sec:perf_rank_1}, the stationary phase analysis of the two point migration function, carried out in Appendix~\ref{app:stat_phase}, suggests that the peaks of $\mathcal{I}(\mb y_k,\mb y_{k'})$ are all the points $(\mb y_k,\mb y_{k'})=(\mb y^t_i,\mb y^t_j)$, where $\mb y^t$ is the collection of all scatterer locations. In the single scatterer case we would have only one peak. We analyze the behavior of $\mathcal{I}(\mb y_k,\mb y_{k'})$ around the peaks, as $\theta_{\text{rot}}$ changes, to see in greater detail the effect of rotation, as illustrated in  Figure~\ref{fig:single_cross}. Since for a two dimensional image domain $\mathcal{I}(\mb y_k,\mb y_{k'})$ would be four dimensional, we look at all the possible planar cross sections of $\mathcal{I}(\mb y_k,\mb y_{k'})$ at the peak- 6 in total. Since $\mathcal{I}(\mb y_k,\mb y_{k'})$ is Hermitian, there are only 4 distinct cross sections (up to conjugation). We can see that indeed as $\theta_{\text{rot}}$ decreases towards $\pi/2$ the anisotropy in the peak increases, with the directions corresponding to the diagonal unchanged. These simulations are in accordance with the analysis of Appendix~\ref{app:stat_phase}.

\begin{figure}
	\centering
		\begin{subfigure}[t]{0.32\textwidth}
		\includegraphics[width=\textwidth]{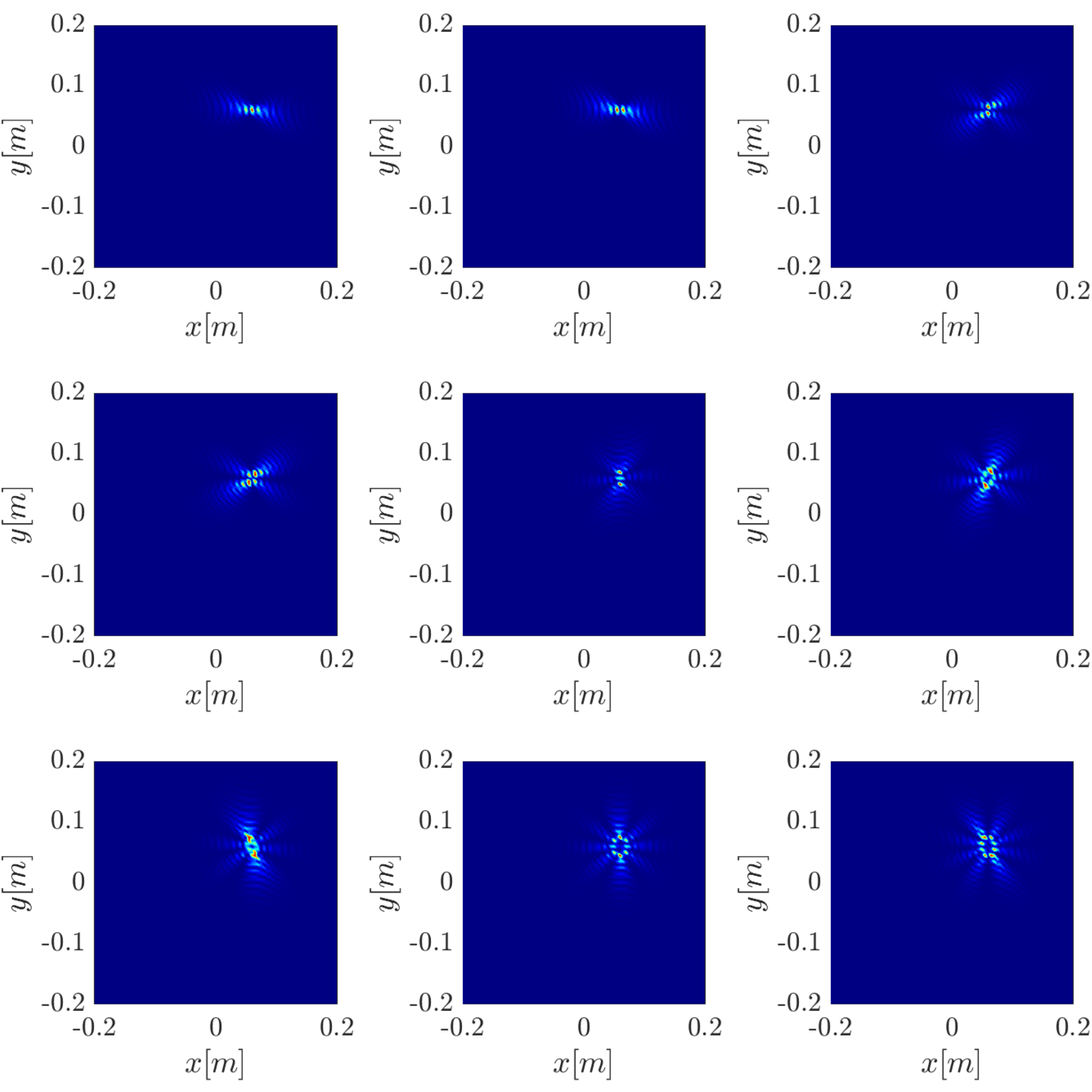}
		\caption{}
	\end{subfigure}
	\hspace{3em}
	\begin{subfigure}[t]{0.40\textwidth}
		\includegraphics[width=\textwidth]{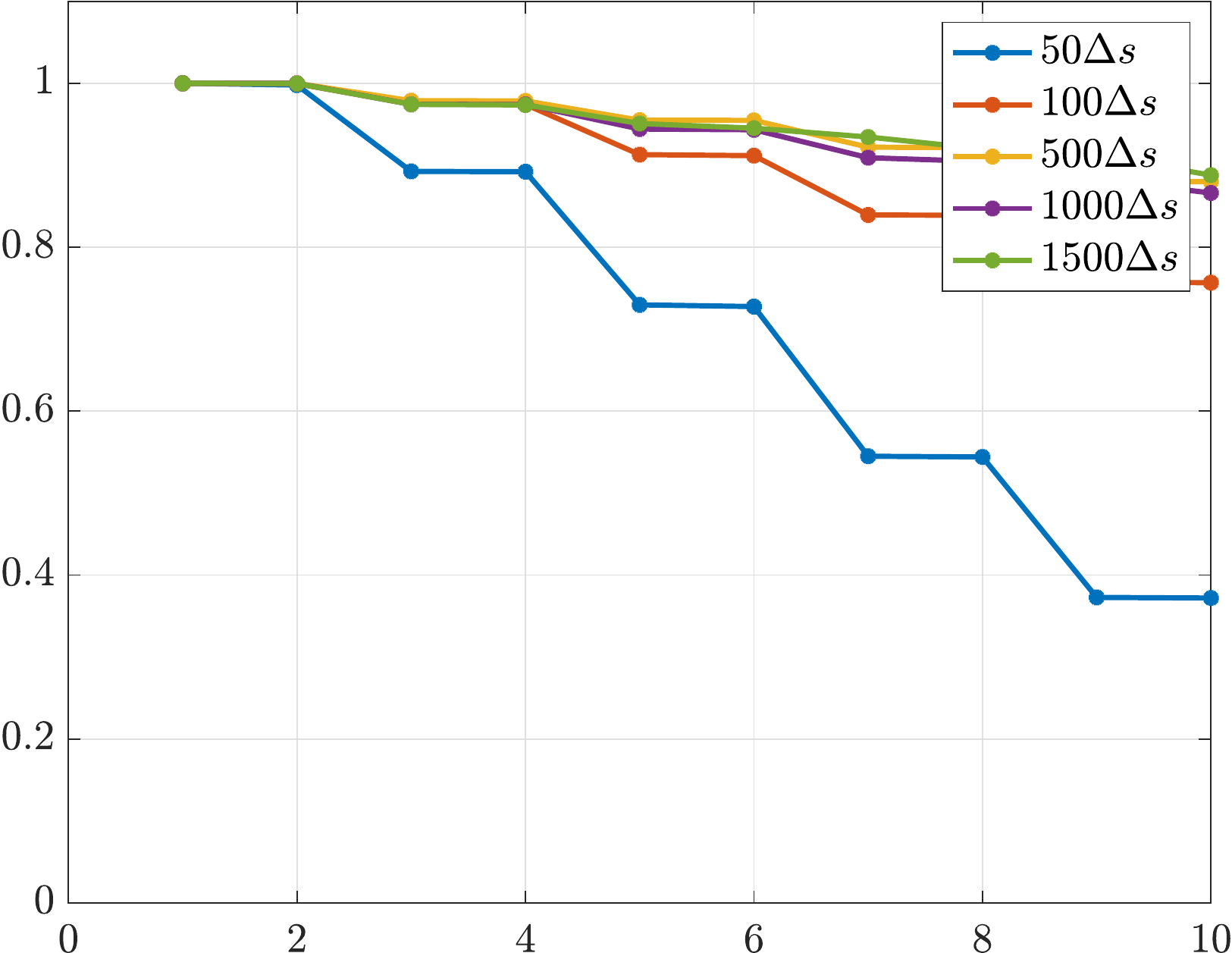}
		\caption{}
	\end{subfigure}
	\begin{subfigure}[t]{0.75\textwidth}
	\includegraphics[width=\textwidth]{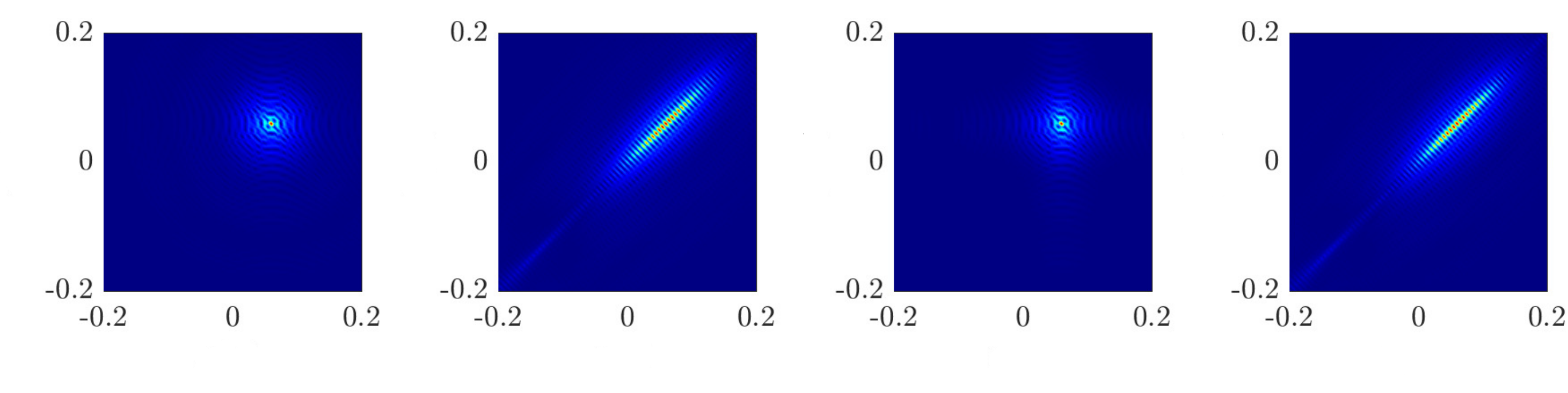}
	\caption{}
\end{subfigure}
\caption{The two point interference function for $\theta_{\text{rot}}=5\pi/8$ and a single rotating target. $(a)$ The first 9 modes. $(b)$ The first 10 eigenvalues. $(c)$ Cross sections of $\tilde{\mb X}$ at the peak. We can see that while the modes are localized, the spectrum's slow decay as well as the slower decay of the peak, can affect the achieved resolution of the rank-1 image.}
\label{fig:theta_5pi_8_eig_modes}
\end{figure}

We next investigate the deterioration in resolution, observed in Section~\ref{sec:numerical_simulations} when the rotation angle is decreased below $3\pi/4$ (see Figure \ref{fig:theta_deter}). When looking at the results for $\theta_{\text{rot}}=5\pi/8$, illustrated in Figure~\ref{fig:theta_5pi_8_eig_modes}, for which the performance of the rank-1 image deteriorates, we notice that the spectrum decay becomes slower, with the first two eigenvalues being very close to each other. We observe also that the first two eigenvectors seem to have almost the same support. Moreover, looking at the two point interference pattern for the single target in Figure~\ref{fig:theta_5pi_8_eig_modes}, we see that while the main lobe becomes narrower, the decay away from it is slower. The resolution of the rank-1 image for a single point scatterer is still well localized, which suggests that the deterioration observed in Figure~\ref{fig:theta_deter} arises from the interaction between scatterers. As explained in Appendix~\ref{app:stat_phase}, the rotation induced resolution takes the form of a Bessel function (see \eqref{eq:B_eff_approx}), which has a slower decay rate with distance ($1/\sqrt{r}$) compared to the array induced resolution ($1/r$).

\begin{figure}[htbp]
	\centering
	\begin{subfigure}[t]{0.3  \textwidth}
		\includegraphics[width=\textwidth]{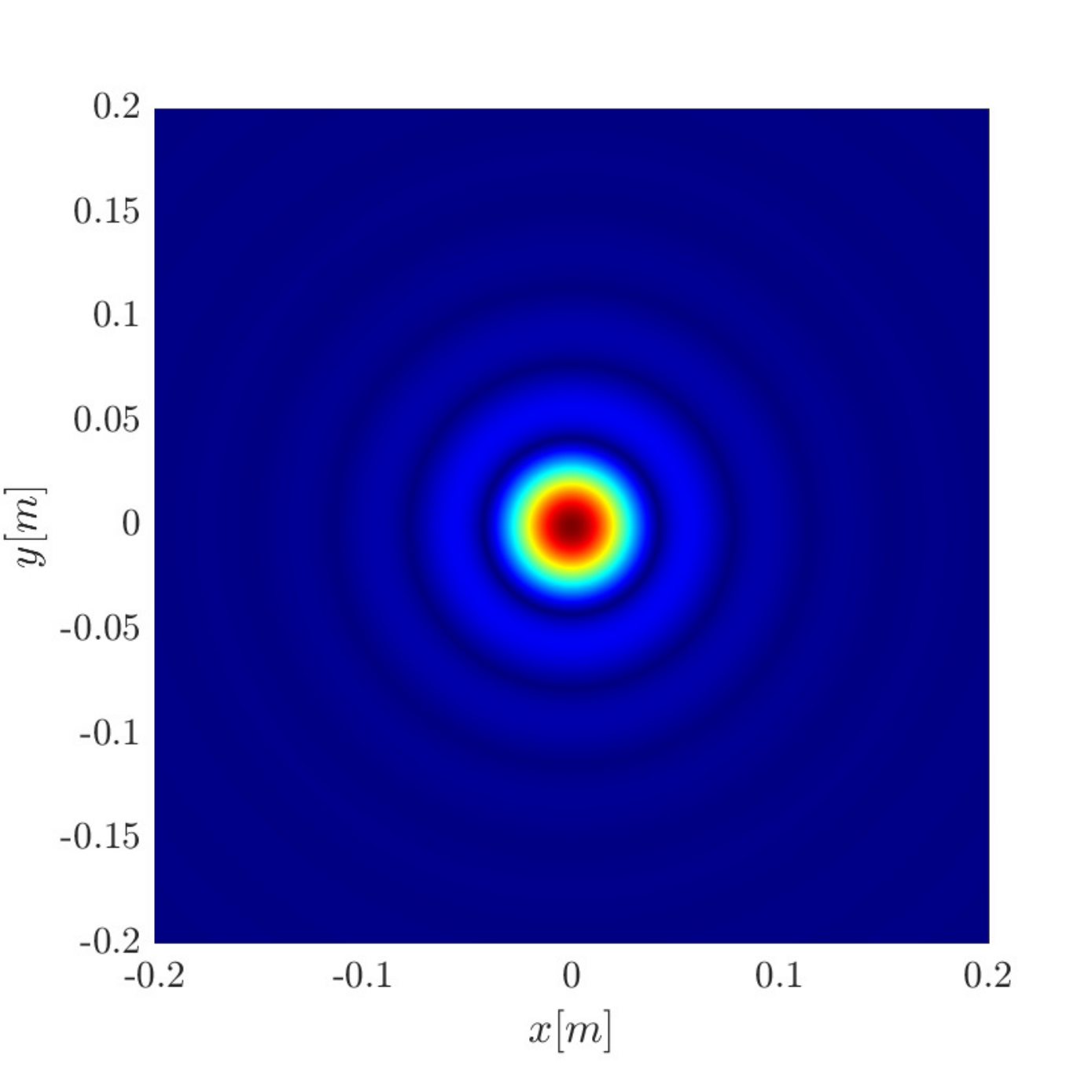}
		\caption{}
	\end{subfigure}
	\begin{subfigure}[t]{0.3  \textwidth}
		\includegraphics[width=\textwidth]{LPT_Figures_accepted/B_eff_3pi_4-eps-converted-to.pdf}
		\caption{}
	\end{subfigure}
	\begin{subfigure}[t]{0.3\textwidth}
		\includegraphics[width=\textwidth]{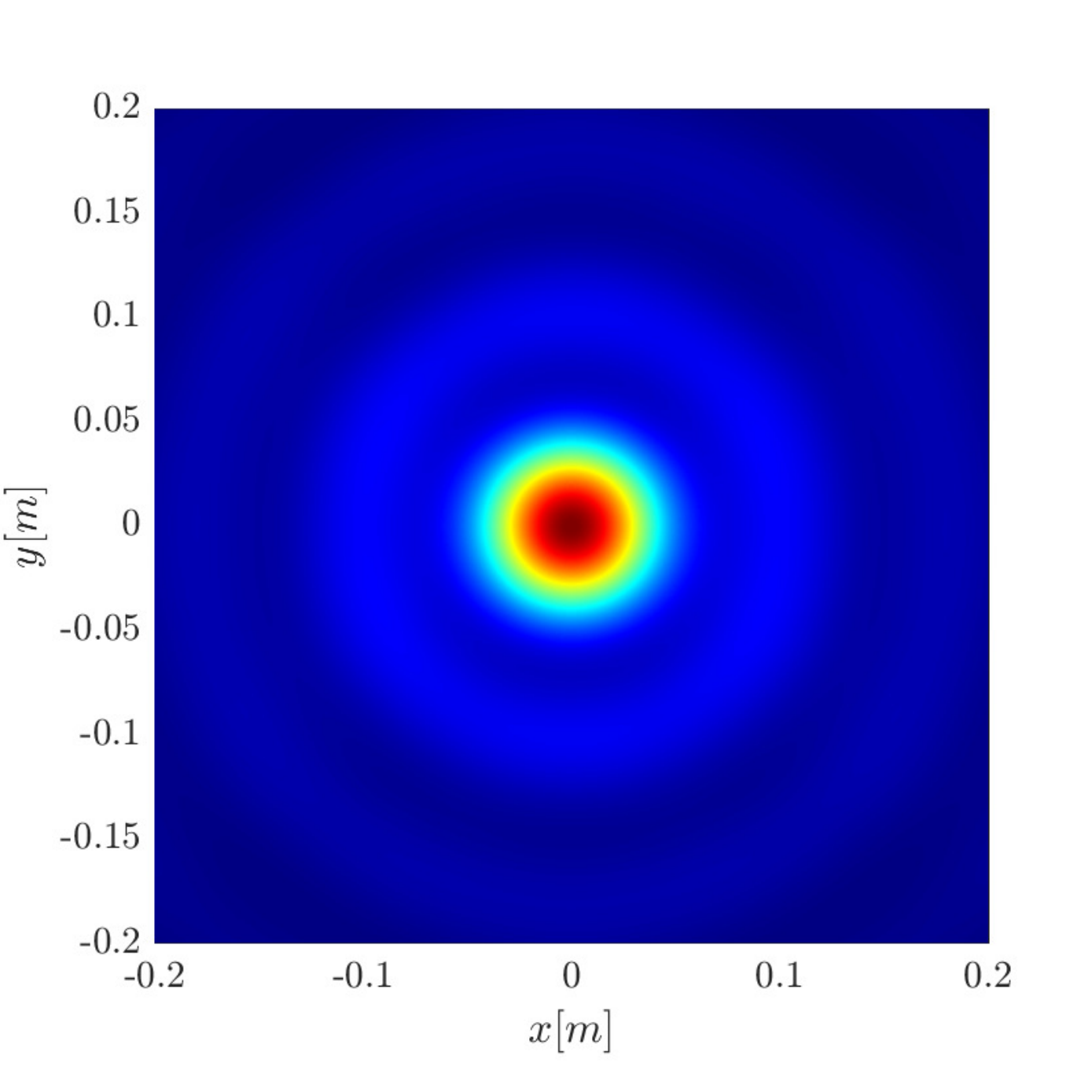}
		\caption{}
	\end{subfigure}
	\begin{subfigure}[t]{0.45\textwidth}
		\includegraphics[width=\textwidth]{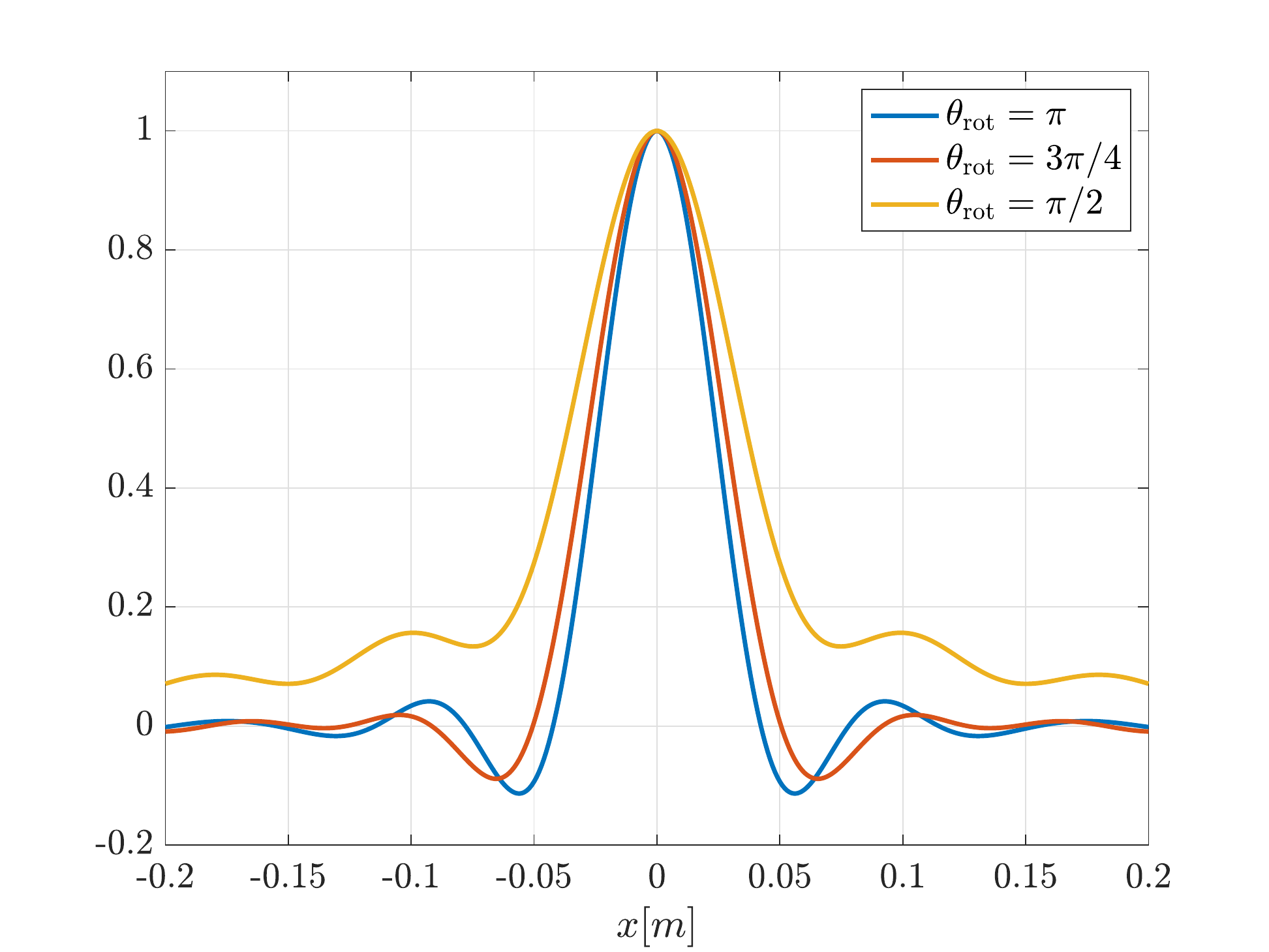}
		\caption{}
	\end{subfigure}
	\caption{$\mathcal{B}_\text{eff}(\mb x)$ as a function of $\theta_{\text{rot}}$. $(a)$ $\theta_{\text{rot}}=\pi$; $(b)$ $\theta_{\text{rot}}=3\pi/4$ $(c)$ $\theta_{\text{rot}}=\pi/2$. $(d)$ 1D cross section. As $\theta_{\text{rot}}$ decreases, the main lobe retains its width, while the decay away from it gets slower, which can affect the resolution as illustrated in Figure~\ref{fig:theta_deter}.}
	\label{fig:B_eff}
\end{figure}

When $\theta_{\text{rot}}$ becomes smaller, the array induced resolution also decays at a slower rate, as illustrated in Figure~\ref{fig:B_eff}. As a result, the performance of the rank-1 image is affected, as illustrated in Figure~\ref{fig:theta_deter}. Thus, even though the main lobe  of the peak has a smaller width, the slower decay rate means that we would see interference between two peaks at a greater distance. As a result, we would violate the condition of \eqref{eq:loc_cond}, the eigenvector will not be the sum of the local eigenvectors, and the rank-1 image would deteriorate. 

\textcolor{black}{
\subsection{Effect of errors in the estimation of the rotation parameters}
Noisy data affect the image performance in two ways: in estimating the rotational parameters and in the migration scheme itself. It was shown in \cite{leibovich2020generalized} that the rank-1 image is robust to additive noise. We also demonstrate here a robustness with respect to error in the estimation of the rotation parameters. As illustrated in Figure~\ref{fig:migration_noise}, an error in estimating the rotation parameters can affect the resolution. When imaging moving targets there is an inherent trade off in resolution between taking a smaller aperture, which mitigates motion estimation errors but limits the resolution, and taking a larger aperture, which improves resolution but is more sensitive to errors in the estimation of the parameters. We observe in Figure~\ref{fig:migration_noise} that the rank-1 image remains stable for a longer period of time and is more robust compared to the linear KM image. 
\begin{figure}[htbp]
	\centering
	\begin{subfigure}[t]{0.3  \textwidth}
		\includegraphics[width=\textwidth]{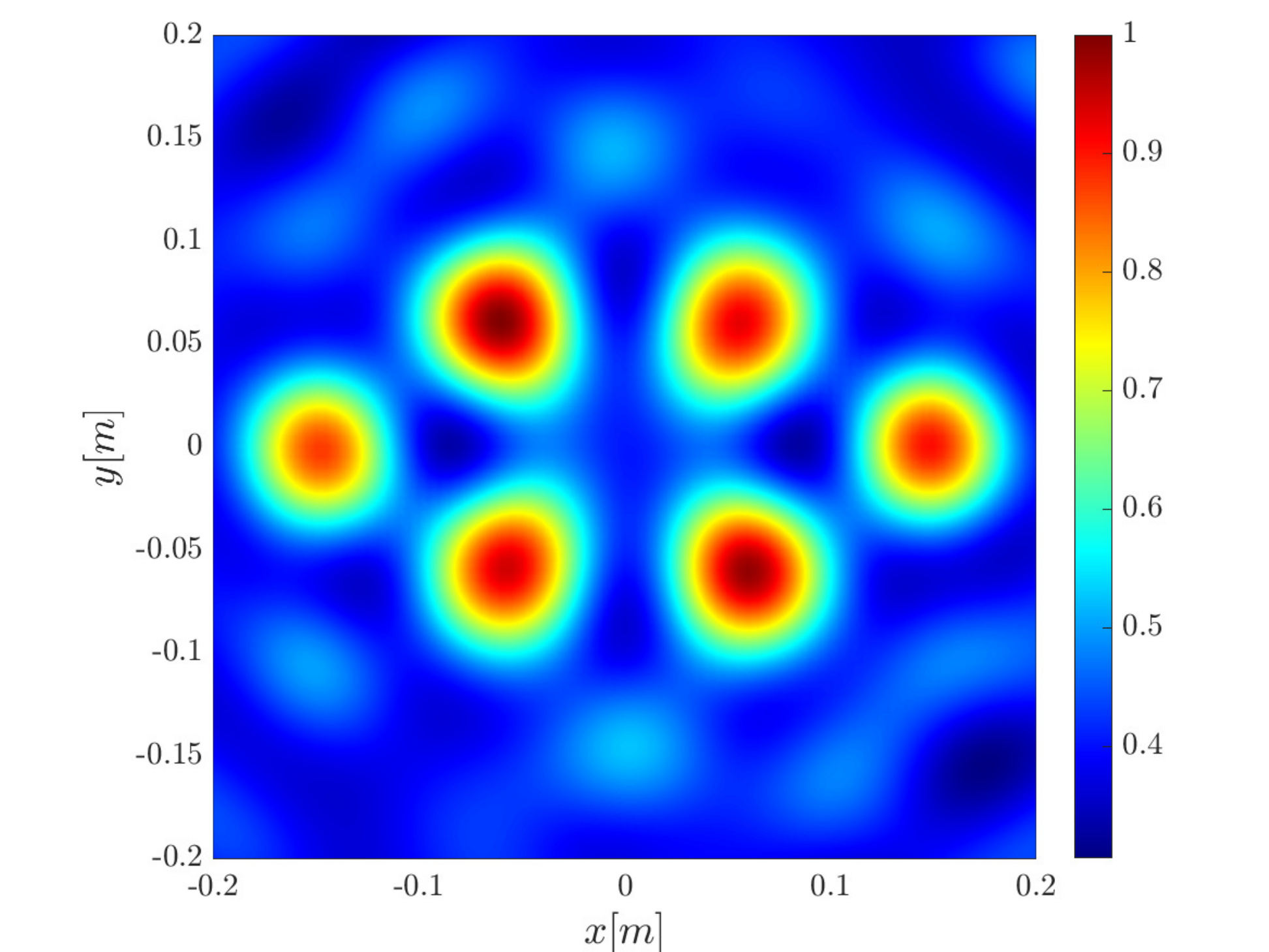}
		\caption*{}
	\end{subfigure}
	\begin{subfigure}[t]{0.3  \textwidth}
		\includegraphics[width=\textwidth]{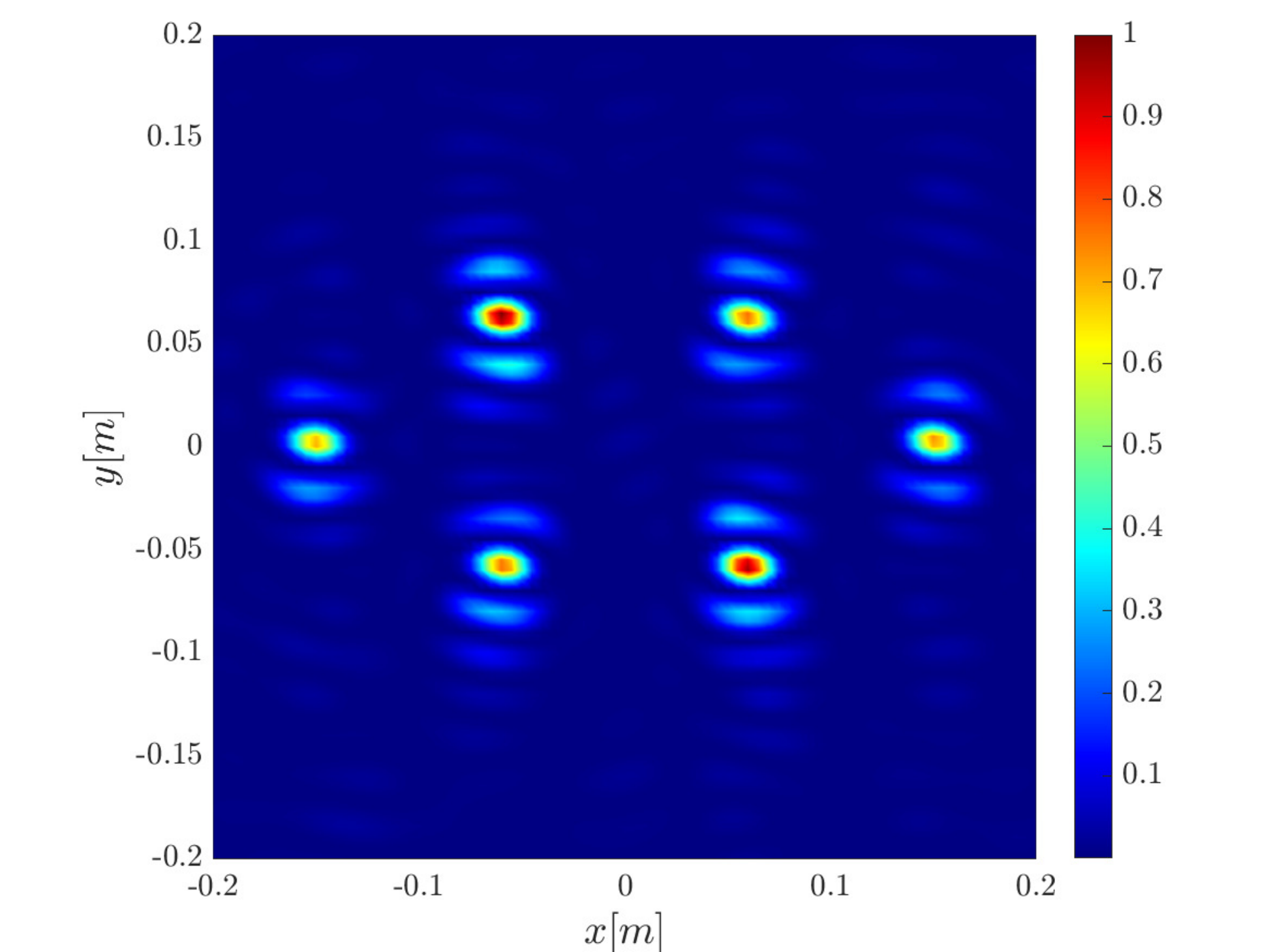}
		\caption*{}
	\end{subfigure}
	\begin{subfigure}[t]{0.3\textwidth}
		\includegraphics[width=\textwidth]{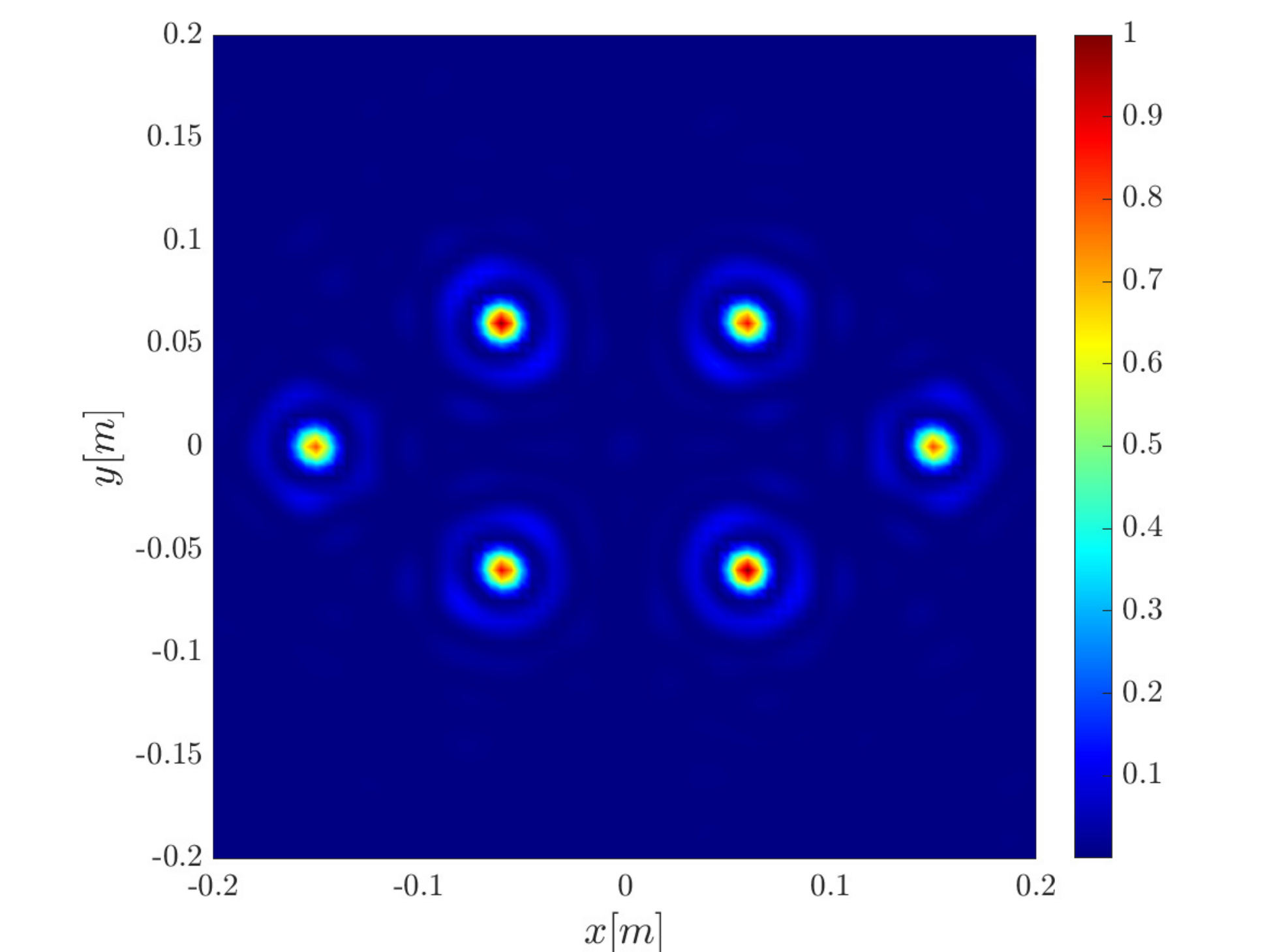}
		\caption*{}
	\end{subfigure}
	\begin{subfigure}[t]{0.3 \textwidth}
	\includegraphics[width=\textwidth]{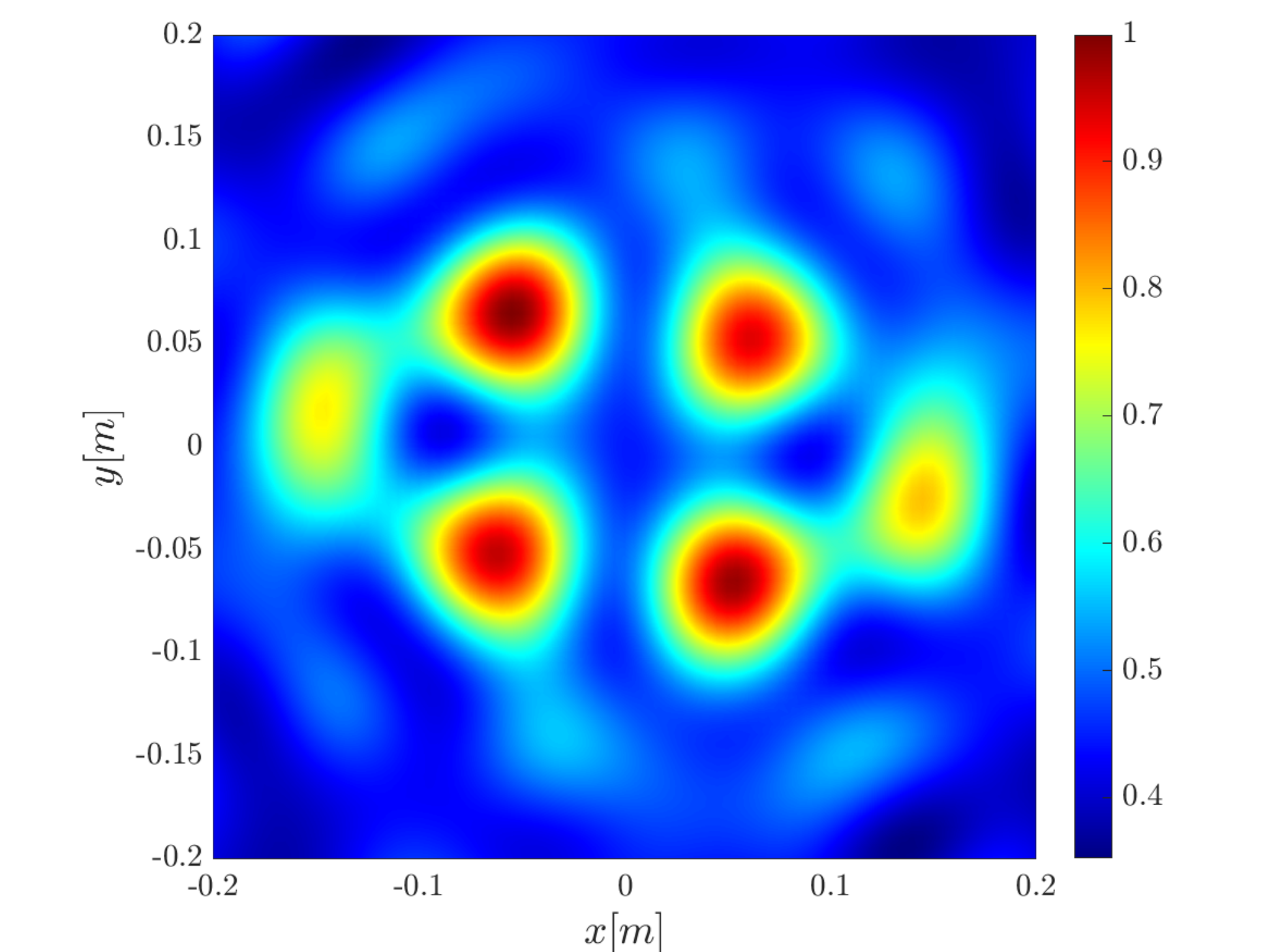}
	\caption{}
\end{subfigure}
\begin{subfigure}[t]{0.3  \textwidth}
\includegraphics[width=\textwidth]{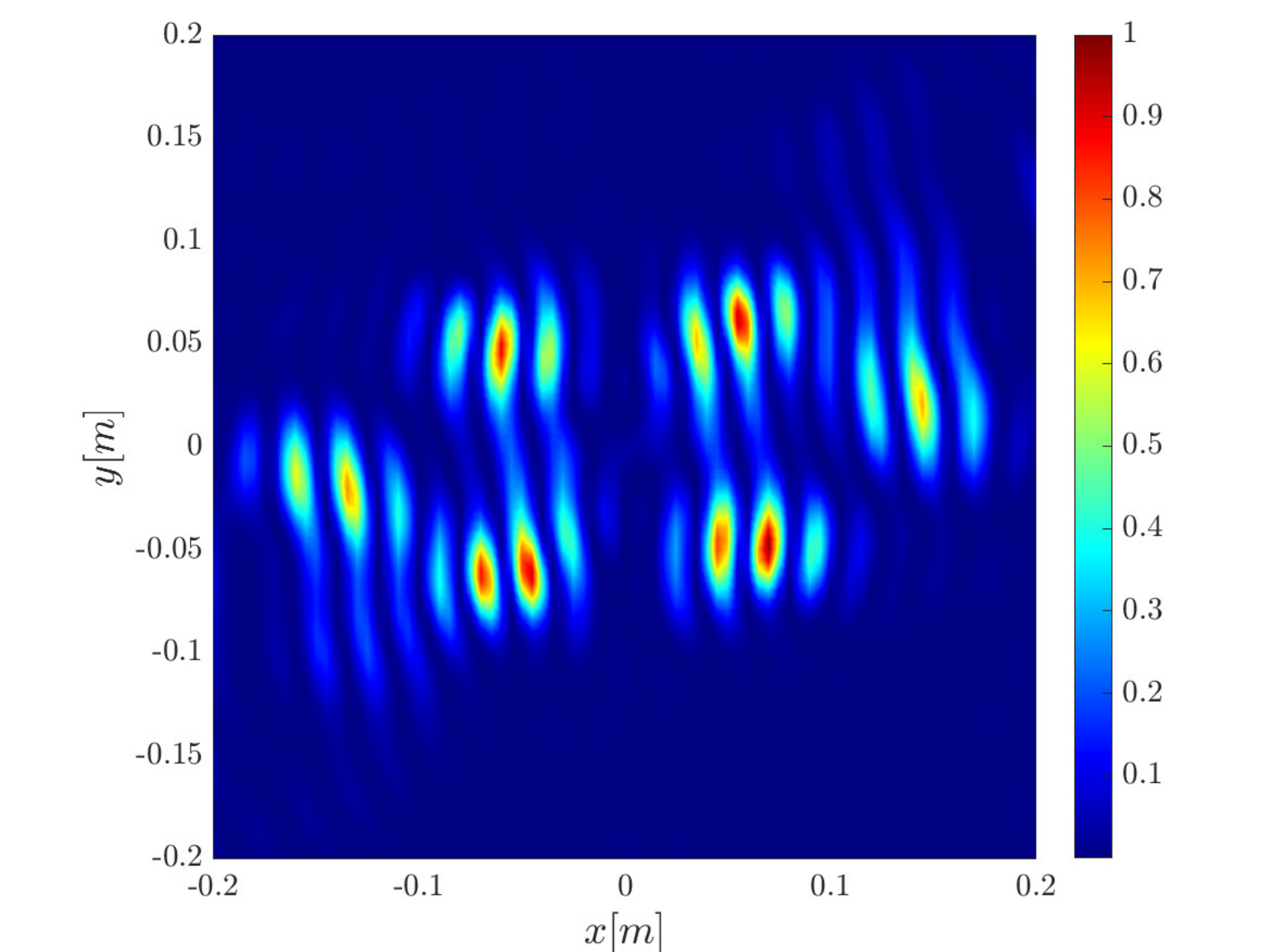}
\caption{}
\end{subfigure}
\begin{subfigure}[t]{0.3\textwidth}
\includegraphics[width=\textwidth]{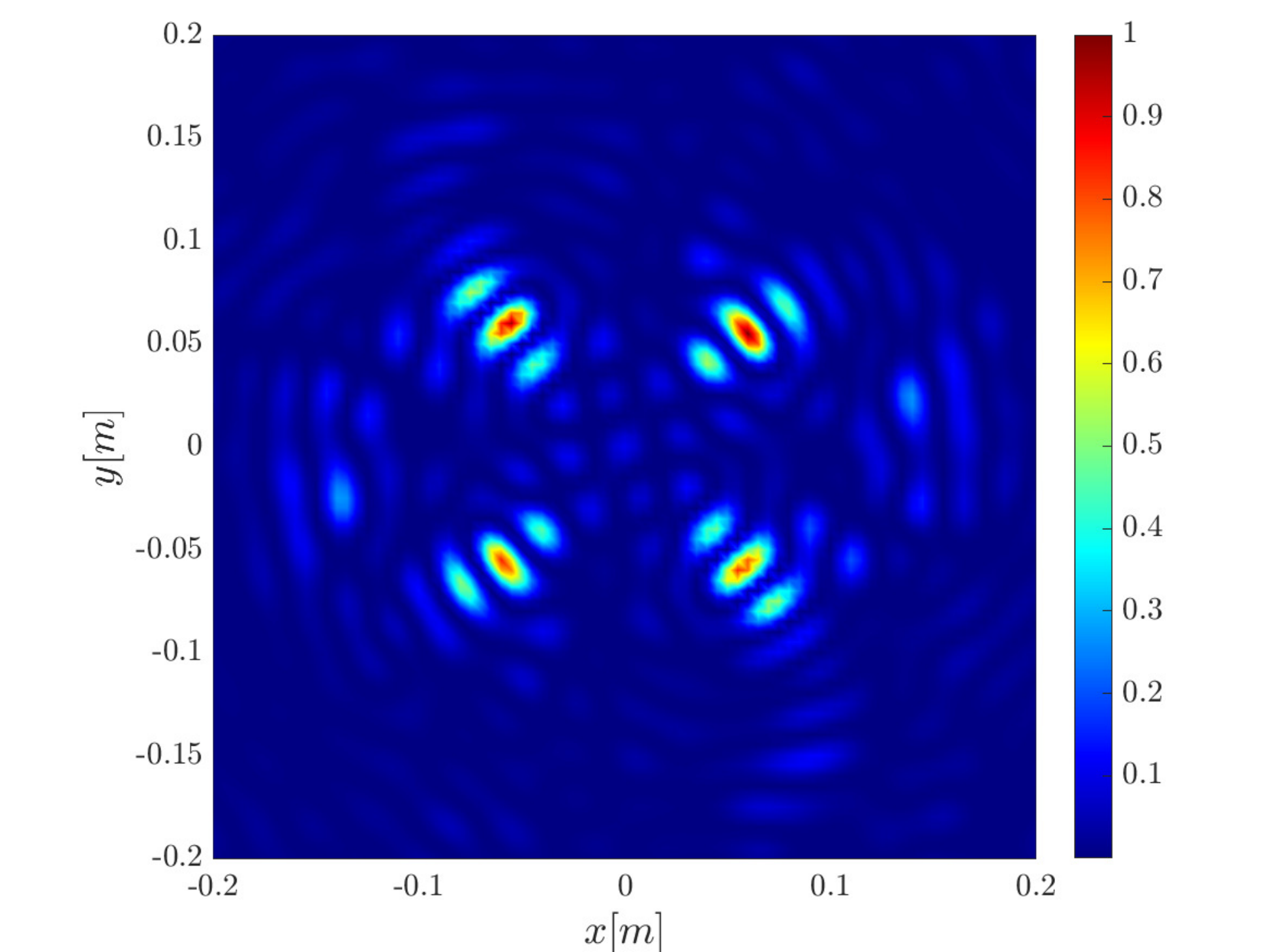}
\caption{}
\end{subfigure}
	\caption{Effect of errors in the estimation of the rotational parameters. The data are noisy with $0dB$ SNR. The aperture size is $1000\Delta s$.  Top row: migration with accurate rotation parameters. Bottom row: migration when $\omega_r, \theta_{\text{rot}},\phi_{\text{rot}}$ have a 3\% error. $(a)$ Single point migration $(b)$ Rank-1 image $(c)$ Kirchhoff migration. While there is deterioration, we can see that the rank-1 image is more robust to errors in the estimation of the rotational parameters than Kirchhoff migration, which loses the outer most targets.}
	\label{fig:migration_noise}
\end{figure}
}

\section{Summary and Conclusions}
\label{sec:summary}

	\textcolor{black}{In this paper we considered the problem of imaging fast moving objects, which are also rotating as a rigid body around a fixed axis, such as satellites in low earth orbit.  The imaging system consists of receivers flying over the turbulent atmosphere and asynchronous sources located on the ground. The objective is to estimate the object’s rotation parameters and obtain a high resolution image of the object in orbit.}
	
	\textcolor{black}{To mitigate distortions in the image due to ambient medium fluctuations induced by atmospheric effects, we use an imaging method that relies on the data structure, $C_{\mb R\mb R'}(s,\tau)$, obtained by cross-correlating the data at different receivers, ${\mb R}$, ${\mb R'}$ and adequately compensating for the Doppler effect.} 	
	
	\textcolor{black}{The first step of the proposed methodology consists in estimating the rotational parameters, namely the axis of rotation and angular velocity. Assuming that the rotating object is eccentric enough, we show %in Section \ref{sec:rot_estimate} 
	that the rotation parameters can be inferred from the autocorrelation data at the receivers by solving a regression problem.} 
	
	\textcolor{black}{The cross-correlation data structure $C_{\mb R\mb R'}(s,\tau)$  suggests the construction of the two-point interference pattern matrix $\mathcal{I}^{GCC}(\mb y_k,\mb y_{k'})$ defined by \eqref{eq:2_point_mat_imag} as a natural extension of the linear Kirchhoff migration imaging. %In Section~\ref{sec:generalized_migration}, 
	 We review two possible ways to extract an image from the interference pattern matrix $\mathcal{I}^{GCC}(\mb y_k,\mb y_{k'})$. Specifically, the single point migration image $\mathcal{I}^{CC}(\mb y_k)$ which is the diagonal of $\mathcal{I}^{GCC}(\mb y_k,\mb y_{k})$, and our proposed algorithm, the rank-1 image $\mathcal{I}^{R1CC}(\mb y_k)$, which is the first eigenvector of the interference pattern matrix.} 
 
 \textcolor{black}{We then demonstrate with analysis and numerical simulations that the rank-1 image has superior resolution compared to the single point migration when imaging rotating objects. We show, in particular, that the rank-1 image provides optimal diffraction limited resolution images for rotating objects assuming that data during one period of rotation is measured. Moreover, our numerical simulations illustrate that the rank-1 method is robust to additive measurement noise as well as to errors in the estimation of the rotational parameters.}

\section{Acknowledgments}
The work of M. Leibovich and G. Papanicolaou was partially supported by AFOSR FA9550-18-1-0519. The work of C. Tsogka was partially supported by AFOSR FA9550-17-1-0238 and AFOSR FA9550-18-1-0519.

\appendix
\setcounter{equation}{0}
\renewcommand{\theequation}{\Alph{section}.\arabic{equation}}

\section{Auto-correlation for rotation estimation}
\label{app:rot_est}
We are looking at the autocorrelation %for different receivers. 
\begin{equation}
C_{\mathbf{R}}(\tau,s)=\int dt \tilde u_{\mathbf{R},\mb x _{\mb L}(s)}(t) \tilde u_{\mathbf{R},\mb x _{\mb L}(s)}(t+\tau).
\end{equation}
Within the point scatterer model of \eqref{eq:point_scatter}, we can model $\tilde u_{\mb R,\mb x_{\mb L}(s)}(s,\tau)$ as
\begin{equation}
\tilde u_{\mb R,\mb x_{\mb L}(s)}(s,\tau)\approx\frac{1}{(4\pi|\mb x_{\mb T}-\mb x_\mb R|)^2}\sum\limits_{i=1}^{N_T} -\rho_i f''(t-t_\mb R^i(s)),
\end{equation}
with the travel time $t_\mb R^i$ defined in \eqref{eq:t_rot}.
Denote $\mathcal G (\tau)$ the auto-correlation of $f''(t)$,
\begin{equation}
\mathcal G (\tau)=\int dt f''(t)f''(t+\tau), 
\end{equation}
which is also localized.
then 
\begin{equation}
C_{\mb R}(s,\tau)\propto \sum\limits_{i,j=1}^{N_T} \mathcal G(\tau-\Delta \tau_{ij}), 
\end{equation}
with 
$$\Delta \tau _{ij}= t_\mb R^{i}(s)-t_\mb R^j(s).$$
Since $f''(t)$ is a localized signal, $\mathcal G(\tau)$ would be localized as well, and we expect the total support of the signal in $\tau$ to be $\approx 2\max\limits_{i,j} \Delta \tau_{ij}$. \textcolor{black}{This is the main observation on which the derivation of the proposed loss function is based as we explain in detail below.  Specifically, the total support of the signal is measured for different receivers and a model that expresses this support as a function of the unknown rotational parameters is derived. By minimizing the distance between the model and the measurements the rotational parameters are determined.}

\textcolor{black}{To derive the model  for the total support of the signal, we use Taylor approximation and approximate} $\Delta \tau_{ij}$ to first order as, 
\begin{equation}
\Delta \tau _{ij}\approx \frac{1}{c_0}(\mb x_{\mathcal R}^i(s)-\mb x_{\mathcal R}^j(s))^T\left[\frac{\mb x_{\mb L}(s)-\mb x_{\mb E}}{|\mb x_{\mb L}(s)-\mb x_{\mb E}|}+\frac{\mb x_{\mb L}(s)-\mb x_{\mb R}}{|\mb x_{\mb L}(s)-\mb x_{\mb R}|}\gamma_{\mathbf{R}}(\mb x_\mb L(s),\mb x_{\mb E},\mb v_\mb L)\right].
\end{equation}
Notice that 
\textcolor{black}{
\begin{equation}
(\mb x_{\mathcal R}^i(s)-\mb x_{\mathcal R}^j(s))^T=(\mb x_i-\mb x_j)^T\mathcal R_{z}^T(\omega_r s)\mathcal{R}_{\Omega}^T(\phi_{\rm rot},\theta_{\rm rot}),
\end{equation}}
where we model the rotation matrices, as $\mathcal R_{\Omega}(\phi,\theta)$, taking the frame of reference from the rotation plane to the $xyz$ space 
\begin{equation}
\mathcal{R}_{\Omega}(\phi,\theta)=\begin{pmatrix} \cos\phi&-\sin\phi&0 \\\sin\phi&\cos\phi&0\\0&0&1\end{pmatrix} 
\begin{pmatrix}\cos\theta&0&-\sin\theta\\ 0&1&0\\ \sin\theta&0&\cos\theta \end{pmatrix}
\label{eq:rot_mat}
\end{equation}
and $\mathcal{R}_z({\omega_rs})$, the in-plane rotation
\begin{equation}
\mathcal{R}_{z} (\varphi) =\begin{pmatrix} \cos\varphi&-\sin\varphi&0 \\\sin\varphi&\cos\varphi&0\\0&0&1\end{pmatrix}. 
\end{equation}
Denoting 
\begin{equation}
\mb d_\mb R(s)=\frac{\mb x_{\mb L}(s)-\mb x_{\mb E}}{|\mb x_{\mb L}(s)-\mb x_{\mb E}|}+\frac{\mb x_{\mb L}(s)-\mb x_{\mb R}}{|\mb x_{\mb L}(s)-\mb x_{\mb R}|}\gamma_{\mathbf{R}}(\mb x_\mb L(s),\mb x_{\mb E},\mb v_\mb L), 
\end{equation}
then we can rewrite the inner product as
\begin{equation}
\Delta\tau_{ij}\approx (\mathcal R_{z}(\omega_r s)(\mb x_i-\mb x_j))^T\textcolor{black}{{\mathcal R}}_{\Omega}^T(\phi_{\rm rot},\theta_{\rm rot}) \mb d_\mb R(s).
\end{equation}
If the object is rotationally invariant, or has multiple directions in which it is extended to the same distance, it would be hard to identify a peak with a specific aspect of the object. However, if the object has nontrivial eccentricity, i.e, there is a certain direction in which the object is further spread out from the center of rotation, there will be a single pair $i^*,j^*$ for which $\Delta \tau _{i,j} $ achieves the maximum value, i.e.
\begin{equation}
(i^*,j^*)=\arg\max\limits_{i,j,s} 
(\mathcal R_{z}(\omega_r s)(\mb x_i-\mb x_j))^T\textcolor{black}{{\mathcal R}}_{\Omega}^T(\phi_{\rm rot},\theta_{\rm rot}) \mb d_\mb R(s),
\end{equation}
and we can assign peaks to a specific orientation of this pair as follows.\\ 
We can write,
\begin{equation}
(\mb x_{i^*}-\mb x_{j^*})=(r\cos\varphi,r\sin\varphi\textcolor{black}{,0}), 
\end{equation}
then
\begin{equation}
\mathcal R_{z}(\omega_r s)(\mb x_{i^*}-\mb x_{j^*})=(r\cos(\omega_rs+\varphi),r\sin(\omega_r s+\varphi)\textcolor{black}{,0}).
\end{equation}
We define $\tau_{\mb R,\text{supp} (s)}$, the support of $C_\mb R(s,\tau)$ with respect to $\tau$, 
\begin{equation}
\tau_{\mb R,\text{supp} }(s)=2 \max\left\{\tau \hspace{0.2em} \Big|\hspace{0.2em} |C_{\mb R }(s,\tau)|\ge \textcolor{black}{\alpha} \max \limits_{\tau}|C_{\mb R }(s,\tau)|\right\},
\end{equation}
and look for all the points of local maxima, $s^*_\mb R$, as was illustrated in Figure~\ref{fig:time_smoot}.
\textcolor{black}{The value of $\alpha$ is adapted to the ambient noise level of the received signal. For SNR of $0$ dB we used $\alpha=0.075$. For noiseless measurements $\alpha$ can be much smaller, we used $\alpha=0.001$ in the noiseless case.}

In those points the inner product is maximal, i.e., operating with the transpose (inverse) rotation matrix on $\mb d_\mb R(s^*)$ would yield a projection in the 2D plane, parallel to $\mathcal R_{z}(\omega_r s^*_\mb R)(\mb x_{i^*}-\mb x_{j^*})$, i.e.,
\begin{equation}
\label{eqrot1}
\textcolor{black}{{\mathcal R}}_{\Omega}^T \mb d_\mb R(s^*_\mb R)=\alpha (\cos(\omega_r s^*_\mb R+\varphi),\sin(\omega_r s^*_\mb R+\varphi),\beta),\quad \alpha,\beta\in\mathbb{R}.
\end{equation}
We use this observation to define the objective in \eqref{eq:rot_loss_obj}. \textcolor{black}{From the equation \eqref{eqrot1} by dividing the second component of the vector $\textcolor{black}{{\mathcal R}}_{\Omega}^T \mb d_\mb R(s^*_\mb R)$ with the first we get
\begin{equation}
\frac{(\textcolor{black}{{\mathcal R}}_{\Omega}^T \mb d_\mb R(s^*_\mb R))_2}{(\textcolor{black}{{\mathcal R}}_{\Omega}^T \mb d_\mb R(s^*_\mb R))_1}=\tan(\omega_rs^*_\mb R+\varphi).
\label{eq:tan_ratio}
\end{equation} 
On the other hand, using the definition of the rotation matrix \eqref{eq:rot_mat}, we get the expression
\begin{equation}
g(\theta,\phi,\mathbf{d}_\mb R):= 
\frac{(\textcolor{black}{{\mathcal R}}_{\Omega}^T \mb d_\mb R(s^*_\mb R))_2}{(\textcolor{black}{{\mathcal R}}_{\Omega}^T \mb d_\mb R(s^*_\mb R))_1} 
=
\frac{-(d_{\mb R})_1 \sin\phi + (d_{\mb R})_2 \cos \phi }{(d_{\mb R})_1 \cos\theta\cos\phi +(d_{\mb R})_2 \cos\theta\sin\phi + (d_{\mb R})_3 \sin\theta  }
\end{equation}
Assuming we collect the times from different receivers so that we have a set $(s_{\mb R_i}^*,\mathbf{d}_{\mb R_i})$, we can build an objective function for $\omega_r,\theta,\phi$ by taking the $\tan^{-1}$ on both sides of \eqref{eq:tan_ratio} and taking the difference between two consecutive data points to eliminate the phase $\varphi$, i.e.,
\begin{equation}
\tan^{-1}g(\theta,\phi,\mathbf{d}_{\mb R_i}(s_{\mb R_i}^*))-\tan^{-1}g(\theta,\phi,\mathbf{d}_{\mb R_{i-1}}(s_{\mb R_{i-1}}^*))=\omega_r (s_{\mb R_i}^*-s_{\mb R_{i-1}}^*).
\label{eqrot2}
\end{equation}
We can use \eqref{eqrot2} to define the loss objective of \eqref{eq:rot_loss_obj}
\begin{equation}
L(\theta,\phi,\omega_r)=\sum\limits_i \left(\tan^{-1} g(\theta,\phi,\mathbf{d}_{\mb R_i}(s_{\mb R_{i}}^*))-\tan^{-1} g(\theta,\phi,\mathbf{d}_{\mb R_{i-1}}(s_{\mb R_{i-1}}^*))-(\omega_r (s^*_{\mb R_{i}}-s^*_{\mb R_{i-1}}))\right)^2,
\label{eq:loss_obj}
\end{equation}
and determine the rotational parameters by solving the optimization problem
\begin{equation}
\left( \theta^*,\phi^*,\omega_r^*\right)=\arg\min\limits_{\theta,\phi,\omega_r}L(\theta,\phi,\omega_r).
\end{equation}}

\section{Approximate evaluation of the two point interference pattern for rotating objects}
\label{app:stat_phase}
\setcounter{equation}{0}
In this appendix we analyze the expression of the two point interference pattern \eqref{eq:M_tilde}, assuming the rotation parameters have been resolved. We follow the steps of the analysis carried in \cite{leibovich2020generalized}. As in \cite{leibovich2020generalized}, the structure of the interference pattern plays a determining role in the resolution of the rank-1 image of the rotating object.

We look at plane images, that is, images with a fixed $z$ coordinate (height), so that the image coordinate $\mb y_k\in \mathbb{R}^2$. We assume a point scatterer model. The scatterers are located at $\mb y_i, i=1,\dots,M$ with respect to the center of the image window, which is also the axis of rotation, $\mb x_\mb T(s)$, that is, 
$$ \mb y_i(s)=\mb x_\mb T(s)+\mb y_i ^{\mathcal R}(s),\quad \mb y_i ^{\mathcal R}(s)=\mathcal{R}(s)\mb y_i$$
$\mathcal{R}(s)=\mathcal{R}_{\Omega}\mathcal{R}_{z,\omega_r s}$ is the rotation operator, defined in \eqref{eq:rot_op}. $\mathcal{R}_{z,\omega _r s}$ rotates the point scatterers in plane with a rotational velocity $\omega_r$. The $x-y$ plane of rotation is then transformed to the real coordinates by $\mathcal{R}_\Omega$, transforming the axis of rotation.

With the reflectivities of the targets $\rho_i$ \eqref{eq:M_tilde} takes the form
\begin{equation}
\hat{C}_{\mb R \mb R'}(s,\omega)\approx |\xi(s,\omega)|^2\sum\limits_{i,j=1}^M\rho_i\rho_je^{i\omega (t_\mb R^i(s)-t_{\mb R'}^j(s)-(t_\mb R(s)-t_{\mb R'}(s)))} 
%\\
%\Delta \tau^s_\mathbf{RR'}=&t_{\mb R'}(\mb x_\mb T+s\mb v_\mb T, \mb x_ \mb E, \mb v_\mb T)-t_\mb R(\mb x_\mb T+s \mb v_\mb T, \mb x_ \mb E, \mb v_\mb T)
\end{equation}
with 
$$
\begin{array}{ll} 
t_\mathbf{R}^{i}(s)& \displaystyle =\frac{|\mb y^{\mathcal{R}}_i(s) -\mb x_\mathbf{E}|}{c_0}+\frac{|\mb y^{\mathcal{R}}_i(s) -\mb x_\mathbf{R}|}{c_0}\gamma_{\mathbf{R}}(\mb x_\mb L(s),\mb x_\mathbf{E},\mathbf{v}_T), \\[12pt]
t_\mathbf{R}(s) & \displaystyle=\frac{|\mb  x_\mb L(s) -\mb x_\mathbf{E}|}{c_0}+\frac{|\mb x_\mb L(s) -\mb x_\mathbf{R}|}{c_0}\gamma_{\mathbf{R}}(\mb x_\mb L(s),\mb x_\mathbf{E},\mathbf{v}_T). 
\end{array}
$$
Let us define, 
\begin{equation}
\begin{split}
%\mb x^{\omega_r}(s)&=\mathcal{R}(s)\mb x\\
t_\mathbf{R}^{\mb x}(s)=&\frac{|\mb x_\mb L(s)+\mathcal{R}(s)\mb x-\mb x_\mathbf{E}|}{c_0}+\frac{| \mb x_\mb L(s)+\mathcal{R}(s)\mb x-\mb x_\mathbf{R}|}{c_0}\gamma_{\mathbf{R}}(\mb x_\mb L(s) ,\mb x_\mathbf{E},\mathbf{v}_T).\\
\end{split}
\end{equation}
The two point migration translates the cross-correlation data to a pair of points $(\mb x,\mb y)$ in the image window, and then sums over all receiver pairs $\mb R,\mb R'$, pulses $s$, and frequencies $\omega$. 
The interference pattern $\tilde{\mb X}$ in \eqref{eq:M_tilde} then has the form
\begin{equation}
\begin{split}
\tilde{\mb X}_{\mb x,\mb y}&=\sum\limits_{s,\omega,\mb R,\mb R'}\hat{\mb C}_{\mb R\mb R'}(s,\omega)e^{i\omega (t_\mb R^\mb x(s)-t_\mb R(s))}e^{-i\omega (t_{\mb R'}^\mb y(s)- t_{\mb R'}(s))}\\
&\approx\sum\limits_{s,\omega,\mb R,\mb R'}  |\xi(s,\omega)|^2\sum\limits_{i,j=1}^M\rho_i\rho_je^{i\omega (t_\mb R^{\mb x}(s)-t_\mb R^i(s)-(t_{\mb R'}^\mb y(s)-t_{\mb R'}^j(s)))}
\end{split}
\end{equation}

We approximate the sum over pulses and frequencies with integrals. The sum over receiver pairs is replaced by a double integral over the physical aperture spanned by the receivers. This is well justified if we assume that the receiver positions are uniformly distributed over an area of $200 \text{km} \times 200\text{km}$ as we do here. By making the change $\sum\limits_{s,\mb R,\mb R'}\rightarrow \int ds \int d\mb x_\mb R\int d\mb x_{\mb R'}$, the interference pattern $\tilde{\mb X}$ has the form
\begin{equation}
\begin{split}
\tilde{\mb X}_{\mb x,\mb y}&\approx\int d\mb x_\mb R d\mb x_{\mb R'} ds d\omega |\xi(s,\omega)|^2\sum\limits_{i,j=1}^M\rho_i\rho_je^{i\omega (t_\mb R^{\mb x}(s)-t_\mb R^i(s)-(t_{\mb R'}^\mb y(s)-t_{\mb R'}^j(s)))}\\&=\sum\limits_{i,j=1}^M\rho_i\rho_j\int d\mb x_\mb R d\mb x_{\mb R'} ds d\omega |\xi(s,\omega)|^2 e^{i\omega (t_\mb R^{\mb x}(s)-t_\mb R^i(s)-(t_{\mb R'}^\mb y(s)-t_{\mb R'}^j(s)))}\\
&=\sum\limits_{i,j=1}^M\rho_i\rho_j\int d\omega \int ds |\xi(s,\omega)|^2 \int d\mb x_\mb R e^{i\omega (t_\mb R^{\mb x}(s)-t_\mb R^i(s))}\int d\mb x_{\mb R'}e^{-i\omega (t_{\mb R'}^\mb y(s)-t_{\mb R'}^j(s))}
\end{split}
\label{eq:M_w}
\end{equation}
We see that the integrals over the physical receiver aperture separate.
We can approximate the travel time difference in the exponent using $|\mb x_\mb T(s)-\mb x_\mb E|,|\mb x_\mb T(s)-\mb x_\mb R|\gg |\mb x|,|\mb y|$. We have to first order
$$|\mb z+ \mb w|\approx |\mb z|+\frac{\mb z}{|\mb z|} \cdot \mb w +O(|\mb w|^2) .$$
Thus, also taking $\gamma_{\mathbf{R}}\approx 1$, we can approximate the argument of the exponent as
\begin{equation}
\begin{split}
t_\mathbf{R}^{\mb x}(s)-t_\mathbf{R}^{i}(s)=&\frac{|\mb x_\mb T(s)+\mathcal{R}(s)\mb x -\mb x_\mathbf{E}|}{c_0}+\frac{| \mb x_\mb T(s)+\mathcal{R}(s)\mb x -\mb x_\mathbf{R}|}{c_0}\gamma_{\mathbf{R}}(\mb x_\mb T(s)+\mathcal{R}(s)\mb x ,\mb x_\mathbf{E},\mathbf{v}_T)\\
-&\frac{|\mb x_\mb T(s)+\mathcal{R}(s)\mb y_i -\mb x_\mathbf{E}|}{c_0}-\frac{| \mb x_\mb T(s)+\mathcal{R}(s)\mb y_i-\mb x_\mathbf{R}|}{c_0}\gamma_{\mathbf{R}}(\mb x_\mb T(s)+\mathcal{R}(s)\mb y_i,\mb x_\mathbf{E},\mathbf{v}_T)\\
&=\frac{1}{c_0}\mathcal{R}(s)(\mb x-\mb y_i)\cdot \frac{\mb x_\mb T(s)-\mb x_\mb E}{|\mb x_\mb T(s)-\mb x_\mb E|}+ \frac{1}{c_0}\mathcal{R}(s)(\mb x-\mb y_i)\cdot \frac{\mb x_\mb T(s)-\mb x_\mb R}{|\mb x_\mb T(s)-\mb x_\mb R|}.
\end{split}
\end{equation}
We can then write the integral over the physical aperture as
\begin{equation}
\begin{split}
&\int d\mb x_\mb R \textcolor{black}{\exp}{\left(i\omega (t_\mb R^{\mb x}(s)-t_\mb R^i(s))\right)}\approx \int d\mb x_\mb R \textcolor{black}{\exp}{\left(i \omega \left(\frac{1}{c_0}\mathcal{R}(s)(\mb x-\mb y_i)\cdot \frac{\mb x_\mb T(s)-\mb x_\mb E}{|\mb x_\mb T(s)-\mb x_\mb E|}+ \frac{1}{c_0}\mathcal{R}(s)(\mb x-\mb y_i)\cdot \frac{\mb x_\mb T(s)-\mb x_\mb R}{|\mb x_\mb T (s)-\mb x_\mb R|}\right)\right)}\\
&= \textcolor{black}{\exp}\left({i \omega \left(\frac{1}{c_0}\mathcal{R}(s)(\mb x-\mb y_i)\cdot \frac{\mb x_\mb T(s)-\mb x_\mb E}{|\mb x_\mb T(s)-\mb x_\mb E|}\right)}\right)\int d\mb x_\mb R 
\textcolor{black}{\exp}\left({i \omega \left(\frac{1}{c_0}\mathcal{R}(s)(\mb x-\mb y_i)\cdot \frac{\mb x_\mb T(s)-\mb x_\mb R}{|\mb x_\mb T(s)-\mb x_\mb R|}\right)}\right).
\end{split}
\label{eq:k_w_ex}
\end{equation}
This is a classical result in array imaging \cite{bleistein}. The physical receiver array induces an integral over phases linear in the difference between the positions of the scatterer and the search point $\mb x- \mb y_i$. 
If we further note 
$$\mathcal{R}(s)(\mb x-\mb y_i)\cdot \frac{\mb x_\mb T(s)-\mb x_\mb R}{|\mb x_\mb T(s)-\mb x_\mb R|}=(\mb x-\mb y_i)\cdot \mathcal{R}(s)^T\frac{\mb x_\mb T(s)-\mb x_\mb R}{|\mb x_\mb T(s)-\mb x_\mb R|},$$
and define
 $$\mb k_\mb R^s=\frac{\omega}{c_0}\mathcal{R}(s)^T\frac{\mb x_\mb T(s)-\mb x_\mb R}{|\mb x_\mb T(s)-\mb x_\mb R|},\quad  |\mb k _{\mb R }^s|=\frac{\omega}{c_0},$$ 
the integral is in fact one over a domain $\Xi^s$ in $\mb k$ space, which is a subdomain of the 2-sphere with radius $\omega/c_0$, as illustrated in Figure~\ref{fig:k_w_1},
\begin{equation}
\begin{split}
\textcolor{black}{\exp}\left({i \omega \left(\frac{1}{c_0}\mathcal{R}(s)(\mb x-\mb y_i)\cdot \frac{\mb x_\mb T(s)-\mb x_\mb E}{|\mb x_\mb T(s)-\mb x_\mb E|}\right)}\right)\int\limits_{\Xi^s} d \mb k_{\mb R}^s|J|
e^{i \mb k_{\mb R}^s\cdot (\mb x-\mb y_i)},\quad \Xi^s\subset S^2(\omega/c_0).
\end{split}
\end{equation}
The Jacobian $J$ and the integral can be evaluated in a more general setting using the Green-Helmholtz identity \cite{garnier16}.
We can then  write \eqref{eq:M_w} as
\begin{equation}
\begin{split}
\sum\limits_{i,j=1}^M\rho_i\rho_j&\int d\omega \int ds|\xi(s,\omega)|^2\textcolor{black}{\exp}\left({i \omega \left(\frac{1}{c_0}(\mathcal{R}(s)(\mb x-\mb y_i)-\mathcal{R}(s)(\mb y-\mb  y_j))\cdot \frac{\mb x_\mb T(s)-\mb x_\mb E}{|\mb x_\mb T(s)-\mb x_\mb E|}\right)}\right)\\
\times &\int\limits_{\Xi^s\times\Xi^s} d \mb k_{\mb R}^sd \mb k_{\mb R'}^s|J||J'|
\textcolor{black}{\exp}\left({i \left(\mb k_{\mb R}^s\cdot(\mb x-\mb y_i)- \mb k_{\mb R'}^s\cdot(\mb y-\mb y_j)\right)}\right).
\end{split}
\label{eq:k_w_int}
\end{equation}
Similar expressions have also been analyzed in \cite{bleistein}. Notice the domain itself is dependent on the fast time as the result of the rotation. A qualitative explanation for the improved resolution of rotating objects, illustrated in Figure~\ref{fig:k_w_1} is that by rotation an effective larger subdomain $\Xi$ is mapped
$$\Xi=\bigcup\limits_s \Xi ^s.$$
$\Xi^s$ is proportional to the physical aperture size. The larger $\Xi^s$ the better the resolution, which means that the resolution is inversely proportional to it. $\Xi$ can cover a much larger subdomain, bringing the resolution close to the optimal value of $\lambda/2$ which we get for integrating over the entire sphere. 
\subsection{Approximate closed form expression for $\tilde{\mb X}$}
We can make a further approximation to simplify the calculation. Assume $$|\mb x_\mb T(s)-\mb x_\mb R|,|\mb x_\mb T(s)-\mb x_\mb E|\approx H_\mb T, $$ where $H_\mb T$ is the height of the targets. This is well justified for uniformly distributed ground receivers and an inverse aperture of a size comparable to the ground array ($|\mb v_T S|\gtrsim a$,  $a$ the diameter of the receiver array). In this approximation we also have that $\xi(s,\omega)\approx \xi(\omega)$. We can rewrite the integral \eqref{eq:k_w_int} as
\begin{equation}
\label{eq:inter_res}
\begin{split}
& \sum\limits_{i,j=1}^M\rho_i\rho_j\int d\omega |\xi(\omega)|^2 \int \limits_{-S/2}^{S/2}ds \textcolor{black}{\exp}\left({i  \frac{\omega}{c_0}\frac{\mb x_\mb T(s)-\mb x_\mb E}{H_\mb T}\cdot\mathcal{R}(s)\left((\mb x-\mb y_i)-(\mb y-\mb  y_j)\right)}\right)\\&\times\int d\mb x_\mb R \textcolor{black}{\exp}\left({i \frac{\omega}{c_0}\frac{\mb x_\mb T(s)-\mb x_\mb R}{H_\mb T}\cdot\mathcal{R}(s)(\mb x-\mb y_i)}\right)\int d\mb x_\mb {R'} \textcolor{black}{\exp}\left({-i \frac{\omega}{c_0}\frac{\mb x_\mb T(s)-\mb x_\mb {R'}}{H_\mb T}\cdot\mathcal{R}(s)(\mb y-\mb y_j^{\omega_r}(s))}\right)
\\
& \sum\limits_{i,j=1}^M\rho_i\rho_j\int d\omega |\xi(\omega)|^2 \int \limits_{-S/2}^{S/2}ds \textcolor{black}{\exp}\left({i  \frac{\omega}{c_0}\frac{2\mb x_\mb T(s)-\mb x_\mb E}{H_\mb T}\cdot\mathcal{R}(s)\left((\mb x-\mb y_i)-(\mb y-\mb  y_j)\right)}\right)\\&\times\int d\mb x_\mb R \textcolor{black}{\exp}\left({-i \frac{\omega}{c_0}\frac{\mb x_\mb R}{H_\mb T}\cdot\mathcal{R}(s)(\mb x-\mb y_i)}\right)\int d\mb x_\mb {R'} \textcolor{black}{\exp}\left({i \frac{\omega}{c_0}\frac{\mb x_\mb {R'}}{H_\mb T}\cdot\mathcal{R}(s)(\mb y-\mb y_j)}\right)
\\
&\approx  C\sum\limits_{i,j=1}^M\rho_i\rho_j\int d\omega |\xi(\omega)|^2 \int \limits_{-S/2}^{S/2}ds \textcolor{black}{\exp}\left({i \frac{\omega }{c_0}\frac{2\mb x_\mb T(s)-\mb x_\mb E-H_\mb R \hat{z}}{H_\mb T}\cdot\mathcal{R}(s)\left((\mb x-\mb y_i)-(\mb y-\mb  y_j)\right)}\right)\\
&\times\mathcal{B}_{A}(\mathcal{R}(s)(\mb x-\mb y_i))\mathcal{B}_{A}^*(\mathcal{R}(s)(\mb y-\mb y_j)),
\end{split}
\end{equation}
where we assume a rectangular grid $\mb x_\mb R=(x_1,x_2,H_\mb R), x_i\in[-a/2,a/2]$. Then $B_A(\mb x)$ defines the point spread function induced by the aperture
\begin{equation}
\mathcal{B}_A(\mb x-\mb y_i)=\int d\mb x_\mb R \textcolor{black}{\exp}\left({-i \frac{\omega}{c_0}\frac{\mb x_\mb R-H_\mb R\hat{z}}{H_\mb T}\cdot(\mb x-\mb y_i)}\right)=a^2\text{sinc}\left(\frac{\omega}{c_0}\frac{a}{2H_\mb T}(x_1-y_{i,1})\right)\text{sinc}\left(\frac{\omega}{c_0}\frac{a}{2H_\mb T}(x_2-y_{i,2})\right).
\label{eq:k_w_cart}
\end{equation}
Recall that
\begin{equation}
\mathcal{R}(s)=\begin{pmatrix} \cos\phi&-\sin\phi&0 \\\sin\phi&\cos\phi&0\\0&0&1\end{pmatrix} 
\begin{pmatrix}\cos\theta&0&-\sin\theta\\ 0&1&0\\ \sin\theta&0&\cos\theta \end{pmatrix}\begin{pmatrix} \cos\omega_r s&-\sin\omega_r s&0 \\\sin\omega_r s&\cos\omega_r s&0\\0&0&1\end{pmatrix} .
\end{equation}
Since we assume $\mb x=(x_1,x_2,0)$, that means all our vectors are planar, including the receiver positions $\mb x_\mb R$. Since $\mathcal{B}_A$ is real in the continuum limit $\mathcal{R}(s)$ does not  affect the phase, and the main lobe itself remains on the order of $\lambda H_\mb T/a$ in size. This is verified by simulations as can be seen in Figure~\ref{fig:phase_comp}. Thus, we can substitute $\mathcal{B}_A(\mathcal{R}(s)\mb x)$ for an effective, slow time independent aperture $\mathcal{B}_\text{eff}(\mb x)$, which we define as
\begin{equation}
\mathcal{B}_\text{eff}(\mb x)\approx\int ds \mathcal{B}_A(\mathcal{R}(s)x).
\end{equation}
We can then approximate \eqref{eq:inter_res} as
\begin{equation}
\begin{split}
& C\sum\limits_{i,j=1}^M\rho_i\rho_j\int d\omega |\xi(\omega)|^2 \int \limits_{-S/2}^{S/2}ds \textcolor{black}{\exp}\left({i \frac{\omega }{c_0}\frac{2\mb x_\mb T(s)-\mb x_\mb E-H_\mb R \hat{z}}{H_\mb T}\cdot\mathcal{R}(s)\left((\mb x-\mb y_i)-(\mb y-\mb  y_j)\right)}\right)\\
&\times\mathcal{B}_{A}(\mathcal{R}(s)(\mb x-\mb y_i))\mathcal{B}_{A}^*(\mathcal{R}(s)(\mb y-\mb y_j))\\
&\approx  \sum\limits_{i,j=1}^M\rho_i\rho_j\int d\omega |\xi(\omega)|^2 \mathcal{B}_{\text{eff}}(\mb x-\mb y_i)\mathcal{B}_{\text{eff}}^*(\mb y-\mb y_j)\int \limits_{-S/2}^{S/2}ds \textcolor{black}{\exp}\left({i \frac{\omega }{c_0}\frac{2\mb x_\mb T(s)-\mb x_\mb E}{H_\mb T}\cdot\mathcal{R}(s)\left((\mb x-\mb y_i)-(\mb y-\mb  y_j)\right)}\right)
\end{split}
\label{eq:eff_pattern}
\end{equation}

Since $\mb x_\mb T=(0,0,H_\mb T )$, $|\mb x_\mb E|,H_\mb R,|\mb v_\mb T S|\ll |\mb x _\mb T|$, we neglect those terms and further approximate $\mb x_\mb T(s)\approx \mb x_\mb T$ .
Denoting 
$$(\mb x-\mb y_i)-(\mb y-\mb  y_j)=|(\mb x-\mb y_i)-(\mb y-\mb  y_j)|(\cos \phi,\sin \phi)$$
and using the expressions for the rotation matrix we have
\begin{equation}
\frac{2\mb x_\mb T}{H_\mb T}\cdot\mathcal{R}(s)\left((\mb x-\mb y_i)-(\mb y-\mb  y_j)\right)=2 \sin \theta_\text{rot} |(\mb x-\mb y_i)-(\mb y-\mb  y_j)|\cos(\omega_r s+\phi).
\end{equation}
Plugging this into \eqref{eq:eff_pattern} we get
\begin{equation}
\begin{split}
&\sum\limits_{i,j=1}^M\rho_i\rho_j\int d\omega |\xi(\omega)|^2 \mathcal{B}_{\text{eff}}(\mb x-\mb y_i)\mathcal{B}_{\text{eff}}^*(\mb y-\mb y_j)\int \limits_{-S/2}^{S/2}ds \textcolor{black}{\exp}\left({i \frac{\omega }{c_0}2 \sin \theta_\text{rot}|(\mb x-\mb y_i)-(\mb y-\mb  y_j)|\cos(\omega_r s+\phi)}\right)\\
\propto&\sum\limits_{i,j=1}^M\rho_i\rho_j\int d\omega |\xi(\omega)|^2 \mathcal{B}_{\text{eff}}(\mb x-\mb y_i)\mathcal{B}_{\text{eff}}^*(\mb y-\mb y_j)J_0\left( \frac{\omega }{c_0}2 \sin \theta_\text{rot} |(\mb x-\mb y_i)-(\mb y-\mb  y_j)|\right)
\end{split}
\end{equation}
Where we used 
\begin{equation}
\int\limits_0^{2\pi/\omega}\textcolor{black}{\exp}\left({i x \cos(\omega t)}\right)dt=\frac{2\pi}{\omega}J_0(a),
\end{equation}
and $J_0(a)$ is the zeroth order Bessel function of the first kind.
As illustrated in  Figures~\ref{fig:phase_comp} and \ref{fig:k_w_1}, the approximations made above are well justified in the context of our numerical simulations.

We see that $\mathcal{B}_\text{eff}(\mb x-\mb y_i)$ is, as a function of $\mb x$, localized around the scatterer location $\mb y_i$. Hence $\tilde{\mb X}_{\mb x,\mb y}$ has peaks at points $(\mb x,\mb y)=(\mb y_i,\mb y_j)\in \mathbb{R}^4$, where $\mb y_i,\mb y_j$ are scatterer positions. $\mathcal{B}_\text{eff}(\mb x)$ has an effective resolution $\lambda \frac{H_\mb T}{a}$.
% This is the classical cross range resolution obtained with an array of size $a$, imaging a point at range $H_\mb T$. 
On the other hand, after a period of rotation, the synthetic aperture produces the term
$J_0 \left( \frac{\omega }{c_0}2\sin\theta_{\text{rot}}|(\mb x-\mb y_i)-(\mb y-\mb y_j)|\right)$ in (\ref{eq:inter_res}), which has resolution $\sim\lambda/(2\sin\theta_{\text{rot}})$. As a result, anisotropy appears, as illustrated in Figure~\ref{fig:single_cross}. Since the argument of the Bessel function $J_0$ is $|\mb x-\mb y_i-(\mb y-\mb y_j)|$, the direction in which the argument is constant will be unaffected, while the orthogonal direction experiences the strongest decay. We note again that the single point migration is equivalent to taking $\mb x=\mb y$ with $\mb x-\mb y=0$ and so its resolution does not benefit from the synthetic aperture. 

Note that the assumption that $\mathcal{B}_{\text{eff}}$ has resolution of $\lambda \frac{H_\mb T}{a}$ completely independent of $\theta_{\text{rot}}$ is only approximate. As illustrated in Figure~\ref{fig:B_eff}, as $\theta_{\text{rot}}$ decreases the decay from the main lobe is slower, which affects the performance of the rank-1 image, as was shown in Section~\ref{sec:prop_filt}.
We have thus shown that we can approximate the form of $\tilde{\mb X}$ in \eqref{eq:M_tilde} as a collection of localized peaks, located at pairs of scatterer positions. When considering the extended domain of the interference pattern the peaks exhibit anisotropy with principal widths that are aligned with the directions $\mb x+\mb y=\text{constant}$, $\mb x-\mb y=\text{constant}$. 

The rank-1 image takes as an alternative image the top eigenvector of $\tilde{\mb X}$. In \cite{leibovich2020generalized} the top eigenfunction of kernels that have a similar anisotropic form were analyzed. It was shown that in general the resolution of the eigenfunction is better than the maximal width associated with single point migration. This explains the superior performance of the rank-1 image for rotating objects.

%------------------
\bibliographystyle{plain} \bibliography{SBIR_SAT}
% ---------------------

\end{document}